\documentclass[acmsmall,screen,nonacm,svgnames,x11names]{acmart}
\pdfoutput=1
\setcopyright{none}
\usepackage{booktabs}   
\usepackage{subcaption} 
                        
\usepackage{adjustbox}
\usepackage{cleveref}
\usepackage{amsmath,amsfonts}
\usepackage{csquotes}
\usepackage{marvosym}
\usepackage{hyperref}
\usepackage{graphicx}
\usepackage{nameref}
\usepackage{stmaryrd}
\usepackage{xcolor}
\usepackage{xfrac}
\usepackage{xspace}
\usepackage{cancel}
\usepackage{tikz-cd}
\usepackage{proof}
\usepackage{wrapfig}
\usepackage{thmtools}
\usepackage{thm-restate}
\usepackage{thm-autoref}
\usepackage{tabularx}
\usepackage{tabto}

\usepackage{caption}
\usepackage{listings,multicol}

\usepackage{tensor}
\usepackage{tikz}
\usetikzlibrary{patterns,decorations.pathreplacing,arrows,arrows.meta,decorations.pathmorphing,positioning,fit,trees,shapes,shadows,automata,calc}
\usepackage{pgfplots}

\usepackage{caption}
\usepackage{proof}
\usepackage{mathtools}
\usepackage[textwidth=13mm,disable]{todonotes}                        
\AtBeginDocument{\theoremstyle{acmdefinition}\newtheorem{remark}[theorem]{Remark}}

\makeatletter
\def\@acmplainindent{0pt}
\def\@acmdefinitionindent{0pt}
\def\@proofindent{\noindent}
\makeatother

\usepackage{enumitem}
\setlist{leftmargin=*}

\newcommand{\var}{\textsf{Var}}
\newcommand{\cov}{\textsf{Cov}}
\newcommand{\hyperpsi}{\psi\hspace*{-0.44em}\psi}
\newcommand{\hyperphi}{\phi\hspace*{-0.44em}\phi}
\newcommand{\hyperf}{f\hspace*{-0.39em}f}
\newcommand{\hyperg}{g\hspace*{-0.34em}g}
\newcommand{\hyperh}{h\hspace*{-0.25em}h}

\newcommand{\HL}{\textsf{HL}\xspace}
\newcommand{\IL}{\textsf{IL}\xspace}
\newcommand{\OL}{\textsf{OL}\xspace}
\newcommand{\HHL}{\textsf{HHL}\xspace}

\newcommand{\lzcomment}[1]{\todo[color=red!30,size=\scriptsize,fancyline,author=Linpeng]{#1}\xspace}

\newcommand{\lz}[1]{\lzcomment{#1}}

\newcommand{\noam}[1]{\todo[inline,color=blue!30,size=\footnotesize,fancyline,author={\textbf{Noam}}]{#1}\xspace}

\newcommand{\todoin}[1]{\todo[inline]{#1}}

\newcommand{\qedtriangle}{\hfill\raisebox{-.15ex}{\rotatebox{90}{$\triangle$}}}

\newcommand{\hhl}[3]{\models_{\textnormal{hh}}\left\{ \,{#1}\vphantom{#3}\, \right\} \mathrel{#2} \left\{ \, {#3}\vphantom{#1} \, \right\}}
\newcommand{\tcl}[3]{\models_{\textnormal{atc}}\left\{ \,{#1}\vphantom{#3}\, \right\} \mathrel{#2} \left\{ \, {#3}\vphantom{#1} \, \right\}}
\newcommand{\dtcl}[3]{\models_{\textnormal{dtc}}\left\{ \,{#1}\vphantom{#3}\, \right\} \mathrel{#2} \left\{ \, {#3}\vphantom{#1} \, \right\}}
\newcommand{\pcl}[3]{\models_{\textnormal{pc}}\left\{ \,{#1}\vphantom{#3}\, \right\} \mathrel{#2} \left\{ \, {#3}\vphantom{#1} \, \right\}}
\newcommand{\apcl}[3]{\models_{\textnormal{apc}}\left\{ \,{#1}\vphantom{#3}\, \right\} \mathrel{#2} \left\{ \, {#3}\vphantom{#1} \, \right\}}
\newcommand{\til}[3]{\models_{\textnormal{ti}}\left[ \,{#1}\vphantom{#3}\, \right] \mathrel{#2} \left[ \, {#3}\vphantom{#1} \, \right]}
\newcommand{\pil}[3]{\models_{\textnormal{pi}}\left[ \,{#1}\vphantom{#3}\, \right] \mathrel{#2} \left[ \, {#3}\vphantom{#1} \, \right]}

\newcommand{\anonl}[3]{\models\left\{ \,{#1}\vphantom{#3}\, \right\} \mathrel{#2} \left\{ \, {#3}\vphantom{#1} \, \right\}}

\newcommand{\sfsymbol}[1]{\textsf{\upshape {#1}}}

\newcommand{\ttsymbol}[1]{\texttt{\upshape {#1}}}

\newcommand{\whpsymbol}{\sfsymbol{whp}}
\newcommand{\boldwhpsymbol}{\textbf{\sfsymbol{whp}}}
\newcommand{\whp}[2]{\whpsymbol\,\llbracket#1\rrbracket\left(#2\right)}
\newcommand{\boldwhp}[2]{\boldwhpsymbol\,\boldsymbol{\llbracket#1\rrbracket\left(#2\right)}}
\newcommand{\whpC}[1]{\whpsymbol\llbracket#1\rrbracket}

\newcommand{\wpsymbol}{\sfsymbol{wp}}
\newcommand{\boldwpsymbol}{\textbf{\sfsymbol{wp}}}
\renewcommand{\wp}[2]{\wpsymbol\,\llbracket#1\rrbracket\left(#2\right)}
\newcommand{\boldwp}[2]{\boldwpsymbol\,\boldsymbol{\llbracket#1\rrbracket\left(#2\right)}}

\newcommand{\wpC}[1]{\wpsymbol\llbracket#1\rrbracket}

\newcommand{\suppsymbol}{\sfsymbol{supp}}
\newcommand{\supp}[1]{\suppsymbol\left(#1\right)}

\newcommand{\invsem}[2]{\ensuremath{\llbracket {#1} \rrbracket}^{{-}1}(#2)}

\newcommand{\spsymbol}{\sfsymbol{sp}}
\newcommand{\boldspsymbol}{\textbf{\sfsymbol{sp}}}
\renewcommand{\sp}[2]{\spsymbol\,\llbracket#1\rrbracket\left(#2\right)}
\newcommand{\boldsp}[2]{\boldspsymbol\,\boldsymbol{\llbracket#1\rrbracket\left(#2\right)}}
\newcommand{\spC}[1]{\spsymbol\llbracket#1\rrbracket}

\newcommand{\slpsymbol}{\sfsymbol{slp}}

\newcommand{\slp}[2]{\slpsymbol\llbracket#1\rrbracket\left(#2\right)}

\newcommand{\supwlpsymbol}{\sfsymbol{wlp}_{\sup}}

\newcommand{\supwlp}[2]{\supwlpsymbol\llbracket#1\rrbracket\left(#2\right)}

\newcommand{\infwpsymbol}{\sfsymbol{wp}_{\inf}}

\newcommand{\infwp}[2]{\infwpsymbol\llbracket#1\rrbracket\left(#2\right)}

\newcommand{\dwpsymbol}{\sfsymbol{dwp}}

\newcommand{\dwp}[2]{\dwpsymbol\llbracket#1\rrbracket\left(#2\right)}

\newcommand{\wlpsymbol}{\sfsymbol{wlp}}

\newcommand{\wlp}[2]{\wlpsymbol\llbracket#1\rrbracket\left(#2\right)}

\newcommand{\awlpsymbol}{\sfsymbol{awlp}}

\newcommand{\awlp}[2]{\awlpsymbol\llbracket#1\rrbracket\left(#2\right)}

\newcommand{\conditionalPair}[2]{{\let\oldarraystretch\arraystretch}\renewcommand{\arraystretch}{1}~\holter{~\raisebox{.5ex}{${#1}$}~}{~\raisebox{.125ex}{${#2}$}~}~\renewcommand{\arraystretch}{\oldarraystretch}}

\newcommand{\annotate}[1]{\boldsymbol{\annocolor{\!\fatslash\!\!\!\fatslash~~\vphantom{G'} {#1}}}}

\newcommand{\sndannotate}[1]{\boldsymbol{\annocolor{\hspace{1em} {#1}}}}

\newcommand{\eqannotate}[1]{\annocolor{\!\!\hspace{.55ex}{}^{\annocolor{{=}}}{\!\!\!{\fatslash}\!\!{\fatslash}~~\hspace{.5ex}\vphantom{G'} {\boldsymbol{#1}}}}}

\newcommand{\guard}{\ensuremath{\varphi}} 
\newcommand{\iguard}{\ensuremath{\iverson{\guard}}} 
\newcommand{\inguard}{\ensuremath{\iverson{\neg\guard}}} 
\newcommand{\ee}{\ensuremath{e}} 

\newcommand{\SKIP}{\ttsymbol{skip}}

\newcommand{\DIVERGE}{\ensuremath{\textnormal{\texttt{diverge}}}}

\newcommand{\ASSUME}[1]{\ensuremath{\textnormal{\texttt{assume}} ~ {#1}}}

\newcommand{\AssignSymbol}{\coloneqq}
\newcommand{\ASSIGN}[2]{\ensuremath{#1 \AssignSymbol #2}}
\newcommand{\ASSIGNNONDET}[1]{\ensuremath{#1 \AssignSymbol \textnormal{\texttt{nondet()}}}}

\newcommand{\AVAILLOC}[1]{\PosNats}
\usepackage{actuarialangle}

\newcommand{\COMPOSE}[2]{\ensuremath{{#1}\,{\fatsemi}\,~ {#2}}}
\newcommand{\PCHOICE}[3]{\ensuremath{\left\{\, {#1} \,\right\}\mathrel{\left[\,#2\,\right]}\left\{\, {#3} \,\right\}}}
\newcommand{\NDCHOICE}[2]{\ensuremath{\left\{\, {#1} \,\right\}\mathrel{\Box}\left\{\, {#2} \,\right\}}}

\newcommand{\STAR}[1]{\ensuremath{ {#1}^\star}}
\newcommand{\LOOP}[3]{\ensuremath{{#1}^{\langle {#2},{#3}\rangle}}}
\newcommand{\WEIGHT}[1]{\ensuremath{ {\odot}\,#1}}
\newcommand{\AWEIGHT}{\WEIGHT{a}}

\newcommand{\IFSYMBOL}{\ensuremath{\textnormal{\texttt{if}}}}
\newcommand{\IF}[1]{\ensuremath{\IFSYMBOL\,\left(\, {#1} \,\right)\,\{}}
\newcommand{\ELSESYMBOL}{\ensuremath{\textnormal{\texttt{else}}}}
\newcommand{\ELSE}{\ensuremath{\}\,\ELSESYMBOL\,\{}}
\newcommand{\ITE}[3]{\ensuremath{\IFSYMBOL\,\left(\, {#1} \,\right)\,\left\{\, {#2} \,\right\}\,\ELSESYMBOL\,\left\{\, {#3} \,\right\}}}

\newcommand{\WHILESYMBOL}{\ensuremath{\textnormal{\texttt{while}}}}

\newcommand{\WHILEDO}[2]{\ensuremath{\WHILESYMBOL \left(\, {#1} \,\right)\left\{\, {#2} \,\right\}}}

\newcommand{\wrcl}{\textnormal{\sfsymbol{wReg}}\xspace}   
\newcommand{\rcl}{\textnormal{\sfsymbol{Reg}}\xspace}   
\newcommand{\Vars}{\ensuremath{\mathsf{Vars}}\xspace}   

\newcommand{\R}{\ensuremath{\mathbb{R}}}
\newcommand{\Nats}{\ensuremath{\mathbb{N}}\xspace}
\newcommand{\PosNats}{\ensuremath{\mathbb{N}_{>0}}\xspace}

\newcommand{\Bool}{\textsf{Bool}}
\newcommand{\Probs}{\textsf{Prob}}
\newcommand{\Tropical}{\textsf{Tropical}}
\newcommand{\Reals}{\mathbb{R}}
\newcommand{\PosReals}{\mathbb{R}_{\geq 0}}

\newcommand{\PosRealsInf}{\mathbb{R}_{\geq 0}^\infty}
\newcommand{\minfty}{{-}\infty}
\newcommand{\pinfty}{{+}\infty}
\newcommand{\Rinf}{\ensuremath{\R^{{\pm}\infty}}}

\newcommand{\E}{\mathbb{E}}
\newcommand{\A}{\mathbb{A}}
\newcommand{\B}{\mathbb{B}}
\newcommand{\hyperB}{\mathbb{B}\hspace*{-.48ex}\mathbb{B}}
\newcommand{\hyperA}{\mathbb{A}\hspace*{-.55ex}\mathbb{A}}

\newcommand{\iverson}[1]{\left[ {#1} \right]}

\newcommand{\iversone}[1]{\boldsymbol{1}_{#1}}

\newcommand{\indicator}[1]{\left[ {#1} \right]}

\newcommand{\subst}[2]{\left[ {#1} \middle/ {#2}\right]}
\newcommand{\statesubst}[2]{\left[ {#1} \mapsto {#2}\right]}

\DeclareMathAlphabet{\dsser}{U}{DSSerif}{m}{n}
\newcommand{\bb}[1]{\dsser{#1}}
\newcommand{\bone}{\bb{1}}
\newcommand{\bzero}{\bb{0}}

\newcommand{\eval}[1]{\ensuremath{\llbracket {#1} \rrbracket}}

\newcommand{\sem}[2]{\ensuremath{\llbracket {#1} \rrbracket}(#2)}

\newcommand{\States}{\Sigma}

\newcommand{\true}{\mathsf{true}}
\newcommand{\false}{\mathsf{false}}
\newcommand{\mydot}{\text{{\Large\textbf{.}}~}}

\newcommand{\qiff}{\quad\textnormal{iff}\quad}
\newcommand{\qqiff}{\qquad\textnormal{iff}\qquad}

\newcommand{\qand}{\quad\textnormal{and}\quad}
\newcommand{\qqand}{\qquad\textnormal{and}\qquad}

\newcommand{\qqimplies}{\qquad\textnormal{implies}\qquad}

\newcommand{\morespace}[1]{~{}#1{}~}
\newcommand{\qmorespace}[1]{\quad{}#1{}\quad}

\newcommand{\lland}{\morespace{\land}}
\newcommand{\ppreceq}{~{}\preceq{}~}

\newcommand{\defeq}{\triangleq}
\newcommand{\ddefeq}{\morespace{\defeq}}
\newcommand{\qdefeq}{\qmorespace{\defeq}}
\newcommand{\eeq}{~{}={}~}

\newcommand{\ooplus}{\morespace{\oplus}}
\newcommand{\oodot}{\morespace{\odot}}

\newcommand{\ccurlyvee}{~{}\curlyvee{}~}
\newcommand{\ccurlywedge}{~{}\curlywedge{}~}

\newcommand{\qeq}{\quad{}={}\quad}
\newcommand{\qqeq}{\qquad{}={}\qquad}
\newcommand{\lleq}{~{}\leq{}~}

\newcommand{\iimplies}{~{}\implies{}~}
\newcommand{\mmodels}{~{}\models{}~}

\newcommand{\setcomp}[2]{\left\{\, {#1} ~\middle|~ {#2} \,\right\}}

\definecolor{webgreen}{rgb}{0,.5,0}

\newcommand{\blue}[1]{\textcolor{DodgerBlue3}{#1}}

\newcommand{\orange}[1]{\textcolor{orange}{#1}}

\newcommand{\annocolor}[1]{\textcolor{webgreen}{#1}}

\newcounter{computationarrowsone}
\newcounter{computationarrowstwo}

\newcounter{sarrow}

\newcommand{\lfp}{\ensuremath{\textnormal{\sfsymbol{lfp}}~}}

\newcommand{\powerset}[1]{\ensuremath{\mathcal{P}(#1)}}

\newcommand{\Diverge}{\ensuremath{{\Uparrow}}}

\newcommand{\Term}{\ensuremath{\Downarrow}}

\begin{document}

\title{Quantitative Weakest Hyper Pre: Unifying Correctness and Incorrectness Hyperproperties via Predicate Transformers}

\author{Linpeng Zhang}
\orcid{0000-0002-1485-327X}
\affiliation{%
    \institution{University College London}
    \city{}
    \country{UK}
}
\email{linpeng.zhang.20@ucl.ac.uk}

\author{Noam Zilberstein}
\email{noamz@cs.cornell.edu}
\orcid{0000-0001-6388-063X}
\affiliation{%
  \institution{Cornell University}
  \country{USA}
}

\author{Benjamin Lucien Kaminski}
\orcid{0000-0001-5185-2324}
\affiliation{%
    \institution{Saarland University}
    \city{}
    \country{Germany}
}
\affiliation{%
    \institution{University College London}
    \city{}
    \country{UK}
}
\email{kaminski@cs.uni-saarland.de}

\author{Alexandra Silva}
\email{alexandra.silva@cornell.edu}
\orcid{0000-0001-5014-9784}
\affiliation{%
  \institution{Cornell University}
  \country{USA}
}


\begin{abstract}
  We present a novel \emph{weakest pre calculus} for \emph{reasoning about quantitative hyperproperties} over \emph{nondeterministic and probabilistic} programs. 
  Whereas existing calculi allow reasoning about the expected value that a quantity assumes after program termination from a \emph{single initial state}, we do so for \emph{initial sets of states} or \emph{initial probability distributions}.
  We thus (i)~obtain a weakest pre calculus for hyper Hoare logic and (ii)~enable reasoning about so-called \emph{hyperquantities} which include expected values but also quantities (e.g. variance) out of scope of previous work. As a byproduct, we obtain a novel strongest post for weighted programs that extends both existing strongest and strongest liberal post calculi. Our framework reveals novel dualities between forward and backward transformers, correctness and incorrectness, as well as nontermination and unreachability.
\end{abstract}

\maketitle
\NumTabs{22}%

\section{Introduction}

 \emph{Hoare Logic} (\HL) \cite{Hoa69} is a proof system for establishing \emph{partial correctness} of programs---properties of {\em individual executions} that will always hold \emph{\underline{i}f} the program terminates. However, certain properties---e.g., establishing that a system is secure via confidentiality, integrity, or authenticity---cannot be expressed in terms of \emph{individual} executions and are therefore beyond the scope of  classical Hoare Logic. This is because attackers may compare several different  traces to infer hidden secrets. \citet{hyperproperties} gave characterizations for this richer class of behaviors, calling them \emph{hyperproperties}. To overcome this limitation of Hoare Logic, \citet{benton2004simple} proposed a {\em relational} extension of Hoare Logic for reasoning about multiple executions and verifying hyperproperties.
 
 The common element of Hoare Logic and its relational counterparts is that they apply only to properties over \emph{all} executions (in the case of relational logics, all pairs of executions). 
 \citet{OHearn19} refers to this class of logics as {\em overapproximate} and argues that it hinders their application in establishing the presence of bugs, advocating for the development of a new generation of program logics that focus on bug-finding. 
 \citet{OHearn19} proposed \emph{Incorrectness Logic} (\IL) (independently proposed by \citet{VK11} under the name \emph{reverse Hoare logic}) as an analogue of Hoare Logic for developing the formal theory of bug-finding. Subsequently, other similar logics and extensions of IL were proposed \cite{IncorrectnessSeparationLogic,Moller21}.
IL can witness the reachability of particular bad outcomes but cannot make guarantees about all the possible outcomes.

The aforementioned theories of incorrectness diverge significantly from theories of correctness (such as \HL), meaning that entirely separate analysis algorithms must be used for verification vs bug-finding. 
To overcome this limitation, new theories for \emph{unified} reasoning about both correctness and incorrectness have been proposed \cite{Bruni2021ALF,maksimovi_c2023exact,Zilberstein2023,Dardinier2023hyper,zilberstein2024outcome,zilberstein2024relatively}. These include logics not only for individual program traces but also on hyperproperties \cite{Dardinier2023hyper}. 

We build on two such developments---Outcome Logic (\OL) \cite{Zilberstein2023,zilberstein2024outcome,zilberstein2024relatively} and Hyper Hoare Logic (\HHL) \cite{Dardinier2023hyper}---which advocate that a single logic can be used to prove (or disprove) a wide variety of properties, including hyperproperties, and we present a novel \emph{(quantitative) weakest pre calculus} perspective. 
Weakest precondition calculi date back to the 1970's when \citet{DBLP:journals/cacm/Dijkstra75,Dijkstra76} introduced them as predicate transformer semantics for imperative programs.
Given a command $C$ and a postcondition $Q$, the \emph{weakest liberal precondition} is the weakest assertion $P$ such that running $C$ in any state satisfying $P$ will terminate in a state satisfying $Q$ or not terminate at all.
\citet{pratt1976semantical} observed that these calculi have a close connection to Hoare Logic and they were later used in a completeness proof for Hoare Logic \cite{clarke1979programming}.\footnote{Although the original relative completeness proof of \citet{Cook1978} used the \emph{strongest postcondition}, a later, simplified proof by \citet{clarke1979programming} used the weakest liberal precondition.}

 Weakest liberal preconditions have been generalized to probabilistic programs to allow for reasoning about expected values of random variables in a program that terminates from a \emph{single initial state}.  
 The core idea in these quantitative calculi~\cite{DBLP:journals/jcss/Kozen85,McIverM05,benni_diss,ZK22} is that one can replace predicates over states by real-valued functions. 
 All these calculi, classical and quantitative, offer predicate transformers that have two key benefits over program logics: 
 First, they discover the \emph{most precise} assertions to make a triple valid. Second, they provide a calculus with a clear path towards mechanizability. 

In this paper, we present a novel \emph{weakest pre calculus} ($\whpsymbol$) for \emph{reasoning about quantitative hyperproperties} over \emph{programs with effects} that cause the program execution to branch such as nondeterminism or probabilistic choice, in the style of weighted programming \cite{batz2022weighted} or \OL \cite{zilberstein2024relatively} (\Cref{se:semantics}).
We generalize existing work on quantitative weakest pre calculi \cite{ZK22} by considering program termination from \emph{initial sets of states} or \emph{initial probability distributions} rather than single initial states. 
We thus obtain weakest preconditions for \HHL and enable reasoning about so-called \emph{hyperquantities} (\Cref{sec:hyperpre}),
which include expected values (considered in previous work), but also more general quantities that were not supported before, e.g. variance. 
Unlike Hyper Hoare Logic, our $\whpsymbol$ supports quantitative probabilistic reasoning, employing \emph{hyperquantities} evaluated in probability distributions. Moreover, we show that many existing logics are subsumed by $\whpsymbol$ (\Cref{sec:express}), and how to prove (and disprove) properties in those logics. 
$\whpsymbol$ is hence a single calculus for correctness and incorrectness analysis, which enjoys expected healthiness and duality properties (\Cref{se:healthiness}).
$\whpsymbol$ can be applied in a variety of settings, which we illustrate through a range of examples (\Cref{sec:case-studies}).

 Similarly to how predicate transformers and Hoare-like logics empower programmers to demonstrate correctness, we contend that our framework offers researchers a deeper comprehension of existing logics. Our calculus reveals novel dualities between forward and backward transformers, correctness and incorrectness, as well as nontermination and unreachability.

\section{Overview: Strategies for Reasoning about Hyperproperties}
\label{se:weakest-pre}

We begin our discussion by focusing on noninterference \cite{goguen1982security}---a hyperproperty commonly used in information security applications. More precisely, noninterference stipulates that any two executions of a program with the same \emph{public} inputs (but potentially different \emph{secret} inputs) must have the same public outputs. This guarantees that the program does not \emph{leak} any secret information to unprivileged observers.
As a demonstration, consider the following program, where the variable $\ell$ (for low) is publicly visible, but $h$ (for high) is secret.%
\begin{align*}
	C_{\text{ni}}
	\eeq
	\COMPOSE{\ASSUME{\mathit{h}>0}}{\ASSIGN{\ell}{\ell+\mathit{h}}}
\end{align*}%
 Suppose we aim to prove $C_{\text{ni}}$ satisfies noninterference. Following the approach of logics such as Hyper Hoare Logic (\HHL), one can define $\text{low}(\ell)$ to mean that the value of $\ell$ is equal in any pair of executions, and then attempt to establish the validity of $\hhl{\text{low}(\ell)}{C_{\text{ni}}}{\text{low}(\ell)}$, meaning that if $C_{\text{ni}}$ is executed twice with the same initial $\ell$, then $\ell$ will also have the same value in both executions when (and if) the program finishes---hence, the initial values of $h$ cannot influence $\ell$.
 
\HHL is sound and complete, meaning that any true triples can be proven in it. However, doing so is not always straightforward. For example, although the specification of the triple above does not mention $h$, intermediary assertions required to complete the proof \emph{must} mention $h$, and introducing this information cannot be done in a mechanical way, but rather requires inventiveness.

Furthermore, whereas \HHL (analogously to \OL) can disprove any of its triples~\cite[Theorem 4]{Dardinier2023hyper}, deriving either a positive or negative result---i.e., proving that a program is secure or not---requires one to know a priori which spec they wish to prove, or trying both.
 
The predicate transformer approach we advocate in this paper proves highly advantageous as it only requires a \emph{single} \emph{hyperpostcondition} to determine the most precise \emph{hyperprecondition} that validates (or invalidates) a triple. In that sense, it solves the two aforementioned issues by mechanically working backward from the postcondition, \emph{discovering} intermediary assertions along the way, and finding the \emph{most precise} precondition with respect to the desired spec.

In this paper, we define a novel $\whpsymbol$ calculus, and the validity of $\text{low}(\ell)\subseteq \whp{C_{\text{ni}}}{\text{low}(\ell)}$ is the answer to the noninterference problem, without the risk of attempting to prove an invalid triple. In the case of the above example, our calculus leads us to a simple counterexample; if we have $\ell = 0$ and $h = 1$ in the first execution and $\ell = 0$ and $h=2$ in the second execution, then clearly $\text{low}(\ell)$ holds, but the values of $\ell$ will be distinguishable at the end. This means that the program is insecure.
In the remainder of this section, we will give an overview of the technical ideas underlying our $\whpsymbol$ calculus.


\subsection{Classical Weakest Pre}

Dijkstra's original weakest precondition calculus employs \emph{predicate transformers} of type%
\begin{align*}
	\wpC{C}\colon\quad \B \morespace{\to} \B~, \qquad \textnormal{where}\quad \B \eeq \States\to\{0, 1\}~.
\end{align*}%
The set $\B$ of maps from program states ($\States$) to Booleans~($\{0, 1\}$) can also be thought of as predicates or assertions over program states.
The \emph{angelic} weakest precondition transformer $\wpC{C}$ maps a {post}{\-}condition~$\psi$ to a {pre}con{\-}dition $\wp{C}{\psi}$ such that executing $C$ on an initial state in $\wp{C}{\psi}$ guarantees that $C$~\emph{can}\footnote{$C$ is a \emph{nondeterministic} program. For the \emph{demonic} setting and for deterministic programs, we can replace \enquote{can} by \enquote{will}.}~terminate in a final state in~$\psi$.
Given a semantics function $\eval{C}$ such that $\sem{C}{\sigma,\, \tau}=1$ iff executing $C$ on initial state $\sigma$ can terminate in $\tau$, the angelic $\wpsymbol$ is so defined:
\begin{align*}
		\wp{C}{\psi}\qeq \{\sigma\in\States \mid\exists\tau\mydot \sem{C}{\sigma, \tau}=1 \lland \tau\in\psi\}
\end{align*}
This allows to check if an angelic total correctness triple holds via the well-known fact
\begin{align*}
		\tcl{G}{C}{F} \text{ is valid for angelic total correctness} \qqiff G \implies \wp{C}{F}~.	
\end{align*}
While the above is a set perspective on $\wpsymbol$, an equivalent perspective on $\wpsymbol$ is a map perspective: the predicate $\wp{C}{\psi}$ is a map that takes as input an initial state $\sigma$, determines for each reachable final state $\tau$ the (truth) value $\psi(\tau)$, takes a disjunction over all these truth values, and finally returns the truth value of that disjunction.
More symbolically,%
%
\begin{align*}
	\wp{C}{\psi}(\sigma) \qqeq\quad \smash{\bigvee_{\mathclap{\tau\colon\sem{C}{\sigma,\tau}=1}} \quad \psi(\tau)}~.
\end{align*}%
\normalsize

\subsection{Weakest Pre over Hyperproperties}
\label{se:intro:hyper}
To reason about hyperproperties~\cite{hyperproperties}, we lift our domain of discourse from \emph{sets of states} to \emph{sets of sets of states}, i.e.\ we go
\begin{align*}
	\text{from}\qquad\wpC{C}\colon \quad \B \morespace{\to} \B \qquad\text{to}\qquad\whpC{C}\colon \quad \hyperB \morespace{\to} \hyperB~,
\end{align*}%
where $\B \eeq \States\to\{0, 1\}$, as before, and $\hyperB\eeq \powerset{\States}\to\{0, 1\}$.

Given a postcondition $\psi \in \B$ (i.e.\ a predicate ranging over states), classical angelic $\wp{C}{\psi}$ anticipates for a \emph{single} initial state $\sigma$ whether running $C$ on $\sigma$ can reach~$\psi$.
Given a hyper{\-}post{\-}condition~$\hyperpsi \in \hyperB$ (a predicate ranging over \emph{sets} of states), the weakest hyperprecondition~$\whp{C}{\hyperpsi}$ anticipates for a given \emph{set} of initial states $\phi$ (a precondition), whether the set of states reachable from executing $C$ on every state in $\phi$ satisfies $\hyperpsi$. 
From a set perspective, we have:
\begin{align*}
	\whp{C}{\hyperpsi}\qeq \{\phi\in\powerset{\States} \mid \sp{C}{\phi} \in \hyperpsi \}~,
\end{align*}%
where $\sp{C}{\phi}$ is the classical \emph{strongest postcondition}~\cite{Dijkstra1990} of $C$ with respect to precondition $\phi$; in other words: the set of all final states \emph{reachable} by executing $C$ on any initial state in $\phi$.
From a map perspective, $\whp{C}{\hyperpsi}$ maps a hyperproperty~$\hyperpsi$ over postconditions to a hyperproperty~$\whp{C}{\hyperpsi}$ over preconditions. In other words, we are anticipating whether the strongest postcondition of $\phi$ satisfies the hyperpostcondition $\hyperpsi$:
\begin{align*}
	\whp{C}{\hyperpsi}(\phi) \qeq \hyperpsi(\sp{C}{\phi})~.
\end{align*}%
In particular, executing $C$ on a \emph{precondition} $\phi$ satisfying $\whp{C}{\hyperpsi}$ guarantees that the set of reachable states $\sp{C}{\phi}$ will satisfy $\hyperpsi$.
Reasoning about hyperproperties is strictly more expressive as it relates multiple executions. 
We showcase this in the following examples.%
\begin{example}[Weakest Hyperpreconditions]
\label{ex:weakest-hyperpreconditions}
	Given some precondition $\phi$, if $\phi$ satisfies
	\begin{enumerate}
		\item $\whp{C}{\lambda \rho.~|\rho| = 2}$, then the number of states reachable from $\phi$ by executing $C$ is 2.
		\item $\whp{C}{\lambda \rho.~ \text{Bugs} \subseteq \rho}$, where $\text{Bugs} \subseteq \Sigma$, then \emph{all} states in the set Bugs are reachable by running $C$ on \emph{some} state in $\phi$ (this amounts to Incorrectness Logic~\cite{OHearn19}).
		\item $\whp{C}{\lambda \rho.~ \rho\subseteq \text{Good}}$, where $\text{Good} \subseteq \Sigma$, then starting from $\phi$ only Good can be reached or $C$ does not terminate (this amounts to partial correctness~\cite{Hoa69}).
	\end{enumerate}
	We refer to~\citet{hyperproperties} for more examples of hyperproperties.%
	\qedtriangle%
\end{example}%
\begin{remark}
	Outcome Logic~\cite{Zilberstein2023} and Hyper Hoare Logic~\cite{Dardinier2023hyper} can handle all of \Cref{ex:weakest-hyperpreconditions} via $\anonl{\hyperphi}{C}{\hyperpsi}$ triples, but are agnostic of pre{\-}conditions not satisfying $\hyperphi$ since $\phi\not\in \hyperphi$ does not imply $\sp{C}{\phi} \not\in \hyperpsi$. 
	Predicate transformers, on the other hand, yield the most precise assertions in the sense that $\phi\in\whp{F}{\hyperpsi}$ iff $\sp{C}{\phi}\in \hyperpsi$.%
	\qedtriangle%
\end{remark}%
%
%

\subsection{Quantitative Reasoning over Hyperproperties}
As shown in~\cite{DBLP:journals/jcss/Kozen85,McIverM05,benni_diss}, one can replace predicates over states by real-valued functions, also known as quantities~\cite[Section 3]{ZK22}. 
These quantitative calculi subsume the classical ones by mimicking predicates through the use of \emph{Iverson brackets}~\cite{Knuth1992}.
To design a calculus for quantitative reasoning over hyperproperties, we lift quantities in $\A = \setcomp{f}{f \colon \States \to\PosRealsInf}$, i.e. functions of type $\States \to \PosRealsInf$, to hyperquantities.

\begin{definition}[Hyperquantities]
	\label{def:quantities}
		The set of all \emph{hyper}quan{\-}tities is defined by%
		\begin{align*}
			\hyperA \eeq \setcomp{\hyperf}{\hyperf \colon (\States \to\PosRealsInf)\to \PosRealsInf}~,
		\end{align*}%
		$\hyperA$ is the set of all functions $\hyperf \colon \A\to \PosRealsInf$ associating an \emph{extended real} (i.e.\ either a non-negative real number or~$\pinfty$) to each quantity in $ \A$.
		The point-wise order%
		\begin{align*}
			\hyperf \ppreceq \hyperg &\qqiff \forall\, f \in \A\colon \quad \hyperf(f) \lleq \hyperg(f)
		\end{align*}%
		renders $\langle \hyperA,\, {\preceq}\rangle$ a complete lattice with join $\curlyvee$ and meet $\curlywedge$, given point-wise by%
		\begin{align*}
			\hyperf \curlyvee \hyperg \eeq \lambda f\mydot \max\bigl\{\hyperf(f),\, \hyperg(f) \bigr\}
			\qqand
			\hyperf \curlywedge \hyperg \eeq \lambda f\mydot \min \bigl\{ \hyperf(f),\, \hyperg(f) \bigr\}
			 ~.
		\end{align*}%
		Joins and meets over arbitrary subsets exist. 
		For $a \curlyvee b \curlywedge c$, we assume that $\curlywedge$ binds stronger.
		\qedtriangle%
\end{definition}%
%
%
%
%
\noindent%
Hyperquantities enable \emph{quantitative} reasoning, e.g., measures over probability distributions.%
%
%
%
\begin{example}[Hyperquantities over Distributions]
	\label{ex:hyperquantities-distributions}
	Given a quantity $f\colon\States \to \PosRealsInf \in \A$ (think: random variable $f$), we define hyperquantities
	\[
		\E [f] \defeq \lambda \mu.\sum_\sigma f(\sigma)\cdot\mu(\sigma)\quad 
		\cov[f,g] \defeq \lambda \mu.~ \E[fg](\mu) - \E [f](\mu)\cdot\E [g](\mu) \quad	
		\var[f] \defeq \cov[f,f] 
	\]
	that take as input quantities (interpreted as probability distributions) $\mu \colon \States \to \PosRealsInf$. 
	The above hyperquantities are then respectively \emph{expected value}, \emph{variance} and \emph{covariance} of $f$ (and $g$) over $\mu$.%
	\qedtriangle%
\end{example}%
\noindent%
We now present as an example an adaptation of~\cite[Example 3]{Dardinier2023hyper} -- showcasing how Boolean Hyper Hoare Logic (\HHL) would deal with statistical properties.%
\begin{example}[Mean Number of Requests]
	\label{ex:meandb}
	Consider a program $C_{\text{db}}$ where after termination the variable $n$ represents the number of database requests performed. For a final set of states $\rho\subseteq\States$, we define its mean number of requests by $\text{mean}_n(\rho)= \sum_{\sigma\in\rho} \frac{\sigma(n)}{|\rho|}$.
	
	%
	%
	\HHL allows to bound $\text{mean}_n$ by a \emph{specific number}, say $2$, by taking as hyperpostcondition $Q=\lambda \rho. \text{mean}_n(\rho) \leq 2$.
	Proving the \HHL triple $\hhl{\true}{C_{\text{db}}}{Q}$ then ensures that for every 
	initial set of states, the \emph{mean number} of performed requests after the execution of $C_{\text{db}}$ is at most $2$.%
	\qedtriangle%
\end{example}%
\begin{example}[Quantitative Information Flow]
	\label{ex:hintqif}
	Consider a program, $C_{\text{qif}}$ containing lowly and highly sensitive variables. As outlined in~\cite[Section 8.1]{ZK22}, we will demonstrate in~\Cref{ex:qif}, how our framework also enables to determine, for instance, the maximum initial value allowable for the secret variable $h$ based on observing a specific final value for $l$. \HHL allows reasoning only about the existence of some information flow or about a bound over $h$.
\end{example}%
%

\noindent%
Using instead quantitative weakest hyper pre has two main advantages over using \HHL:

\paragraph*{Beyond Decision Problems}
While \HHL and Outcome Logic (\OL) are capable of statistical reasoning, our quantitative calculus can directly \emph{measure} quantities of interest, such as the information flow. 

\paragraph*{Probability Distributions}
Reasoning about means is restrictive, especially for infinite sets. 
As shown in~\Cref{ex:hyperquantities-distributions}, hyperquantities assign numerical values such as expected values to distributions. 
For example, $\whp{C_{\text{db}}}{\E[n]}(\mu)$ maps \emph{every distribution} $\mu$ to the \emph{expected number of requests} after executing $C_{\text{db}}$ on some initial state drawn from $\mu$.

\section{Syntax and Semantics}
\label{se:semantics}
%
We introduce a language of commands $\wrcl$, which encompasses nondeterministic imperative constructs similar to those found in the Guarded Command Language \cite{Dijkstra76}. 
Furthermore, we adopt the weighting assertion as in \cite{batz2022weighted,zilberstein2024relatively}, which enables representation of general weights over states. This includes reasoning of expected values over probability distributions, as studied in~\cite{benni_diss,McIverM05}.

\subsection{Algebraic Preliminaries for Weights}
We begin by reviewing some algebraic structures, starting with the weights of computation traces.%
\begin{definition}[Naturally Ordered Semirings]
	\label{def:natural-order}
	A \emph{monoid} $\langle U,\, {\oplus},\, \bzero \rangle$ consists of a set $U$, an associative binary operation $\oplus\colon U \times U \to U$, and an identity element $\bzero \in U$ (with $u \oplus \bzero = \bzero \oplus u = u$). 
	The monoid is \emph{partial} if $\oplus\colon U \times U \rightharpoonup U$ is partial, and \emph{commutative} if $\oplus$ is commutative (i.e.\ $u \oplus v = v \oplus u$).%

	A \emph{semiring} $\langle U,\, {\oplus},\, {\odot},\, \bzero,\, \bone \rangle$ is an algebraic structure such that \mbox{$\langle U,\, {\oplus},\, \bzero \rangle$} is a commutative monoid, $\langle U,\, {\odot},\, \bone \rangle$ is a monoid, and the following additional properties hold:%
	\begin{enumerate}
		\item 
			Distributivity: \tab$u \oodot (v \oplus w) \eeq u \odot v \ooplus u \odot w$ \qand $(u \oplus v) \oodot w \eeq u \odot w \ooplus v \odot w$
		\item 
			Annihilation: \tab$\bzero \odot u \eeq u \odot \bzero \eeq \bzero$%
	\end{enumerate}%
	The semiring is \emph{partial} if $\langle U,\, \oplus,\, \bzero \rangle$ is a partial monoid (but $\odot$ is total).%
%

	On a (partial) semiring $\langle U , \oplus, \odot, \bzero, \bone\rangle$, we define a relation $\leq$ by
	%
		$u \leq v$ iff $\exists w \mydot u \oplus w = v$.	
	%
	The semiring is called \emph{naturally ordered} if $\leq$ is a complete partial order.\qedtriangle%
\end{definition}%
\noindent%
As shown later in~\Cref{def:semantics}, semirings will serve as the structure from which we draw weights of computation traces in our semantics. 
To this end, we extend the definition of quantities~\cite[Definition 3.1]{ZK22} to any semiring, similar to~\citet[Definition 2.3]{zilberstein2024relatively}.
\begin{definition}[Quantities]
	\label{def:weightedquantities}
	Given a partial semiring $\mathcal{A} = \langle U,\, {\oplus},\, {\odot},\, \bzero,\, \bone \rangle$, the set $\A_\mathcal{A}(X)$ of all \emph{quantities} is defined as the set of all functions $f\colon X \to U$, i.e.\
		\begin{align*}
			\A_\mathcal{A}(X) \eeq \setcomp{f}{f\colon X \to U}
		\end{align*}%
\end{definition}%
\noindent%
We will write $\A$ instead of $\A_\mathcal{A}(X)$ when $\mathcal{A}$ and $X$ are clear from context. 
Semiring addition, scalar multiplication, and constants are lifted pointwise to quantities as follows:%
\begin{align*}
	(m_1 \oplus m_2)(x) \ddefeq m_1(x) \oplus m_2(x), \qquad (u \odot m)(x) \ddefeq u \odot m(x), \qqand u(x) \ddefeq u
\end{align*}%
For example, by taking $X$ as the set of program states $\States$ and the semiring $\langle \Rinf,\, {\max},\, {\min},\, \minfty,\, \pinfty\rangle$ one can represent the quantities of \citet[Definition 3.1]{ZK22}. 
Other instances of semirings encode other computations.
For example:%
\begin{itemize}
	\item Nondeterministic computation employs the Boolean semiring $\Bool = \langle \{0,1\},\, {\lor},\, {\land},\, 0,\, 1\rangle$.
	\item Randomization adopts probabilities in the partial semiring $\Probs = \langle [0, 1],\, {+},\, {\cdot},\, 0,\, 1\rangle$, where $x+y$ is undefined if $x + y > 1$.
	\item Optimization problems (e.g., the path with minimum weight) can be encoded via the tropical semiring $\Tropical = \langle [0, \pinfty],\, {\min},\, {+},\, \pinfty,\, 0\rangle$ which utilises non-negative real-valued weights with minimum and addition operations.
\end{itemize}%
We refer to~\cite[Table 1]{batz2022weighted},~\cite[Section 2]{zilberstein2024relatively} for more examples and details.

\subsection{Program States and Quantities}

A state~$\sigma$ is a function that assigns a natural-numbered value to each variable. 
To ensure that the set of states is countable, we restrict to a finite set of program variables $\Vars$.
The set of program states is given by %
	$\States = \setcomp{\sigma}{\sigma\colon\Vars\to\Nats}$.
%
The semantics of an arithmetic, boolean or weight expression~$\ee$ is denoted by $\eval{e}\colon \States \to \Nats \cup U$ and is obtained in a state $\sigma$, by evaluating $\ee$ after replacing all occurrences of variables~$x$ by~$\sigma(x)$.
Moreover, we denote by $\sigma\subst{x}{v}$ a new state obtained from $\sigma$ by setting the valuation of~$x\in\Vars$ to~$v\in\Reals$. 
Formally: 
$\sigma\subst{x}{v}(y) = v$, if $y = x$; and $\sigma(y)$, otherwise.
%
%

A particular useful quantity is the Iverson bracket~\cite{Knuth1992}: denoted as $\iverson{\guard}$ for a given predicate $\guard$, it takes as input a state $\sigma$ and evaluates to $1$ if the statement is true and $0$ if the statement is false. We generalise it to arbitrary semirings, subsuming other quantitative generalisations such as~\cite[Definition 3.5]{ZK22}.

\begin{definition}[Iverson Brackets]
	\label{def:iverson}
		For any semiring $\mathcal{A}=\langle U,\, {\oplus},\, {\odot},\, \bzero,\, \bone \rangle$ and a predicate $\varphi$ over program states $\States$, the Iverson bracket $\iverson{\guard}\colon \States\to U$ is defined as%
		\begin{align*}
			\iverson{\guard}(\sigma) \ddefeq 
			\Bigl\{
				\bone, \quad \text{if } \sigma \mmodels \guard; \qqand
				\bzero, \quad\text{otherwise~.}
			\tag*{{\qedtriangle}}
		\end{align*}%
\end{definition}%
%
%
%

\subsection{Weighted Programs}
Throughout the paper, we denote $\mathcal{A} = \langle U,\, {\oplus},\, {\odot},\, \bzero,\, \bone\rangle$ as a naturally ordered, complete, Scott continuous, partial semiring with a top element $\top\in U$ such that $\top\geq u$ for all $u\in U$. We assign meaning to \wrcl-statements in terms of a denotational semantics, taking as input an \emph{initial state} $\sigma$ and a \emph{final state} $\tau$, and returning the sum of the weights of all paths starting from $\sigma$ and terminating in $\tau$ after the execution of $C$. 
The syntax of the \emph{weighted regular command language} ($\wrcl$) is below:%
\begin{align*}
    	\begin{array}{r@{\quad}l@{\quad}r@{\quad}l@{\quad}r@{\quad}l@{\quad}r}
		C \quad{}\Coloneqq{}	& \ASSIGN{x}{\ee}			&\textnormal{(assignment)}	&{}\mid{}~ \ASSIGNNONDET{x}	&\textnormal{(nondet.~assign.)}	&{}\mid{}~ \WEIGHT{e}		&\textnormal{(weighting)}		\\
							&{}\mid{}~ \COMPOSE{C}{C}	&\textnormal{(sequencing)}	&{}\mid{}~ \NDCHOICE{C}{C}		&\textnormal{(nondet.~choice)}		&{}\mid{}~ \LOOP{C}{e}{e'}	&\textnormal{(iteration)}		\\
	\end{array}
\end{align*}%
where $\WEIGHT{e}$ weights the current computation branch. 
Similarly to~\cite{ZK22,batz2022weighted}, we do not provide an explicit syntax for weights because we focus on semantic assertions.
Our weighting construct is more expressive than~\citet{batz2022weighted, zilberstein2024relatively}: not only we can represent values $u\in U$ and Boolean tests (via Iverson brackets), but we also reason about intensional properties of the computation.
The iteration $\LOOP{C}{e}{e'}$, introduced in~\cite{zilberstein2024relatively}, terminates with weight $e'$ or executes the body $C$ with weight $e$. This construct simplifies the representation of while loops with $\WHILEDO{\guard}{C}$, probabilistic iterations using $\LOOP{C}{p}{1-p}$, and Kleene's star as $\LOOP{C}{\bone}{\bone}$. Its usefulness is evident, especially in partial semirings where loops via Kleene star may not be well-defined due to its nondeterministic nature~\cite[Footnote~2]{zilberstein2024relatively}.
Many common constructs, such as tests, branchings and loops are syntactic sugar, for instance:%
\begin{align*} 
	\ASSUME{\guard} &\ddefeq \WEIGHT{\guard}\qquad \quad\ 
	\DIVERGE \ddefeq \WEIGHT{\bzero}\\
	\ITE{\guard}{C_1}{C_2} &\ddefeq \NDCHOICE{\COMPOSE{\ASSUME{\guard}}{C_1}}{\COMPOSE{\ASSUME{\neg\guard}}{C_2}} \\
	\PCHOICE{C_1}{p}{C_2} &\ddefeq \NDCHOICE{\COMPOSE{\WEIGHT{p}}{C_1}}{\COMPOSE{\WEIGHT{1-p}}{C_1}} \\
	\WHILEDO{\guard}{C} &\ddefeq \LOOP{C}{\guard}{\neg\guard}\qquad 
	\STAR{C} \ddefeq \LOOP{C}{\bone}{\bone}
\end{align*}%
\normalsize%
\begin{figure}
\begin{align*}
    \eval{\ASSIGN{x}{\ee}}  (\sigma,\tau)\tag{assignment} \ddefeq & 
    \iverson{\sigma\subst{x}{\sigma(e)}=\tau} \\
	\eval{\ASSIGNNONDET{x}}  (\sigma,\tau)\tag{nondeterministic assignment} \ddefeq &
	\bigoplus_{\alpha\in\Nats}   \iverson{\sigma\subst{x}{\alpha}=\tau}\\
    \eval{\WEIGHT{e}}(\sigma,\tau) \ddefeq & \sem{e}{\sigma}\odot \iverson{\sigma=\tau}  \tag{weighting} \\
    \eval{\COMPOSE{C_1}{C_2}}(\sigma,\tau) \ddefeq & 
		\bigoplus_{\iota\in\States}~ \sem{C_1}{\sigma,\iota}\odot \eval{C_2}(\iota,\tau) \tag{sequential composition} \\
	\eval{\NDCHOICE{C_1}{C_2}}(\sigma,\tau) \ddefeq & 
	\sem{C_1}{\sigma,\tau} \ooplus \sem{C_2}{\sigma,\tau}\tag{nondeterministic choice} \\
	\eval{\LOOP{C}{e}{e'}}(\sigma,\tau) \ddefeq &
	(\lfp X\mydot \quad \Phi_{C, e, e'}(X)) 
	(\sigma, \tau)  \tag{iteration}\\ 
	\text{where} \quad 
	\Phi_{C, e, e'}(X)(\sigma, \tau)&\eeq\sem{e}{\sigma}\odot\left(\bigoplus_{\iota\in\States}~ 
	\eval{C}(\sigma,\iota)\odot
	 X (\iota, \tau)
	\right) \ooplus \sem{e'}{\sigma}\odot\iverson{\sigma=\tau}\quad\qedtriangle
\end{align*}
\caption{Denotational semantics $\eval{C} \colon (\States\times\States )\to U$ of $\wrcl$ programs, where \mbox{$\mathcal{A} = \langle U,\, {\oplus},\, {\odot},\, \bzero,\, \bone\rangle$} is a semiring and the least fixed point is defined via point-wise extension of the natural order $\leq$ such that $f\leq f'$ iff $f(\sigma_1,\sigma_2)\leq f'(\sigma_1,\sigma_2)$ for all $\sigma,\sigma'\in\States$.}
	\label{def:semantics}
\end{figure}%
%
%
%
The semantics is shown in \Cref{def:semantics} and is described below.

\subsubsection*{\textbf{Assignment:}}
The semantics for assignment asserts that the weight of transitioning from $\sigma$ to $\tau$ after executing $\ASSIGN{x}{\ee}$ is $\bone$ if $\tau$ is equal to $\sigma$ with the value of $x$ updated to $\sigma(e)$, or $\bzero$ otherwise.

\subsubsection*{\textbf{Nondeterministic Assignment:}}
The denotational semantics for $\ASSIGNNONDET{x}$,
indicates that the weight of transitioning from initial state $\sigma$ to final state $\tau$ after executing $\ASSIGNNONDET{x}$ is $\bone$ if $\sigma$ and $\tau$ differ only in the value of $x$, and $\bzero$ otherwise. This is achieved by treating $\bigoplus$ akin to an existential quantifier. Specifically, given $\sigma$, we consider all possible values that $x$ may take after the execution of $\ASSIGNNONDET{x}$.

\subsubsection*{\textbf{Assume/Weighting:}}
The semantics for $\ASSUME{\guard}$ indicates that the weight of transitioning from~$\sigma$ to $\sigma$ is determined by the evaluation of $\guard$ in $\sigma$. If $\tau \neq \sigma$, then the weight of the transition is $\bzero$.

The intuition of the weighting statement in~\citet{batz2022weighted} is to weight arbitrary constant values $u\in U$, which does not generalize $\ASSUME{\guard}$ (but only $\ASSUME{\true}$ and $\ASSUME{\false}$). 
In our setting, weight can be any expression, so $\WEIGHT{e}$ is a proper generalization of the assume rule and is defined as%
$	\eval{\WEIGHT{e}}(\sigma,\tau) \eeq \sem{e}{\sigma} \odot \iverson{\sigma = \tau}.
$
%
 Here, the weighting rule expresses that the weight of transitioning from $\sigma$ to itself after a weighting operation is determined by the weight $\sem{e}{\sigma}$.

\subsubsection*{\textbf{Sequential Composition:}}
The semantics for $\COMPOSE{C_1}{C_2}$ calculates the weight of transitioning from $\sigma$ to $\tau$ after executing a sequence of $C_1$ followed by $C_2$, considering all possible intermediate states $\sigma'$.

\subsubsection*{\textbf{Nondeterministic Choice:}}
The semantics for $\NDCHOICE{C_1}{C_2}$ captures the weight of transitioning from $\sigma$ to $\tau$ after executing either $C_1$ or $C_2$, with the weight being the sum of the individual weights.

\subsubsection*{\textbf{Iteration:}}
The intended meaning of $\LOOP{C}{e}{e'}$ is to be equal to $\NDCHOICE{\COMPOSE{\WEIGHT{e}}{\COMPOSE{C}{\LOOP{C}{e}{e'}}}}{\WEIGHT{e'}}$. Replacing the recursive instance of $\LOOP{C}{e}{e'}$ with $X$, we get $\Phi_{C,e,e'}(X)$, and so by Kleene's fixpoint theorem, the least fixed point corresponds to iterating on the least element of the complete partial order $\bzero$, which yields an ascending chain of unrollings. This process can be demonstrated through the following sequence:
\begin{align*}
    \Phi_{C, e, e'}(\bzero)(\sigma,\tau) &\eeq \sem{\NDCHOICE{
		\COMPOSE{\WEIGHT{e}}{\DIVERGE}
	}{\WEIGHT{e'}}}{\sigma,\tau
	} \\
	\Phi_{C, e, e'}^2(\bzero)(\sigma,\tau) &\eeq \sem{
		\NDCHOICE{
			\COMPOSE{\WEIGHT{e}}{
				\COMPOSE{C}{
					\NDCHOICE{
		\COMPOSE{\WEIGHT{e}}{\DIVERGE}
	}{\WEIGHT{e'}}
				}
			}
		}
		{	\WEIGHT{e'}}}{\sigma,\tau
	}\\
	\Phi_{C, e, e'}^3(\bzero)(\sigma,\tau) &\eeq \sem{
		\NDCHOICE{
			\COMPOSE{\WEIGHT{e}}{
				\COMPOSE{C}{
					\NDCHOICE{
		\COMPOSE{\WEIGHT{e}}{
			\COMPOSE{C}{\NDCHOICE{
				\COMPOSE{\WEIGHT{e}}{\DIVERGE}
			}{\WEIGHT{e'}}}}
	}{\WEIGHT{e'}}
				}
			}
		}
		{	\WEIGHT{e'}}}{\sigma,\tau
	}
\end{align*}
and so on, whose supremum is the least fixed point of $\Phi_{C,e,e'}$.


\subsection*{Well-definedness of the Denotational Semantics}
We argue that the semantics of iteration loops is well-defined in~\Cref{prop:fixpoint-esistence-semantics}, assuming that $\Phi_{C, e, e'}(X)$ is a total function. This is always the case for any total semirings (such as \Bool,\Tropical), rendering our semantics more general than several others~\cite{Dardinier2023hyper,batz2022weighted,ZK22}.
For partial semi-rings, extra caution is necessary as $\oplus$ may not always be well-defined. 
Hence:%
\begin{enumerate}
    \item We restrict the  assignment \ASSIGNNONDET{x}, Kleene's star \STAR{C} and nondeterministic choices \NDCHOICE{C_1}{C_2} to total semi-rings only.
    \item We allow only nondeterministic choices of the form \NDCHOICE{\COMPOSE{e}{C_1}}{\COMPOSE{e}{C_2}} and loops \LOOP{C}{e}{e'} where the expressions are compatible~\cite[Section A.3]{zilberstein2024relatively}, that is, $\sem{e_1}{ \sigma} \oplus \sem{e_2}{ \sigma}$ is defined for any $\sigma \in \States$.
\end{enumerate}%
Restricting to compatible expressions allows the use of \ITE{\guard}{C_1}{C_2} and the guarded loop \WHILEDO{\guard}{C} for \emph{every} semiring. Additionally, the probabilistic choice \PCHOICE{C_1}{p}{C_2} remains well-defined for the partial semiring \Probs.
For the remainder of the paper, we assume that programs are constructed in this manner, ensuring they are always well-defined. Proofs of well-definedness are in~\Cref{app:wd-semantics}.

\section{Quantitative Weakest Hyper Pre}\label{sec:hyperpre}

\subsection{A Quantitative Strongest Post for Weighted Programs}
As hinted in~\Cref{se:intro:hyper}, we want our calculus to anticipate the so-called strongest post. Therefore, we define a novel \emph{quantitative strongest post} transformer for $\wrcl$.

\begin{definition}[Quantitative Strongest Post]
	\label{def:sp}
		The \emph{strongest post transformer} $\spsymbol\colon \wrcl \to (\A \to \A)$ %
		%
		is defined inductively according to the rules in \Cref{tab:rules} on p.~\pageref{tab:rules}, middle column.%
		\qedtriangle%
\end{definition}%
\noindent%
Let us show what $\spsymbol$ computes semantically, before providing some intuitions on the rules.%
\begin{restatable}[Characterization of $\spsymbol$]{theorem}{spsoundness}%
	\label{thm:sp-soundness}%
	For all programs $C\in\wrcl$ and final states $\tau\in\States$,
	\begin{align*}
		\sp{C}{\mu}(\tau) \qeq\quad \bigoplus_{\sigma\in \States}\quad
		\mu(\sigma)\odot \eval{C}(\sigma,\tau)~. 
	\end{align*}
\end{restatable}%
\noindent%
\Cref{thm:sp-soundness} guarantees the correct behavior of $\spsymbol$\footnote{It is essential to note that our formulation of $\spsymbol$ differs from the one disproven by~\cite[p. 135]{claire90}. The latter focuses on identifying the most precise assertion for the triples defined in~\cite[p. 124]{claire90}.} by asserting that it appropriately maps initial quantities to final quantities, including probability distributions and program sets of states.
In particular,~\Cref{tab:sp:subsumption} shows that by instantiating our calculus with different semirings we subsume several existing strongest post calculi. Additionally, similarly to~\cite[Table 1]{batz2022weighted}, weighted strongest post can handle optimization and combinatorial problems as well, with the main difference to be our calculus moving forward instead of backward. 
\begin{table}[b]%
	\small%
	\centering%
	\renewcommand{\arraystretch}{1.15}
	\begin{adjustbox}{max width=.99\textwidth}%
		\begin{tabular}{ll}       
			\hline
			\hline
			\textbf{Calculus} 										& \textbf{Semiring} 
			\\
			\hline
			\textsf{Strongest Postcondition~\cite{Dijkstra1990}} 				& $\langle \{0,1\}, \lor, \land, 0, 1\rangle$
			\\
			\textsf{Strongest Liberal Postcondition~\cite{ZK22}} 				& $\langle \{0,1\}, \land, \lor, 1, 0\rangle$ 
			\\
			\textsf{Quantitative Strongest Post~\cite{ZK22}} 			& $\langle \Rinf, \max, \min, \minfty, \pinfty\rangle$
			\\
			\textsf{Quantitative Strongest Liberal Post~\cite{ZK22}}	& $\langle \Rinf,  \min,\max, \pinfty, \minfty\rangle$\\
			\hline\hline
		\end{tabular}%
	\end{adjustbox}%
	\normalsize%
	\medskip%
	\caption{Existing strongest post calculi subsumed via our quantitative strongest post.}%
	\label{tab:sp:subsumption}%
\end{table}

We contend that our definition of $\spsymbol$ is inherently intuitive, extending the classical concept of "reachable sets" to final distributions where the binary notion of reachability is substituted with real values. This inherent intuitiveness is additionally justified by the close connection between weakest pre and strongest post in our framework. To underscore this point, we revisit Kozen's duality between forward transformers and $\wpsymbol$.%
\begin{theorem}[\citet{DBLP:journals/jcss/Kozen85} Duality]
	\label{thm:kozen}
	For all programs $C$, probability distributions $\mu\colon\States\to [0,1]$, and all functions~$f \in \A$, we have 
	%
	$	 \wp{C}{f}(\sigma)  = \sum_{\tau \in \States}~\sem{C}{\sigma,\tau} \cdot f(\tau) 
		.
		$
	%
\end{theorem}%
\noindent%
We now prove a more general version of the duality above for weighted programming.%
\begin{restatable}[Extended Kozen Duality For Weighted Programming]{theorem}{wpsoundness}%
	\label{thm:wp-soundness}%
	For all programs $C\in\wrcl$ and final states $\tau\in\States$,  with $\wpsymbol$ for \wrcl as defined in~\textnormal{\Cref{table:wp}}, the following equality holds:
	\begin{align*}
		\wp{C}{f}(\sigma) \qeq \bigoplus_{\tau\in \States }\quad
		 \eval{C}(\sigma,\tau)\odot f(\tau)~. 
	\end{align*}
\end{restatable}%
\noindent%
We can also prove that the following more symmetrical duality between our $\spsymbol$ and $\wpsymbol$ holds:
\begin{restatable}[Weighted $\spsymbol$-$\wpsymbol$ Duality]{theorem}{forwardbackward}%
	\label{thm:forwardbackward}
	For all programs $C$ and all functions $\mu, g \in \A$, we have%
  \begin{align*}
    \bigoplus_{\tau \in \States}~ \sp{C}{\mu}(\tau) \odot g(\tau) \qeq \bigoplus_{\sigma \in \States}~ \mu(\sigma) \odot \wp{C}{g}(\sigma)~.
  \end{align*}%
\end{restatable}%
\noindent%
In essence,~\Cref{thm:forwardbackward} establishes a novel equivalence between forward and backward transformers. An intuition for the probabilistic semiring $\Probs$ is that computing the expectation of a quantity~$g$ after the program execution---captured in the final distribution $\sp{C}{\mu}$---is analogous to calculating the expected value through $\wp{C}{g}(\sigma)$ but with the added nuance of being weighted by the initial distribution $\mu$.
In the case of other semirings, the idea is that on the left-hand side all terminating traces originating from $\mu$ are aggregated and then $g$ appended. Conversely, on the right-hand side, the process is reversed: we initiate from $g$ and move backward until we reach $\mu$.%
\begin{example}
	Consider the semiring of formal languages $\mathcal{A}=\langle \powerset{\{a,b\}^*}, \cup,\odot, \emptyset, \{\epsilon\} \rangle$ and the program $C=\NDCHOICE{\WEIGHT{\{a\}}}{\WEIGHT{\{b\}}}$. Let $\mu=\lambda \sigma\mydot \{a\}$ and $g=\lambda \sigma\mydot \{b\}$ represent the prequantity we aim to prepend and the postquantity we intend to append at the end of the execution, respectively. This results in the following language:
	\begin{align*}
		\bigoplus_{\sigma \in \States}~ \mu(\sigma) \odot \wp{C}{g}(\sigma) 
		&\eeq \bigoplus_{\sigma \in \States}~  \{a\} \odot (\wp{\WEIGHT{\{a\}}}{g}(\sigma) \oplus\wp{\WEIGHT{\{b\}}}{g}(\sigma))\\
		&\eeq \{a\} \odot (\{ab\} \oplus\{bb\}) \eeq \{aab, abb\}
	\end{align*}
	which is exactly
	\begin{align*}
		\bigoplus_{\tau \in \States}~ \sp{C}{\mu}(\tau) \odot g(\tau) 
		&\eeq \bigoplus_{\tau \in \States}~  (\sp{\WEIGHT{\{a\}}}{\mu}(\sigma) \oplus\sp{\WEIGHT{\{b\}}}{\mu}(\sigma))\odot\{b\}\\
		&\eeq  (\{aa\} \oplus\{ab\})\odot \{b\} \eeq \{aab, abb\} \tag*{\qedtriangle}
	\end{align*}
\end{example}%
\noindent%
Let us explain the rules in \Cref{tab:rules} individually.
\subsubsection*{\textbf{Assignment:}}
The quantitative strongest post $\sp{\ASSIGN{x}{\ee}}{f}$ is calculated by considering all possible values $\alpha$ that $x$ could have had before the assignment and summing all evaluations of quantity $f$ under those possible $\alpha$. 

\subsubsection*{\textbf{Nondeterministic Assignment:}}
The statement $\ASSIGNNONDET x$ is analogous to $\ASSIGN{x}{\ee}$, but without any restriction on the initial value of $x$, since the assignment is entirely nondeterministic and hence the original value of $x$ cannot be retrieved.

\subsubsection*{\textbf{Assume/Weighting:}}
In the assume statement, the strongest post is given by $\iguard \cdot f$, where $\iguard$ acts as a filter, nullifying states for which the predicate does not hold.

The weighting statement $\WEIGHT{a}$ extends the assume rule by allowing any weighting function $a$. The strongest post for weighting involves scaling the initial quantity $f$ by the weight $a$.

\subsubsection*{\textbf{Sequential Composition:}}
The quantitative strongest post for sequential composition $\COMPOSE{C_1}{C_2}$ is obtained by evaluating the second program $C_2$ starting from the strongest post of the first program $C_1$. The quantity $\sp{C_1}{f}$ represents the possible states reached with associated weights after executing $C_1$, and $C_2$ is then executed from these states.

\subsubsection*{\textbf{Nondeterministic Choice:}}
For the nondeterministic choice $\NDCHOICE{C_1}{C_2}$, the strongest post is the sum of the strongest posts of $C_1$ and $C_2$. This accounts for the possibility of either program being executed, resulting in a combination of the quantities reached by each.

\subsubsection*{\textbf{Iteration:}}
The strongest post for the iteration $\LOOP{C}{e}{e'}$ is an extension to the one in~\cite[Definition 4.1]{ZK22}, but generalised to arbitrary weights $e, e'$ instead of predicates. It is thus obtained via loop unrollings
\begin{align*}
	\Psi_f(\bzero)\odot\eval{e'} &\eeq \sp{\NDCHOICE{
		\COMPOSE{\WEIGHT{e}}{\DIVERGE}
	}{\WEIGHT{e'}}}{f}\\
	\Psi_f^2(\bzero)\odot\eval{e'} &\eeq \sp{\NDCHOICE{
		\COMPOSE{\WEIGHT{e}}{
			\COMPOSE{C}{
				\NDCHOICE{
	\COMPOSE{\WEIGHT{e}}{\DIVERGE}
}{\WEIGHT{e'}}
			}
		}
	}
	{	\WEIGHT{e'}}}{f}\\
	\Psi_f^3(\bzero)\odot\eval{e'}&\eeq \sp{
		\NDCHOICE{
			\COMPOSE{\WEIGHT{e}}{
				\COMPOSE{C}{
					\NDCHOICE{
		\COMPOSE{\WEIGHT{e}}{
			\COMPOSE{C}{\NDCHOICE{
				\COMPOSE{\WEIGHT{e}}{\DIVERGE}
			}{\WEIGHT{e'}}}}
	}{\WEIGHT{e'}}
				}
			}
		}
		{	\WEIGHT{e'}}}{f
	}
\end{align*}
which converge to the least fixed point of $\Psi_f(X)=f \oplus \sp{C}{X\odot \eval{e}}$, yielding the rule%
\begin{align*}
   \sp{\LOOP{C}{e}{e'}}{f}\eeq \big(\lfp  X\mydot  f \oplus \sp{C}{X\odot \eval{e}}\big) \odot \eval{e'}.
\end{align*}%
%


\subsection{Quantitative Weakest Hyper Pre}

First of all, we show in which sense we can represent hyperproperties via functions. We have already seen that predicates can be encoded via Iverson brackets (~\Cref{def:iverson}), and decoded by the support set, since every quantity $f\colon\States\to U$ can be seen as a set of states via $\supp{f}=\{\sigma\colon f(\sigma)\neq \bzero\}$. For example, the set of reachable states starting from $\phi\subseteq\States$ is $\supp{\sp{C}{\iverson{\phi}}}$.
To encode and decode hyperpredicates, we need to introduce hyper Iverson brackets.%

\begin{definition}[Hyper Iverson Brackets]
	\label{def:hyper-iverson}
		Given a semiring $\mathcal{A}=\langle U, \oplus,\odot, \bzero, \bone\rangle$, for a hyperpredicate $\hyperphi\colon \powerset{\powerset{\States}}$ we define the hyper Iverson bracket $\iverson{\hyperphi}\colon(\States\to U) \to\PosRealsInf$ by%
		\begin{align*}
			\iverson{\hyperphi}(f) \eeq 
			\Bigl\{
				\pinfty \quad\text{if $\supp{f}\in \hyperphi$}; \qqand
				0 \quad\text{otherwise.}
			\tag*{\qedtriangle}
		\end{align*}%
		\normalsize%
\end{definition}%
\noindent%
For a hyperquantity $\hyperf$, its corresponding hyperpredicate is defined by $\supp{\hyperf}=\{f\colon \hyperf(f)>0\}$. We shall remark that hyperpredicates in our setting can represent predicates over quantities, including hyperproperties and predicates over probability distributions.
\begin{table*}[!t]%
	\begin{center}%
		\begin{adjustbox}{max width=0.99\linewidth}%
			\renewcommand{\arraystretch}{1.25}%
			\begin{tabular}{l@{\qquad}l@{\qquad}l}%
				\hline\hline
				$\boldsymbol{C}$			& $\boldsp{C}{f}$ 												& $\boldwhp{C}{\hyperf}$																		\\
				\hline
				$\ASSIGN{x}{\ee}$			& 
				$\bigoplus_{\alpha}f\subst{x}{\alpha} \odot\iverson{x= \ee\subst{x}{\alpha}}$	& 
				$\hyperf\subst{x}{\ee}$					\\
				$\ASSIGNNONDET{x}$			& 
				$\bigoplus_{\alpha} f\subst{x}{\alpha}$	& 
				$\lambda f\mydot \hyperf(\bigoplus_{\alpha} f\subst{x}{\alpha})$					\\
				$\WEIGHT{w}$			& $f\odot w$													& $ \hyperf \odot w$	 \\
				$\COMPOSE{C_1}{C_2}$		& $\sp{C_2}{\sp{C_1}{f}}$											& $\whp{C_1}{\vphantom{\big(}\whp{C_2}{\hyperf}}$\\
				$\NDCHOICE{C_1}{C_2}$	& $\sp{C_1}{f} \oplus \sp{C_2}{f}$										& $\bigoplus_{\nu_1,\nu_2}~\hyperf(\nu_1\oplus\nu_2)\odot \whp{C_1}{\iverson{\nu_1}} \odot \whp{C_2}{\iverson{\nu_2}}$	\\
				$\LOOP{C}{e}{e'}$				& $\big(\lfp  X\mydot  f \oplus \sp{C}{X\odot \eval{e}}\big) \odot \eval{e'}$ 			 & $\lambda f\mydot \hyperf\big(\big(\lfp  X\mydot  f \oplus \sp{C}{X\odot \eval{e}}\big) \odot \eval{e'}\big)$	\\
				\hline\hline
			\end{tabular}%
		\end{adjustbox}%
	\end{center}%
	\medskip%
	\caption{Rules for defining the quantitative strongest post and weakest hyper pre transformers.}%
	\label{tab:rules}%
\end{table*}
\begin{definition}[Quantitative Weakest Hyper Pre]
	\label{def:whp}
		The \emph{quantitative weakest hyper pre transformer} $\whpsymbol\colon \rcl \to (\hyperA \to \hyperA)$
		%
		is defined inductively according to the rules in \Cref{tab:rules}, right column.%
\end{definition}%

Let us show for some of the rules how the quantitative weakest hyper pre semantics can be developed and understood analogously to Dijkstra's classical weakest preconditions.

\subsubsection*{\textbf{Assignment.}}

The weakest pre\emph{condition} of an assignment is given by %
%
$	\wp{\ASSIGN{x}{e}}{\psi} \eeq \psi\subst{x}{e}$,
%
where $\psi\subst{x}{e}$ denotes the substitution of the variable $x$ in $\psi$ with the expression $e$. From a semantic perspective, this replacement can be expressed as
$
\psi\subst{x}{e} \coloneqq \lambda\sigma\mydot \psi\Bigl( \sigma\statesubst{x}{\sigma(e)} \Bigr). 
$
In simpler terms, the weakest precondition operates by predicting the operational semantics: it examines whether, given an initial state $\sigma$, the final state $\sigma\statesubst{x}{\sigma(e)}$ adheres to the condition $\psi$.

For \emph{quantitative weakest hyper pre}, a similar approach is taken, but we anticipate the strongest post rather than the operational semantics. 
Therefore, the value of $\hyperf$ in the resulting distribution (or set of states) after the execution of $\ASSIGN{x}{e}$ on the initial distribution (or set) $f$ corresponds to $\hyperf$, but evaluated at the final distribution $\sp{\ASSIGN{x}{\ee}}{f}=\bigoplus_{\alpha}f\subst{x}{\alpha} \odot\iverson{x= \ee\subst{x}{\alpha}}$. We thus define the syntactic replacement of the variable $x$ in a hyperquantity $\hyperf$ by $\hyperf \subst{x}{\ee} \coloneqq \lambda f\mydot \hyperf(\sp{\ASSIGN{x}{\ee}}{f})$, yielding the rule $\whp{\ASSIGN{x}{e}}{\hyperf} \eeq \hyperf \subst{x}{\ee}$

\subsubsection*{\textbf{Nondeterministic Assignment:}}
The nondeterministic assignment is analogous to the standard assignment, but now with $x$ ranging over any possible value.

\subsubsection*{\textbf{Assume/Weighting.}}
We have $\wp{\ASSUME{\guard}}{\psi} \eeq \guard \wedge \psi$. 
Indeed, if the initial state $\sigma$ satisfies the combined precondition $\varphi \land \psi$, the execution of $\ASSUME{\guard}$ entails progression through the assumption of $\varphi$. 
Since the assumption itself does not alter the program state, the process concludes in state $\sigma$, which also satisfies the post $\psi$. Conversely, if $\sigma$ fails to meet $\varphi \land \psi$, the execution of $\ASSUME{\guard}$ results in either not progressing through the assumption of $\varphi$ or passing through the assumption but $\sigma$ not satisfying the post $\psi$.
The \emph{quantitative weakest hyper pre} on an initial distribution (set) $f$ anticipates the strongest post, yielding the rule $\whp{\ASSUME{\guard}}{\hyperf} = \lambda f\mydot \hyperf(\iguard \odot f)$.

To simplify the notation, we introduce the product $\odot$ between quantities and hyperquantities as:%
\begin{align*}
	\hyperf \odot w \eeq  \lambda f\mydot \hyperf(f \odot w) \qquad w\odot \hyperf\eeq  \lambda f\mydot \hyperf (w \odot f)   ~,
\end{align*}%
leading to the syntactically simpler rule $\whp{\ASSUME{\guard}}{\hyperf}\eeq \hyperf \odot \iguard$.
For the more general weighting statement, $	\whp{\WEIGHT{w}}{\hyperf} \eeq \hyperf\odot w$ 
is a generalization, where $w$ can be any quantity.

\subsubsection*{\textbf{Nondeterministic Choice.}}

When executing nondeterministic choice $\NDCHOICE{C_1}{C_2}$ on some initial state $\sigma$, operationally \emph{either} $C_1$ \emph{or}~$C_2$ will be executed.
Hence, the execution will reach either a final state in which executing~$C_1$~on~$\sigma$ terminates or a final state in which executing~$C_2$~on~$\sigma$ terminates (or no final state if both computations diverge).

The \emph{angelic} weakest pre\emph{condition} of $\NDCHOICE{C_1}{C_2}$ \mbox{is given by} $\wp{\NDCHOICE{C_1}{C_2}}{\psi} \eeq \wp{C_1}{\psi} \morespace{\vee} \wp{C_2}{\psi}$.
Indeed, whenever an initial state $\sigma$ satisfies the precondition $\wp{C_1}{\psi}$  or $\wp{C_2}{\psi}$, then \mbox{--- either by} executing $C_1$ or~\mbox{$C_2$ ---} it is possible that the computation will terminate in some final state satisfying the postcondition $\psi$.

Moving to hyperquantities, the elimination of nondeterminism occurs because the strongest post $\spC{\NDCHOICE{C_1}{C_2}}$ is deterministic. Consequently, the value of $\hyperf$ in the resulting distribution (or set of states) after executing either $C_1$ or $C_2$ on the initial distribution (or set) $f$ is%
\begin{align*}
	\whp{\NDCHOICE{C_1}{C_2}}{\hyperf} \eeq \bigoplus_{\nu_1,\nu_2}~\hyperf(\nu_1\oplus\nu_2)\odot \whp{C_1}{\iverson{\nu_1}} \odot \whp{C_2}{\iverson{\nu_2}}~.
\end{align*}%
Recalling that the final distribution is the combination of $\sp{C_1}{f}$ and $\sp{C_2}{f}$, identifying $\nu_i$ such that $\nu_i=\sp{C_i}{f}$ makes computing $\hyperf(\nu_1\oplus\nu_2)$ sufficient. By aggregating over every $\nu_i$ for which $\whp{C_i}{\iverson{\nu_1}}(f)$ holds, we ensure that only those $\nu_i$ where $\nu_i=\sp{C_i}{f}$ will contribute, making the sum non-zero. Consequently, $\hyperf(\nu_1\oplus\nu_2)$ precisely equals $\hyperf(\sp{\NDCHOICE{C_1}{C_2}}{f})$.

\begin{remark}
	In the case of $\NDCHOICE{C_1}{C_2}$, \OL and \HHL exhibit forward-style rules that are simpler but not comprehensive. While these rules maintain soundness, completeness necessitates the inclusion of an existential rule. As our approach adopts a weakest pre style calculus aiming for both soundness and completeness, the introduction of the $\bigoplus$ quantification becomes imperative. This quantification mirrors the existential rule utilized in \OL and \HHL, encompassing all relevant cases.
Our rule shares similarities with~\citet[Definition 6.5.2]{hartog}, although they provide multiple rules depending on the structure of the hyperquantity. Since our paper focuses on semantic assertions, we refrain from analyzing the syntactic structure of hyperquantities. However, we later introduce simpler rules for the class of linear hyperquantities, as outlined in \Cref{def:quantities:linear}.
\end{remark}

\subsubsection*{\textbf{Sequential Composition.}}

What is the anticipated value of $\hyperf$ after executing $\COMPOSE{C_1}{C_2}$, i.e.\ the value of $\hyperf$ after first executing $C_1$ and then $C_2$?
To answer this, we first anticipate the value of $\hyperf$ after execution of $C_2$ which gives~$\whp{C_2}{\hyperf}$.
Then, we anticipate the value of the intermediate quantity $\whp{C_2}{\hyperf}$ after execution of $C_1$, yielding $\whp{\COMPOSE{C_1}{C_2}}{\hyperf} = \whp{C_1}{\whp{C_2}{\hyperf}}$.

\subsubsection*{\textbf{Iteration}}
The rule for $\LOOP{C}{e}{e'}$ is obtained by anticipating the execution of $\LOOP{C}{e}{e'}$. It is consistent in the sense that it is a solution of the equation:
\begin{align*}
 \whpC{\LOOP{C}{e}{e'}}
 &\eeq \whpC{\NDCHOICE{\COMPOSE{\COMPOSE{\WEIGHT{e}}{C}}{\LOOP{C}{e}{e'}}}{\WEIGHT{e'}}}\\ 
 &\eeq  \lambda \hyperh\,\lambda f\mydot \bigoplus_{\nu}~\hyperh(\nu \oplus f\odot\eval{e'}) \odot \whp{C}{\whpC{\LOOP{C}{e}{e'}}(\iverson{\nu})}(f\odot\eval{e})
\end{align*}%
Indeed one can show the following.
\begin{restatable}[Consistency of iteration rule]{proposition}{propfixpointuniqueness}%
	Let
	\begin{align*}
		\Phi(\textsf{trnsf})=  \lambda \hyperh\,\lambda f\mydot 
		\bigoplus_{\nu}~\hyperh(\nu \oplus f\odot\eval{e'}) \odot \whp{C}{\textsf{trnsf}(\iverson{\nu})}(f\odot\eval{e})
	\end{align*}
	Then, $\whpC{\LOOP{C}{e}{e'}}$ is a fixpoint of the higher order function $\Phi(\textsf{trnsf})$, that is:
	\begin{align*}
		\Phi(\lambda \hyperf\,\lambda \mu\mydot \hyperf (\sp{\LOOP{C}{e}{e'}}{\mu})) =   \lambda \hyperf\,\lambda \mu\mydot \hyperf (\sp{\LOOP{C}{e}{e'}}{\mu})
	\end{align*}
\end{restatable}

\begin{remark}
	One might attempt a rule for $\LOOP{C}{e}{e'}$ by defining $F(X)=\lambda f \mydot X(f\oplus \sp{C}{f\odot\eval{e}})$. Intuitively, $F$ takes as input a hyperquantity $X$, but instead of applying it on a distribution $f$, it computes one iteration of the loop $\sp{C}{f\odot\eval{e}}$ and then pass all as argument of $X$.
	Recalling that $\Psi_f(X)=f\oplus\sp{C}{ X\odot\eval{e}}$, one can then observe that for every $n\in\Nats$:
	\begin{align*}
		\lambda f\mydot \hyperf (f\odot\eval{e'}) &\eeq \lambda f\mydot \hyperf (
			\Psi_f(\bzero)
			\odot \eval{e'})\\
		F(\lambda f\mydot \hyperf (f\odot\eval{e'}))&\eeq\lambda f\mydot \hyperf (
		\Psi^2_f(\bzero)
		\odot \eval{e'})\\[-.75em]
		&\vdots \\
		F^n(\lambda f\mydot \hyperf (f\odot\eval{e'}))&\eeq\lambda f\mydot \hyperf (
			\Psi^{n+1}_f(\bzero)
			\odot \eval{e'})
	\end{align*}
	However, it's important to note that in general, \( F^n(\lambda f\mydot \hyperf (f\odot\eval{e'})) \) does not form an ascending or descending chain. For example, take $ \hyperf \eeq \iversone{\nu}$, where $\nu$ is a probability distribution. It's very well possible that $\iversone{\nu} (
		\Psi^k_f(\bzero)
		\odot \eval{e'})=\bone$ for some $k,\mu$: that is, we anticipate an incomplete proability distribution and find out that it is equal $\nu$. However, at the $k+1$ iteration, the anticipated probability distribution is refined, so that it could be $\Psi^{k+1}_\mu(\bzero)
			\odot \eval{e'} \neq \nu$, leading to a decreasing iterate. Additionally, it's not always desirable to stop at the first fixpoint - as multiple extra iterations might be needed to compute the correct anticipated probability distribution. That said, it is entirely possible that simpler rules exist when restricting \( \hyperf \), see e.g.~\Cref{table:whp:linear}.
	\lz{maybe: $\hyperf$ monotonic => least fix point? Also if $f\odot\eval{e'}\leq F(\lambda \mu\mydot \hyperf (\mu\odot\eval{e'}))$ the chain should be ascending I believe}
	\qedtriangle
\end{remark}
\noindent%
%
%
%
After having provided an intuition on the rules, let us show that $\whpsymbol$ does actually anticipate $\spsymbol$.
\begin{restatable}[Characterization of \textnormal{$\whpsymbol$}]{theorem}{whpsoundness}%
	\label{thm:whp-soundness}%
	For all programs~$C$, hyperquantities $\hyperf\in\hyperA$ and quantities $f\in\A$: \quad $\whp{C}{\hyperf}(f) = \hyperf(\sp{C}{f})$.
\end{restatable}%
\noindent%
For a given hyperquantity $\hyperf$ and initial quantity $\mu$, $\whp{C}{\hyperf}(\mu)$ represents the value assumed by $\hyperf$ in the final quantity reached after the termination of $C$ on $\mu$. 
Unlike standard~$\wpsymbol$, which distinguishes between terminating and nonterminating states, $\whpsymbol$ does not make this distinction. 
When there are no terminating states, i.e., $\sp{C}{\mu} = \bzero$, the value of $\whp{C}{\hyperf}(\mu)$ is determined by $\hyperf(\bzero)$. 
The assignment of any desired value to the empty set of states $0$ by the hyperquantity $\hyperf$ allows us to express both weakest preconditions and weakest liberal ones.

\section{Expressivity}\label{sec:express}

In the preceding sections, we characterized our quantitative weakest hyper pre calculus. 
In this section, we aim to illustrate the expressive capabilities of the calculus by demonstrating that it subsumes several other logics and calculi.
\subsection{An Overview of Several Hoare-Like Logics}
\noindent%
We subsume Hyper Hoare Logic for non-probabilistic programs (since \HHL is non-probabilistic). 
\begin{restatable}[Subsumption of \HHL]{theorem}{subhyperhoare}%
	For hyperpredicates $\hyperpsi,\, \hyperphi$ and non-probabilistic program $C$:%
	\small%
	\begin{align*}
		\hhl{\hyperpsi}{C}{\hyperphi} \qiff \supp{\indicator{\hyperpsi}} \subseteq  \supp{\whp{C}{\indicator{\hyperphi}}}
	\end{align*}%
	\normalsize%
\end{restatable}%
\noindent%
As a byproduct, $\whpsymbol$ subsumes demonic partial correctness, angelic total correctness, partial incorrectness, and total incorrectness (according to the terminology in~\cite{ZK22}). To highlight this, we will utilize the following modality syntax introduced in~\cite{zilberstein2024relatively}:
        \[
			\Box P \eeq \lambda \rho \mydot [\rho \subseteq P] \qqand
			\Diamond P \eeq \lambda \rho \mydot [P\cap\rho \neq \emptyset]
		\]
When reasoning about hyperproperties, we may omit Iverson brackets and write $\hyperpsi \subseteq \whp{C}{\hyperphi}$ instead of $\supp{\indicator{\hyperpsi}} \subseteq \supp{\whp{C}{\indicator{\hyperphi}}}$. We obtain the relationships in~\Cref{tab:logics}.%
\begin{table}[t]%
	\centering%
	\begin{adjustbox}{max width=.99\textwidth}
		\renewcommand{\arraystretch}{1.25}%
%
		\begin{tabular}{l@{\quad}c@{\quad}c@{\qquad}c}       
			\hline\hline
			\textbf{Logic} 								& \textbf{Syntax}  				&\textbf{Semantics}  					& \textbf{Semantics via $\whpsymbol$} \\
			\hline
			\textsf{Hoare Logic (partial correctness)} 			& \hspace*{-.1em}$\pcl{P}{C}{Q}$ 	&\hspace*{.65em}$P\subseteq \wlp{C}{Q}$	& \hspace*{-2.5em}$\Box P \subseteq \whp{C}{\Box Q}$
			\\
			\textsf{Lisbon Logic (angelic total correctness)} 		& \hspace*{-.25em}$\tcl{P}{C}{Q}$ 				&\hspace*{.55em}$P\subseteq \wp{C}{Q} $	& \hspace*{-2.5em}$\Diamond P \subseteq \whp{C}{\Diamond Q}$
			\\
			\textsf{Partial Incorrectness Logic} 				& $\pil{P}{C}{Q}$ 				&\hspace*{.05em}$Q\subseteq \slp{C}{P}$  			& \hspace*{-2.0em}$\{\neg P\} \subseteq \whp{C}{\Box (\neg Q)}$
			\\
			\textsf{Incorrectness Logic/Reverse Hoare Logic}	& \hspace*{.1em}$\til{P}{C}{Q}$  	&$Q\subseteq \sp{C}{P} $ 			& $\{P\} \subseteq\whp{C}{\lambda \rho.~ Q\subseteq \rho}$ \\
			\hline\hline
		\end{tabular}%
	\end{adjustbox}%
	\medskip%
	\caption{%
		Partial and total (in)correctness using classical predicate transformers and $\whpsymbol$. 
	}%
	\label{tab:logics}%
\end{table}

Arguably, Hoare-like logics are designed to be accessible to programmers to prove correctness, whereas reasoning about $\whpsymbol$ (and \HHL, \OL) enables better understanding of relationships between different program logics, leading to definitions of new logics, as we will show in the following.

\subsection{Disproving Hoare-Like Triples}
\allowdisplaybreaks
For example, we can semantically define new triples by falsifying the triples of \Cref{tab:logics}, see \Cref{tab:disprove-logics}.

\begin{table}[t]
	\renewcommand{\arraystretch}{1.25}%
	\small%
	\centering%
	\begin{adjustbox}{max width=.66\textwidth}
		\begin{tabular}{c@{\qquad}c@{\qquad}c}       
			\hline\hline
			\textbf{Syntax}  & \textbf{Semantics}  & \textbf{Semantics via $\whpsymbol$} \\
	        		\hline
			\hspace*{-.25em}$\not\pcl{P}{C}{Q}$ & \hspace*{-.35em}$P\cap \wp{C}{\neg Q}\neq \emptyset$ & \hspace*{.6em}$\{P\} \subseteq \whp{C}{\Diamond(\neg Q)}$ \\
			\hspace*{-.4em}$\not\tcl{P}{C}{Q}$ & \hspace*{-.4em}$P\cap \wlp{C}{\neg Q}\neq \emptyset$  & \hspace*{-4.7em}$\exists \sigma \in P \mydot  \{\{\sigma\}\}\subseteq \whp{C}{\Box{\neg Q} }$ \\
			\hspace*{-.15em}$\not\pil{P}{C}{Q}$ & $Q\cap \sp{C}{\neg P}\neq \emptyset$  & \hspace*{-1.5em}$\{\neg P\} \subseteq\whp{C}{\Diamond Q}$ \\
			$\not\til{P}{C}{Q}$  & \hspace*{-.1em}$Q\cap \slp{C}{\neg P}\neq \emptyset$ & \hspace*{4em}$\{P\} \subseteq\whp{C}{\lambda \rho.~ Q\cap \neg \rho\neq \emptyset}$\\
			\hline\hline
		\end{tabular}%
	\end{adjustbox}%
	\medskip%
	\caption{%
		Disproving partial and total (in)correctness using classical predicate transformers and $\whpsymbol$. 
	}%
	\label{tab:disprove-logics}%
\end{table}
\begin{itemize}
	\item $\not\pcl{P}{C}{Q}$: there is some state in $P$ that can terminate in $\neg Q$, and hence it is false that every state in $P$ terminates only in $Q$ (if it terminates at all)
	\item $\not\tcl{P}{C}{Q}$: there is some state in $P$ that terminates only in $\neg Q$ (if it terminates at all), and hence it is false that every state in $P$ can terminate in $Q$
	\item $\not\pil{P}{C}{Q}$: there is some state in $Q$ that is reachable from $\neg P$, and hence it is false that every state in $Q$ is reachable only from $P$
	\item $\not\til{P}{C}{Q}$: there is some state in $Q$ that is reachable only from $\neg P$ (if it is reachable at all), and hence it is false that every state in $Q$ is reachable from $P$
\end{itemize}%
It remains to define program logics for the newly defined falsifying triples. To this end, one can prove that the existing program logics are actually falsifying program logics. More precisely:%
\begin{restatable}[Falsifying correctness triples via correctness triples]{theorem}{disprovingTriples}%
	\label{thm:falsify-triples}
	\small%
	\begin{align*}
		\pcl{P}{C}{Q} &\qiff \forall \sigma\in P\mydot \not\tcl{\{\sigma\}}{C}{\neg Q} \\
		\tcl{P}{C}{Q} &\qiff  \forall\sigma\in P\mydot  \not\pcl{\{\sigma\}}{C}{\neg Q}\\
		\pil{P}{C}{Q} &\qiff \forall \sigma\in  Q\mydot \not\til{\neg P}{C}{\{\sigma\}}\\
		\til{P}{C}{Q} &\qiff \forall \sigma\in Q\mydot  \not\pil{\neg P}{C}{\{\sigma\}}
	\end{align*}%
	\normalsize
\end{restatable}%
\noam{Again, I think we need to be really clear that total correctness here is angelic, as that is not the standard that people think of when they hear total correctness in general. Also, I think there's a mixup of notation here where sometimes you use sets to denote predicate functions, i.e., above you use $\{P\}$ to mean $\lambda \rho.[\rho = P]$. Of course, these are isomorphic, but we should be clear about it if we are going to use that notation.}%
\begin{itemize}
	\item $\pcl{P}{C}{Q}$: every state in $P$ can only terminate in $Q$ (if it terminates at all), and hence by starting on any of those state it is false that it can terminate in $\neg Q$
	\item $\tcl{P}{C}{Q}$: every state in $P$ can terminate in $Q$, and hence by starting on any of those states it is false that it can terminates only in $\neg Q$ (if it terminates at all)
	\item $\pil{P}{C}{Q}$: every state in $Q$ is reachable only from $P$, and hence from any of those states it is false that it is reachable from $\neg P$
	\item $\til{P}{C}{Q}$: every state in $Q$ is reachable from $P$, and hence from any of those states it is false that it is reachable only from $\neg P$
\end{itemize}%
\Cref*{thm:falsify-triples} not only demonstrates that existing program logics can generate proofs to falsify other triples but also establishes a crucial "if and only if" relationship. This indicates that not only the current logics are sound, but they are \emph{complete} as well: the existence of an invalid triple implies the presence of a corresponding valid triple that renders the original one invalid. Restating~\Cref*{thm:falsify-triples} from a negative perspective as below might make it more clear how to practically falsify triples.
\begin{corollary}
	\label{cor:falsify-triples}
	\small%
	\abovedisplayskip=-1\baselineskip%
	\begin{align*}
		\not\pcl{P}{C}{Q} &\qiff  \exists \sigma\in P\mydot \tcl{\{\sigma\}}{C}{\neg Q} \\
		\not\tcl{P}{C}{Q} &\qiff  \exists \sigma\in P\mydot \pcl{\{\sigma\}}{C}{\neg Q} \\
		\not\pil{P}{C}{Q} &\qiff \exists \sigma\in Q\mydot \til{\neg P}{C}{\{\sigma\}}\\
		\not\til{P}{C}{Q} &\qiff \exists\sigma\in Q\mydot\pil{\neg P}{C}{\{\sigma\}}
	\end{align*}%
	\normalsize%
\end{corollary}%
\noindent%
As highlighted by \citet[p.~20, "Other Triples"]{ZK22}, the use of the terms "correctness" and "incorrectness" in naming conventions may be imprecise. Correctness triples can be seen as $\forall$-properties over preconditions, whereas incorrectness triples exhibit characteristics of $\forall$-properties over postconditions. Furthermore, it is noteworthy that the falsification of such $\forall$-triples can be interpreted as $\exists$-triples, a result that aligns with the expectation that disproving these properties involves finding at least one counterexample. This perspective concurs with the observation made by \citet[Logic 23]{cousot2024calculational} that Incorrectness Logic provides sufficient (though not necessary) conditions to falsify partial correctness triples, thereby demonstrating its greater-than-needed power.
Let us show how to practically falsify triples.%
\begin{example}[Backward-Moving Assignment Rule for (Total) Incorrectness Logic]
	Consider the triple $\til{y=42}{\ASSIGN{x}{42}}{y=x}$, obtained by taking as precondition the syntactic replacement of $x=42$ from the post. As shown in~\cite{OHearn19} with a counterexample, this is not valid. We can prove it by computing a partial incorrectness triple with precondition $y\neq 42$.

	Using the rules defined in~\cite[Table 2, Column 2]{ZK22}, we have:
	\begin{align*}
	\pil{y\neq 42}{\ASSIGN{x}{42}}{y\neq 42 \lor x\neq 42}
	\end{align*}
	This post clearly contains at least one state with $y=x$ (e.g., take a state where $\sigma(x)=\sigma(y)=0$), which implies $\not\til{y=42}{\ASSIGN{x}{42}}{y=x}$ (by \Cref{cor:falsify-triples}).%
	\qedtriangle
\end{example}%
\noindent%
We conclude the section by observing that we have the following connection.
\begin{proposition}[$\wpsymbol$ / $\spsymbol$ Connection]
\qquad$
	P \cap \wp{C}{ Q}\neq \emptyset \qiff Q\cap \sp{C}{P}\neq \emptyset.
$
\end{proposition}%
\noindent%
A simple consequence of the above is the duality
$\not\pcl{P}{C}{\neg Q}$ iff $\not\pil{\neg P}{C}{Q}$, 
which is not surprising, as the duality $\pcl{P}{C}{Q}$ iff $\pil{\neg P}{C}{\neg Q}$ has already been explored in~\cite[p.22, "Duality"]{ZK22} and again in~\cite{ascari2023sufficient}.

\subsection{Designing (Falsifying) Hoare-Like Logics via Hyperpredicate Transformers}

The observations above indicate that there is no advantage for new program logics to falsify triples from an expressivity point of view, as they can be converted into existing triples via \Cref{thm:falsify-triples}. However, one may wonder whether it is possible to design triples that are more useful in practice. In this regard, we emphasize that the design of program logics should follow predicate transformer reasoning. We provide an intuition on how $\whpsymbol$ aids in reasoning about designing logics (rather than triples). We illustrate this with an example of partial correctness.

\paragraph*{Partial Correctness as Classical Predicate Transformers}
Partial correctness amounts to a logic that takes $Q \subseteq \powerset{\States}$ and proves every $P$ such that $P \subseteq \wlp{C}{Q}$.

\paragraph*{Partial Correctness as a Hyperproperty}
We observe that partial correctness, as a logic, is a hyperproperty. Indeed, $P \subseteq \wlp{C}{Q}$ iff $P \in \{S \mid S \subseteq \wlp{C}{Q}\}$, and this is a predicate over sets of states. 
Also, by Galois connection, this is equivalent to proving $\sp{C}{P} \subseteq Q$ iff $\sp{C}{P} \in \{S \mid S \subseteq Q\}$, explaining why our $\whpsymbol$ captures partial correctness (via $P \in \whp{C}{\lambda \rho \mydot \rho \subseteq Q}$).

\paragraph*{(Dis)proving Partial Correctness, Practically}
One may wonder why partial correctness is much easier than our $\whpsymbol$ calculus. At first glance, it seems that, for a given post $Q$, one may want to find $\{S \mid S \subseteq \wlp{C}{Q}\}$. 
However, the actual logic aims to find just $\wlp{C}{Q}$ since $\wlp{C}{Q}$ fully characterizes the original hyperproperty. 
Even if $\wlp{C}{Q}$ itself is not found, any $S \subseteq \wlp{C}{Q}$ allows soundly proving $\pcl{P}{C}{Q}$ by checking $P \subseteq S$. The same reasoning applies to falsify partial correctness triples. Our key insight is that it is enough to find any $\wlp{C}{Q} \subseteq S$ and then prove $\not\pcl{P}{C}{Q}$ by checking $P \not\subseteq S$. With this in mind, we argue that the most sensible proof system to falsify partial correctness should aim for $\wlp{C}{Q} \subseteq P$.

So we obtain the following sound and complete falsifying partial correctness logic, which is the same as partial correctness except for the following different rules:
\begin{align*}
	\infer[\text{Antecedence\footnotemark}]{
		\anonl{G}{C}{F}
	}{
		G \Longleftarrow G' \quad \anonl{G'}{C}{F'} \quad F' \Longleftarrow F
	}
	\qquad
	\infer[\text{Kleene}]{
		\anonl{\forall n.p(n)}{C^\star}{p(0)}
	}{
		\forall n \mydot \anonl{p(n+1)}{C}{p(n)}
	}
\end{align*}%
\footnotetext{{Which replaces the rule of \emph{consequence}.}}%
We argue that by similar reasoning, it is easy to find falsifying logics for the other triples.

\paragraph*{Do we need falsifying logics?}
It is known from~\cite[p.22]{ZK22} that $\wlp{C}{Q} \subseteq S$ corresponds to the contrapositive of Lisbon Logic, i.e., amounts to $\neg S \subseteq \wp{C}{\neg Q}$. This means that, to prove $\not\pcl{P}{C}{Q}$, one should prove $\tcl{\neg S}{C}{\neg Q}$ (possibly keeping $\neg S$ large) and then check $P \not\subseteq S$. Similar reasoning applies if we want to apply~\Cref{thm:falsify-triples}, and so we argue that reasoning via contrapositive is a lot harder to do for the average programmer.

\subsection{Semantics of Nontermination and Unreachability}

\begin{wraptable}[7]{r}{.66\textwidth}%
	\vspace*{-1\intextsep}%
	\renewcommand{\arraystretch}{1.25}%
	\small%
	\centering%
	\begin{adjustbox}{max width=.99\linewidth}
		\begin{tabular}{@{\hspace*{-.5em}}ccl}       
			\textbf{Triple}  & \textbf{Semantics} & \textbf{Property}\\
			\hline
			\hspace*{2.1em}$\pcl{P}{C}{\false}$	&\hspace*{.1em}$\forall \sigma\in P\mydot\not\exists\tau\mydot \tau\in \sem{C}{\sigma}$	& Must-Nontermination	\\
			\hspace*{1.7em}$\tcl{P}{C}{\true}$ 	&\hspace*{.1em}$\forall \sigma\in P\mydot \exists\tau\mydot  \tau\in\sem{C}{\sigma}$	& May-Termination		\\
			\hspace*{-.4em}$\pil{\false}{C}{Q}$	&$\forall \tau\in Q\mydot \not\exists \sigma \mydot \tau \in \sem{C}{\sigma}$			& Unreachability		\\
			$\til{\true}{C}{Q}$ 				&$\forall \tau\in Q\mydot \exists \sigma \mydot \tau \in \sem{C}{\sigma}$				& Reachability 			\\
		\end{tabular}
	\end{adjustbox}%
	\medskip%
	\caption{$\forall$-properties on nontermination and unreachability.}%
	\label{tab:forall-non-termination}%
\end{wraptable}%
We now demonstrate how existing triples capture properties such as must-non{\-}termination, may-termina{\-}tion, unreachability, and reachability. 
Our initial focus is on illustrating $\forall$-properties, see \Cref{tab:forall-non-termination}.

\begin{wraptable}[8]{r}{.66\textwidth}%
	\vspace*{-.5\intextsep}%
	\renewcommand{\arraystretch}{1.25}%
	\small%
	\centering%
	\begin{adjustbox}{max width=.99\linewidth}
		\begin{tabular}{@{\hspace*{-.5em}}ccl}       
			\textbf{Triple}  & \textbf{Semantics} & \textbf{Property}\\
			\hline
			\hspace*{2.1em}$\not\pcl{P}{C}{\false}$ & $\exists \sigma\in P\mydot\exists\tau\mydot \tau\in \sem{C}{\sigma}$ & May-Termination\\
			\hspace*{1.7em}$\not\tcl{P}{C}{\true}$ &  $\exists \sigma\in P\mydot \not\exists\tau\mydot  \tau\in\sem{C}{\sigma}$ & Must-Nontermination\\
			\hspace*{-.4em}$\not\pil{\false}{C}{Q}$ & $\exists \tau\in Q\mydot \exists \sigma \mydot \tau \in \sem{C}{\sigma}$& Reachability\\
			$\not\til{\true}{C}{Q}$ & $\exists \tau\in Q\mydot \not\exists \sigma \mydot \tau \in \sem{C}{\sigma}$& Unreachability \\
		\end{tabular}
	\end{adjustbox}%
	\medskip%
	\caption{$\exists$-properties on nontermination and unreachability.}
	\label{tab:exists-non-termination}%
\end{wraptable}%
It is noteworthy that the transition from partial to total involves the negation of the properties under consideration. 
Specifically, the negation of may-termination corresponds to must-nontermination, and unreachability is the negation of reachability. 
A useful perspective is to view reachability as the may-termination of backward semantics, while unreachability can be conceptualized as its must-termination.
By examining their falsification, we derive their dual counterparts, characterized as $\exists$-properties, see \Cref{tab:exists-non-termination}.

\subsection{Expressing Quantitative Weakest Pre}
\label{se:expressivity:qwp}
In this section we show that our calculus subsumes several existing calculi. 
We define $\iversone{\sigma}(\tau)=\bone$ if $\tau=\sigma$ and $ \iversone{\sigma}(\tau)=\bzero$ otherwise.

\paragraph*{Nondeterministic Programs}
We start by defining hyperquantities subsuming existing angelic weakest pre and demonic weakest liberal pre~\cite{ZK22}.

\begin{definition}[Hyper Suprema and Infima]
	For a given semiring $\mathcal{A} = \langle U, \oplus, \odot, \bzero, \bone \rangle$ and a quantity $f\colon\States \to U$, we define hyperquantities
	\begin{align*}
		\bigcurlyvee [f]& \defeq \lambda \mu.~\bigcurlyvee_{\sigma\in\supp{\mu}} f(\sigma)\qquad& 
		\bigcurlywedge [f]& \defeq \lambda \mu.~\bigcurlywedge_{\sigma\in\supp{\mu}} f(\sigma)~,
	\end{align*}%
	that take as input quantities $\mu\colon \States \to U$. 
	Intuitively, $\bigcurlyvee [f]$ and $\bigcurlywedge [f]$ map a given $\mu$ to the maximum (minimum) value of $f(\sigma)$ where $\sigma$ is drawn from the support set $\supp{\mu}$.%
	\qedtriangle
\end{definition}%
\begin{restatable}[Subsumption of Quantitative $\wpsymbol$, $\wlpsymbol$ for Nondeterministic Programs~\cite{ZK22}]{theorem}{subwpwlp}%
	Let $\mathcal{A}=\langle \Rinf, \max, \min, \bzero, \bone\rangle$. For any quantities $g, f$ and any program $C$ satisfying the syntax of~\textnormal{\cite[Section 2]{ZK22}}:
	\begin{align*}
		\whp{C}{\bigcurlywedge [f]}(\iversone{\sigma}) \eeq \wlp{C}{f}(\sigma)
	\qqand
		\whp{C}{\bigcurlyvee [f]}(\iversone{\sigma}) \eeq \wp{C}{f}(\sigma)
	\end{align*}
\end{restatable}%
\noindent%
The result follows from the fact that $\whp{C}{\bigcurlywedge [f]}(\iversone{\sigma})$ and $\whp{C}{\bigcurlyvee [f]}(\iversone{\sigma})$ compute respectively the maximum and the minimum value of $f$ in the support of $\sp{C}{\iversone{\sigma}}$, which is the set of reachable states starting from $\sigma$. Our calculus is strictly more expressive than~\cite{ZK22} as our syntax is richer and allows to reason about weighted programs as well.

\paragraph*{Probabilistic Programs}
By employing the expected value hyperquantity, we show how $\whpsymbol$ subsumes $\wpsymbol$ and $\wlpsymbol$ for deterministic and probabilistic programs~\cite{benni_diss} as well.%
\begin{restatable}[Subsumption of Quantitative $\wpsymbol$, $\wlpsymbol$ for probabilistic programs~\cite{benni_diss}]{theorem}{subwp}%
	\label{thm:sub-wp-wlp-prob}
	Let $\Probs = \langle [0, 1], +, \cdot, 0, 1\rangle$. For any quantities $g, f$ and any \underline{non-non}deterministic program $C$:
	\begin{align*}
		\whp{C}{\E [f]}(\iversone{\sigma}) = \wp{C}{f}(\sigma)
\quad\text{ and }\quad
		\whp{C}{\E [f]+1-\E [1]}(\iversone{\sigma}) = \wlp{C}{f}(\sigma).
	\end{align*}
\end{restatable}%
\noindent%
The results stem from our calculus, which computes $\E [f]$ on the final distribution $\sp{C}{\iversone{\sigma}}$ using the expected values hyperquantity, which precisely yields $\wp{C}{f}(\sigma)$. Additionally, it is known~\cite[Theorem 4.25]{benni_diss} that for nondeterministic programs $\wlp{C}{f}(\sigma)$ calculates the expected value of $f$ in the final distribution $\sp{C}{\iversone{\sigma}},$ but adjusted for the probability of nontermination. This latter probability is in our setting the hyperquantity $1-\E [1]$.

\paragraph*{Nondeterminism, Regular Languages, and Schedulers}
While the results above highlight that many existing $\wpsymbol$ are mere specializations of $\whpsymbol$ for single initial pre-states, we claim that there are some limitations as well, particularly in how nondeterminism is resolved. The main reason is that all of our transformers, being related to the strongest post $\spsymbol$, cannot detect whether a program $C$ starting from $\sigma$ diverges for at least one possible execution. Therefore we cannot express demonic $
\wpsymbol$ and angelic $\wlpsymbol$. The closest attempt is to define the following hyperquantities.

\begin{definition}[Demonic Weakest Pre and Angelic Weakest Liberal Pre]
	Let the ambient semiring be $\mathcal{A}=\langle \Rinf, \max, \min, \minfty, \pinfty\rangle$. Given a quantity $f\colon\States \to \Rinf$, we define hyperquantities
	\begin{align*}
		\bigcurlywedge[f]_{\Term} \qdefeq \lambda \mu\mydot\bigcurlywedge_{\mathclap{\sigma\in\supp{\mu}}}\, f(\sigma) \ccurlywedge \bigcurlyvee_{\mathclap{\sigma \in\supp{\mu}}}\, \pinfty 
		\qqand
		\bigcurlyvee[f]_{\Diverge}  \qdefeq \lambda \mu\mydot \bigcurlyvee_{\mathclap{\sigma\in\supp{\mu}}}\, f(\sigma) \ccurlyvee  \bigcurlywedge_{\mathclap{\sigma\in\supp{\mu}}}\, \minfty
		  ~.
	\end{align*}%
One can define two novel transformers:%
\begin{align*}
	\infwp{C}{f}(\sigma) \defeq \whp{C}{\bigcurlywedge[f]_{\Term}}(\iversone{\sigma}) ~\textnormal{~and~}~
	\supwlp{C}{f}(\sigma) \defeq \whp{C}{\bigcurlyvee[f]_{\Diverge} }(\iversone{\sigma}) \tag*{\qedtriangle}
\end{align*}%
\end{definition}%
\noindent%
Intuitively, $\infwp{C}{f}(\sigma)$ operates akin to a demonic weakest pre calculus by determining the minimum value of $f$ after the execution of program $C$ starting from $\sigma$. 
However, unlike the demonic weakest pre calculus in~\cite{benni_diss}, we do not necessarily assign the value bottom $\bzero$ if the program has a single diverging trace; 
instead, we do so only when all traces are diverging. 
Similarly, for $\supwlpsymbol$, our calculus outputs $\bone$ if all traces are diverging. 
In other words, both our $\infwpsymbol$ and angelic $\supwlpsymbol$ attempt to avoid termination whenever possible, mirroring the behavior of the angelic $\wpsymbol$ and demonic $\wlpsymbol$ as discussed in~\cite[Section 6.2]{ZK22}.

To better illustrate, let us demonstrate that our demonic weakest pre ($\infwpsymbol$) and angelic weakest liberal pre ($\supwlpsymbol$) transformers differ from those in~\cite{benni_diss} through an example.

\begin{example}[Comparing Nondeterminism]
	Let $\dwpsymbol$ and $\awlpsymbol$ be the demonic weakest pre and angelic weakest liberal pre in~\cite{benni_diss}, and let $C=\NDCHOICE{\DIVERGE}{\SKIP}$. Then:
	\begin{itemize}
		\item $\dwp{C}{\iverson{\true}}=\iverson{\false} \quad\neq\quad \iverson{\true}=\infwp{C}{\iverson{\true}}$
		\item $\awlp{C}{\iverson{\false}}=\iverson{\true} \quad\neq\quad \iverson{\false}=\supwlp{C}{\iverson{\false}}$\qedtriangle
	\end{itemize}
\end{example}%
\noindent%
Conventional treatment of nondeterministic programs in established weakest pre calculi inherently involve schedulers~\cite[Definition 3.7]{benni_diss} designed to resolve nondeterminism, seeking the maximum or minimum expected value across all possible schedulers.
In contrast, our approach aligns with the Incorrectness Logic literature, using Kleene Algebra and strongest-post-style calculi as program semantics~\cite{OHearn19, Zilberstein2023, Dardinier2023hyper, ZK22}: 
for nondeterministic programs, we treat all choices as if they were executed. 
To further highlight the differences, using a semantics involving schedulers and extending $\dwpsymbol$ in the sense of~\citet{benni_diss} would invalidate the synctactic sugar of branching and loops.

\begin{example}
	Let $\dwpsymbol$ and $\awlpsymbol$ be the demonic weakest pre and angelic weakest liberal pre of~\citet{benni_diss}. 
	We extend both for the assume statement, obtaining:
	\begin{align*}
		\dwp{\ASSUME{\guard}}{f}=\guard \curlywedge f \qquad \text{and}\qquad\awlp{\ASSUME{\guard}}{f}=\inguard \curlyvee f
	\end{align*}
	We have $\dwp{\blue{\ITE{\true}{\SKIP}{\SKIP}}}{\iverson{\true}} = \blue{\iverson{\true}}$, 
	whereas for the seemingly equivalent $\NDCHOICE{\COMPOSE{\ASSUME{\true}}{\SKIP}}{\COMPOSE{\ASSUME{\false}}{\SKIP}}$ we have:
	\begin{align*}
		& \dwp{\orange{\NDCHOICE{\COMPOSE{\ASSUME{\true}}{\SKIP}}{\COMPOSE{\ASSUME{\false}}{\SKIP}}}}{\iverson{\true}} 
		 \eeq  \iverson{\true} \curlywedge \iverson{\false}	 \eeq \orange{\iverson{\false}}
	\end{align*}
	Similarly, $\awlp{\blue{\ITE{\true}{\SKIP}{\SKIP}}}{\iverson{\false}}=\blue{\iverson{\false}}$ but:
	\begin{align*}
		& \awlp{\orange{\NDCHOICE{\COMPOSE{\ASSUME{\true}}{\SKIP}}{\COMPOSE{\ASSUME{\false}}{\SKIP}}}}{\iverson{\false}} 
		=  \iverson{\false} \curlyvee \iverson{\true}	 = \orange{\iverson{\true}} \tag*{\qedtriangle}
	\end{align*}
\end{example}%
\noindent%
Whilst the fact that demonic total correctness is inexpressible in KAT~\cite{Kozen97} because it lacks a way of reasoning about nontermination~\cite{Wright2002FromKA}, here we argue that also angelic partial correctness in the sense of~\cite{benni_diss} is inexpressible. This highlights the fact that regular languages, such as KAT variants, are not equivalent to guarded imperative languages in general.


\section{Properties}
\label{se:healthiness}

Our quantitative hyper transformers enjoy several \emph{healthiness properties}, some of which are analogous to Dijkstra's, Kozen's, or McIver \& Morgan's calculi.
In this section, we argue that there exists only one backward hyper predicate transformer, as $\whpsymbol$ enjoys several properties and dualities that both liberal and non-liberal weakest pre style calculus have.

\subsection{Healthiness Properties}

\begin{restatable}[Healthiness Properties of Quantitative Transformers]{theorem}{wpwlpspslphealthiness}%
	\label{thm:whphealthiness}%
	For all programs $C$, $\whpC{C}$ satisfies the following properties:
	\begin{enumerate}
		\item
		\label{thm:whphealthiness:conjunctive} 
			Quantitative universal conjunctiveness and disjunctiveness: For any set of hyperquantities \mbox{$S \subseteq \hyperA$},%
			\[
				\whp{C}{\prod S} \eeq \prod_{\hyperf \in S}~ \whp{C}{\hyperf}
				\qquad\text{and}\qquad
				\whp{C}{\sum S} \eeq \sum_{\hyperf \in S}~ \whp{C}{\hyperf}
		\]
		\item \label{thm:whphealthiness:strictness}
		$k$-Strictness:\quad For any \mbox{$k\in\PosRealsInf$}, $\whp{C}{\lambda f\mydot k} \eeq \lambda f\mydot k$.
		\item \label{thm:whphealthiness:mono}
		Monotonicity: $	\quad\hyperf \ppreceq \hyperg \qqimplies \whp{C}{\hyperf} \ppreceq\whp{C}{\hyperg}~.$

	\end{enumerate}
\end{restatable}%
\noindent{}%
Quantitative universal conjunctiveness and strictness in the context of $\wpsymbol$, as well as the notions of disjunctiveness and co-strictness for $\wlpsymbol$, serve as quantitative analogues of Dijkstra and Scholten’s original calculi. These properties have been explored in~\cite[Section 5.1]{ZK22}. We demonstrate that $\whpsymbol$ exhibits all these characteristics, as the $k$-strictness of $\whpsymbol$ implies both strictness and co-strictness. This observation aligns with our intuition that $\whpsymbol$ functions as both a liberal and a non-liberal calculus. Monotonicity, a fundamental property, enables the proof of the $\text{Cons}$ rule outlined in~\cite{Dardinier2023hyper}.

Sub- and superlinearity, extensively studied by Kozen, McIver \& Morgan, and Kaminski for probabilistic $\sfsymbol{w(l)p}$ transformers, also find applications in our $\whpsymbol$. Notably, our calculus adheres to linearity and, additionally, exhibits multiplicativity.
\begin{restatable}[Linearity]{theorem}{whplinearity}%
	\label{thm:whp:linearity}%
	For all programs $C$, $\whpC{C}$ is linear, i.e.\ for all \mbox{$\hyperf, \hyperg \in \hyperA$} and \emph{non-negative} constants $r\in\PosReals$, 
	$
		\whp{C}{r\cdot \hyperf +  \hyperg}  = r\cdot\whp{C}{\hyperf} + \whp{C}{\hyperg} ~.
	$%
\end{restatable}%
\noindent{}%

\todoin{I think ththe below does not hold for previous probabilistic transformers. Show an example.}

\begin{restatable}[Multiplicativity]{theorem}{whpmultiplicativity}%
	\label{thm:whp:multiplicativity}%
	For all programs $C$, $\whpC{C}$ is multiplicative, i.e.\ for all \mbox{$\hyperf, \hyperg \in \hyperA$} and \emph{non-negative} constants $r\in\PosReals$, 
$
		\whp{C}{r\cdot \hyperf\cdot  \hyperg} =  r\cdot\whp{C}{\hyperf} \cdot \whp{C}{\hyperg} ~.$
\end{restatable}%
\noindent{}%

\subsection{Relationship between Liberal and Non-liberal Transformers}
Various dualities between $\wpsymbol$ and $\wlpsymbol$ have been explored extensively in the literature. In Dijkstra's classical calculus, the duality relationship is expressed as $\wp{C}{\psi} = \neg\wlp{C}{\neg\psi}$. In quantitative settings, particularly in Kozen's and McIver \& Morgan's work on probabilistic programs, this duality extends to $\wp{C}{f} = 1 - \wlp{C}{1 - f}$ for 1-bounded functions $f$. This concept is further generalized to $\wp{C}{f} = - \wlp{C}{- f}$ in the case of non-probabilistic programs and unbounded quantities, as demonstrated in~\citet[Theorem 5.3]{ZK22}.

In this section, we argue that there exists only a single $\whpsymbol$ calculus that behaves both as a non-liberal and a liberal transformer.
\begin{restatable}[Liberal--Non-liberal Duality]{theorem}{whpduality}
	\label{thm:whp:duality}%
	For any program $C$ and any $k-$bounded hyperquantity $\hyperf$, we have $\whp{C}{\hyperf}\eeq k-\whp{C}{k-\hyperf}$.
\end{restatable}%
\noindent{}%
As a consequence of the liberal--non-liberal duality of Theorem~\ref{thm:whp:duality}, for hyperproperties we have:
\begin{align*}
	\hyperphi \iimplies\whp{C}{\hyperpsi} \qqiff  \whp{C}{\neg{\hyperpsi}} \iimplies \neg{\hyperphi}~.
\end{align*}

\subsection{Linear Hyperquantities}
In this section, we explore a specific category of hyperquantities from which we can deduce simplified rules akin to established $\wpsymbol$ calculi.

\begin{definition}[Linear Hyperquantities]
	\label{def:quantities:linear}
		A hyperquantity $\hyperf\in\hyperA$ is \emph{linear} if for any quantity $f\in\A$
		\begin{align*}
			\hyperf(r\cdot g\oplus f)\qeq r\cdot\hyperf(g)\ooplus \hyperf(f)~.
		\end{align*}%
\end{definition}%
\begin{restatable}[Weakest Hyper Pre for Linear Hyperquantities]{theorem}{hyperlinear}%
	\label{thm:whp-soundness-linear}
	For linear hyperquantities $\hyperf\in\hyperA$, the simpler rules in \textnormal{\Cref{table:whp:linear}} are valid.
\end{restatable}%
\noindent%
\begin{table}[t]
\small
	\renewcommand{\arraystretch}{1.25}
	\begin{tabular}{@{\hspace{.5em}}l@{\hspace{2em}}l@{\hspace{.5em}}}
		\hline\hline
		$\boldsymbol{C}$			& $\boldwhp{C}{\hyperf}$ \\
		\hline
		$\ASSIGN{x}{\ee}$			& $\hyperf\subst{x}{\ee}$ 	\\
		$\ASSIGNNONDET{x}$ & $\lambda f\mydot \hyperf(\bigoplus_{\alpha} f\subst{x}{\alpha})$\\
		$\WEIGHT{w}$		& 	$\hyperf\odot w $\\
		$\COMPOSE{C_1}{C_2}$		& $\whp{C_1}{\vphantom{\big(}\whp{C_2}{\hyperf}}$\\
		$\NDCHOICE{C_1}{C_2}$		& $\whp{C_1}{\hyperf}\oplus \whp{C_2}{\hyperf}$\\
		$\LOOP{C}{e}{e'}$		& $\bigoplus_{n\in\Nats}W_{e}^n(\hyperf\odot\eval{e'} )$ 	\\
		\hline
		$\ITE{\guard}{C_1}{C_2}$		& $\whp{C_1}{\hyperf}\odot\iguard \oplus \whp{C_2}{\hyperf}\odot\inguard$\\
		$\PCHOICE{C_1}{p}{C_2}$		& $\whp{C_1}{\hyperf}\odot p\oplus
		\whp{C_2}{\hyperf}\odot(1-p)$\\		
		$\WHILEDO{\guard}{C}$		& $\bigoplus_{n\in\Nats} W_{\guard}^n(\hyperf\odot\inguard)$ \\
		\hline\hline
	\end{tabular}%
	\renewcommand{\arraystretch}{1}
	\caption{\small Rules for the weakest hyper pre transformer for linear posts $\hyperf$. Here, $W_{e} (X)=\whp{C}{X}\odot \eval{e}$}.
	\label{table:whp:linear}
\end{table}%
We observe that $\whp{\LOOP{C}{e}{e'}}{\hyperf}=\lfp  X\mydot  \hyperf\odot\eval{e'} \oplus \whp{C}{X}\odot \eval{e}$ holds true within the natural order of the provided semiring. When examining the semiring $\langle \Rinf, \max, \min, \minfty, \pinfty\rangle$, our calculus closely resembles the quantitative $\wpsymbol$ as described in~\citet{ZK22}, albeit in a more expressive context. Further, by adopting $\langle \Rinf, \min, \max, \pinfty, \minfty\rangle$, we derive rules analogous to quantitative $\wlpsymbol$ from~\citet{ZK22}. Notably, in the latter semiring, the natural order is reversed compared to the semiring $\langle \Rinf, \max, \min, \minfty, \pinfty\rangle$. In essence, for $\langle \Rinf, \min, \max, \pinfty, \minfty\rangle$, the least fixed point resulting from our iteration rule aligns with the rule of $\wlpsymbol$ defined through the greatest fixed point in~\citet{ZK22}.

Among linear hyperquantities we have all those in~\Cref{ex:hyperquantities-distributions} and of~\Cref{se:expressivity:qwp}. Additionally, we contend that by combining these properties, we can extend our reasoning to encompass other hyperquantities, such as the covariance of a random variable.

\begin{example}[Covariance]%
	\label{ex:cov}%
	\begin{align*}
	\!\!\!\!\!\!\!\!\whp{C}{\cov[f,g]} 
	& \eeq \whp{C}{\E[fg]-\E[f]\cdot\E[g]}\\
	& \eeq \whp{C}{\E[fg]}-\whp{C}{\E[f]\cdot \E[g]}\tag{by~\Cref{thm:whp:linearity}}\\
	& \eeq \whp{C}{\E[fg]}-\whp{C}{\E[f]}\cdot \whp{C}{\E[g]} \tag{by Theorem~\ref{thm:whp:multiplicativity}}
	\end{align*}
\end{example}

\section{Case Studies}
\label{sec:case-studies}

\allowdisplaybreaks

\begin{wrapfigure}[4]{r}{0.1\textwidth}
	\begin{minipage}{.99\linewidth}
	\vspace*{-2.1em}%
	  \footnotesize%
	  \abovedisplayskip=0pt%
	  \belowdisplayskip=0pt%
  \begin{align*}
		&\eqannotate{g'}\\
	  &\annotate{g}\\
	  &C\\
	  &\annotate{f}
  \end{align*}%
  \normalsize%
	\end{minipage}%
\end{wrapfigure}%
  In this section, we demonstrate the efficacy of quantitative weakest hyper pre reasoning. 
  We use the annotation style on the right to express that $g = \whp{C}{f}$ and furthermore that~$g' = g$.

\subsection{Proving hyperproperties}
In this section we show how to prove noninterference~\cite{goguen1982security} and generalized noninterference~\cite{McCullough87,McLean96} within $\whpsymbol$.

\subsubsection*{NI}
Noninterference amounts to proving that any two executions of the program with the same low-sensitivity inputs must have the same low outputs. This can be formalised by defining $\text{low}(l) \defeq \lambda S\mydot \forall \sigma_1, \sigma_2\in S\mydot \sigma_1(l) = \sigma_2(l)$ and proving $\text{low}(l) \subseteq \whp{C}{\text{low}(l)}$. For example consider the program and the $\whpsymbol$ annotations in~\Cref{fig:ex:prove:ni}. The program satisfies NI since $\text{low}(l) \subseteq\lambda S\mydot \forall \sigma_1, \sigma_2\in S \mydot   \sigma_1(h)>0 \land \sigma_2(h)>0 \implies \sigma_1(l)= \sigma_2(l)$.

\subsubsection*{GNI}
Generalized noninterference is a weaker property of NI: it permits two executions of the program with identical low-sensitivity inputs to yield different low outputs, provided that the discrepancy does not arise from their secret input. This concept can be formally expressed by defining $\text{glow}(l) \defeq \lambda S\mydot \forall \sigma_1, \sigma_2\in S\mydot \exists \sigma \in S\mydot \sigma (h) = \sigma_1(h) \land \sigma (l) = \sigma_2(l)$, where $\sigma$ denotes a potential third execution sharing the same secret input as $\sigma_1$ but producing the same low output as $\sigma_2$. GNI can be proved by checking $\text{low}(l) \subseteq \whp{C}{\text{glow}(l)}$. For example consider the program and the $\whpsymbol$ annotations in~\Cref{fig:ex:prove:gni}. The program satisfies GNI since $\text{low}(l) \subseteq  \lambda S\mydot \forall \sigma_1, \sigma_2\in \{ \sigma\subst{y}{\alpha} \mid \sigma \in S\}\mydot \exists \sigma \in x\{ \sigma\subst{y}{\alpha} \mid \sigma \in S\}\mydot  \sigma (h) = \sigma_1(h) \land \sigma (y+h) = \sigma_2(y+h)$.

\begin{center}
	\scriptsize
	\begin{minipage}{.4\textwidth}%
		\abovedisplayskip=0pt%
		\begin{align*}
			& \eqannotate{\lambda S\mydot \forall \sigma_1, \sigma_2\in S \mydot   \sigma_1(h)>0 \land \sigma_2(h)>0}\\ 
			& \sndannotate{\implies \sigma_1(l)= \sigma_2(l)}\\
			& \annotate{\lambda S\mydot  \forall \sigma_1, \sigma_2\in (h>0)(S)\mydot \sigma_1(l) = \sigma_2(l)}\\
			& \ASSUME{\mathit{h}>0}\\
			& \eqannotate{\lambda S\mydot \forall \sigma_1, \sigma_2\in S\mydot \sigma_1(l) = \sigma_2(l)}\\ 
			& \annotate{\lambda S\mydot \forall \sigma_1, \sigma_2\in S\mydot \sigma_1(l+1) = \sigma_2(l+1)}\\ 
			& \ASSIGN{\mathit{l}}{\mathit{l}+1}\\
			& \annotate{\lambda S\mydot \forall \sigma_1, \sigma_2\in S\mydot \sigma_1(l) = \sigma_2(l)}
		\end{align*}%
		\normalsize%
		\captionof{figure}{Proving noninterference}
		\label{fig:ex:prove:ni}
	\end{minipage}%
	\begin{minipage}{.6\textwidth}%
		\abovedisplayskip=0pt%
		\begin{align*}
			& \eqannotate{\lambda S\mydot \forall \sigma_1, \sigma_2\in \{ \sigma\subst{y}{\alpha} \mid \sigma \in S\}\mydot }\\
			& \sndannotate{\exists \sigma \in \{ \sigma\subst{y}{\alpha} \mid \sigma \in S\}\mydot  \sigma (h) = \sigma_1(h) \land \sigma (y+h) = \sigma_2(y+h)}\\
			& \annotate{\lambda S\mydot \forall \sigma_1, \sigma_2\in \exists_{\alpha}~S\subst{y}{\alpha}\mydot }\\
			& \sndannotate{\exists \sigma \in \exists_{\alpha}~S\subst{y}{\alpha}\mydot  \sigma (h) = \sigma_1(h) \land \sigma (y+h) = \sigma_2(y+h)}\\
			& \ASSIGNNONDET{\mathit{y}}\\
			& \annotate{\lambda S\mydot \forall \sigma_1, \sigma_2\in S\mydot \exists \sigma \in S\mydot  \sigma (h) = \sigma_1(h) \land \sigma (y+h) = \sigma_2(y+h)}\\
			& \ASSIGN{\mathit{l}}{\mathit{y}+\mathit{h}}\\
			& \annotate{\lambda S\mydot \forall \sigma_1, \sigma_2\in S\mydot \exists \sigma \in S\mydot  \sigma (h) = \sigma_1(h) \land \sigma (l) = \sigma_2(l)}
		\end{align*}%
	\normalsize%
		\captionof{figure}{Proving generalized noninterference (GNI)}
		\label{fig:ex:prove:gni}
	\end{minipage}%
\end{center}%

\subsection{Disproving hyperproperties}
As pointed in~\Cref{se:weakest-pre}, evaluating whether a program satisfies a specific hyperproperty necessitates proving two \HHL triples. For instance, when tackling noninterference, one must attempt to establish \emph{both} $\hhl{\text{low}(l)}{C_{\text{ni}}}{\text{low}(l)}$ and $\hhl{Q}{C_{\text{ni}}}{\neg\text{low}(l)}$ (for some $Q \Rightarrow \text{low}(l)$). In this section, we illustrate the advantage of our calculus by disproving NI and GNI.

\begin{wrapfigure}{r}{.575\textwidth}
	\scriptsize%
	\begin{minipage}{.53\textwidth}%
		\abovedisplayskip=-1.8em%
	\begin{align*}
		& \eqannotate{\lambda S\mydot \forall \sigma_1, \sigma_2\in S \mydot   \sigma_1(h)>0 \land \sigma_2(h)>0 \implies\sigma_1(l+h) = \sigma_2(l+h)}\\
		& \annotate{\lambda S\mydot  \forall \sigma_1, \sigma_2\in (h>0)(S)\mydot \sigma_1(l+h) = \sigma_2(l+h)}\\
		& \ASSUME{\mathit{h}>0}\\
		& \annotate{\lambda S\mydot \forall \sigma_1, \sigma_2\in S\mydot \sigma_1(l+h) = \sigma_2(l+h)}\\ 
		& \ASSIGN{\mathit{l}}{\mathit{l}+\mathit{h}}\\
		& \annotate{\lambda S\mydot \forall \sigma_1, \sigma_2\in S\mydot \sigma_1(l) = \sigma_2(l)}\\
	\end{align*}%
	\vspace{-.5cm}
		\captionof{figure}{Disproving noninterference}
		\label{fig:ex:disprove:ni}
	\end{minipage}%
	\vspace{-.3cm}
\end{wrapfigure}
\subsubsection*{NI}
Disproving NI amounts to proving $\text{low}(l) \not\subseteq \whp{C}{\text{low}(l)}$, which is true for the program in~\Cref{fig:ex:disprove:ni}. For example, take $S=\{\sigma_1, \sigma_2\}$ such that $\sigma_1(l)=\sigma_2(l)=0$ and $\sigma_1(h)=1\neq \sigma_2(h)=2$. Clearly $S\in\text{low}(l)$ but $S\not\in\whp{C}{\text{low}(l)}$.

\ \\ \vspace{-.35cm}
\subsubsection*{GNI}
Disproving GNI amounts to prove $\text{low}(l) \not\subseteq \whp{C}{\text{glow}(l)}$, which is true for the program in~\Cref{fig:ex:disprove:gni}. For example, take $S=\{\sigma_1, \sigma_2\}$ such that $\sigma_1(l)=\sigma_2(l)=0$ and $\sigma_1(h)=1\neq \sigma_2(h)=100$. Clearly $S\in\text{low}(l)$ but $S\not\in\whp{C}{\text{glow}(l)}$.
\begin{center}
	\scriptsize%
		\abovedisplayskip=-1em%
	\begin{align*}
		& \annotate{\lambda S\mydot \forall \sigma_1, \sigma_2\in A=\{ \sigma\subst{y}{\alpha} \mid \sigma \in S, \alpha\in[0,10]\} \mydot
			\exists \sigma \in A \mydot\sigma (h) = \sigma_1(h) \land \sigma (y+h) = \sigma_2(y+h)}\\
		& \ASSIGNNONDET{\mathit{y}}\\
		& \annotate{\lambda S\mydot \forall \sigma_1, \sigma_2\in A=\{ \sigma\mid \sigma \in S\land \sigma(y)\in[0,10]\} \mydot 
			\exists \sigma \in A \mydot\sigma (h) = \sigma_1(h) \land \sigma (y+h) = \sigma_2(y+h)}\\
		& \ASSUME{0 \leq y\leq 10}\\
		& \annotate{\lambda S\mydot \forall \sigma_1, \sigma_2\in S\mydot \exists \sigma \in S\mydot  \sigma (h) = \sigma_1(h) \land \sigma (y+h) = \sigma_2(y+h)}\\
		& \ASSIGN{\mathit{l}}{\mathit{y}+\mathit{h}}\\
		& \annotate{\lambda S\mydot \forall \sigma_1, \sigma_2\in S\mydot \exists \sigma \in S\mydot  \sigma (h) = \sigma_1(h) \land \sigma (l) = \sigma_2(l)}
	\end{align*}%
			%
			\captionof{figure}{Disproving generalized noninterference}
		\label{fig:ex:disprove:gni}
\end{center}%

\subsection{Quantitative reasoning}
In this section, we demonstrate how $\whpsymbol$ enables quantitative reasoning.

\subsubsection{Quantitative Information Flow}
\label{ex:qif}
Consider the program $C_\text{qif}$ in~\Cref{fig:ex:qif}. Similarly to~\cite[Section 8.1]{ZK22}, we want to infer what is the maximum initial value that the secret variable $h$ can have, by observing a final value $l'$ for the low-sensitive variable $l$. By using $\whpsymbol$, it is sufficient to consider the hyperpostquantity $\hyperf_{l'}=\lambda f\mydot \bigcurlyvee_{\tau}~(\iverson{l=l'} \odot f )(\tau)(h)$. Indeed, $\whp{C_\text{qif}}{\hyperf_{l'}}(h)$ tells, what is the maximum value of $\sp{C_\text{qif}}{h}(\tau)$ among those final states $\tau$ where the value $l'$ has been observed. Since we know from~\cite{ZK22} that $\sp{C_\text{qif}}{f}(\tau)$ produces the maximum initial value of $h$, we have that $\whp{C_\text{qif}}{\hyperf_{l'}}(h)$ correctly yields the maximum initial value of $h$. For example, $\whp{C_\text{qif}}{\hyperf_{80}}(h)=7$, meaning that if we observe $80$ as the value of $l$, we know that initially $h$ would have been at most $7$.

\begin{center}
	\scriptsize
	\begin{minipage}{.5\textwidth}%
		\abovedisplayskip=-1em%
		\begin{align*}
			& \annotate{\lambda f\mydot \curlyvee_{\sigma}~(\iverson{99=l'}  \odot\iverson{h>7}\curlyvee \iverson{80=l'} \odot\iverson{h\leq 7} )(\sigma)\odot f(\sigma)}\\
			& \IF{ \mathit{h} > 7} \quad \annotate{\lambda f\mydot \ccurlyvee_{\tau}~(\iverson{99=l'} \odot f )(\tau)}\\
			& \qquad \ASSIGN{\mathit{l}}{99} \\
			& \ELSE \quad \annotate{\lambda f\mydot \curlyvee_{\tau}~(\iverson{80=l'} \odot f )(\tau)}\\
			& \qquad \ASSIGN{\mathit{l}}{80} \\
			& \} \\
			& \annotate{\lambda f\mydot \bigcurlyvee_{\tau}~(\iverson{l=l'} \odot f )(\tau)}
		\end{align*}%
		%
		\captionof{figure}{Computing quantitative information flow}
	\label{fig:ex:qif}
	\end{minipage}%
\begin{minipage}{.5\textwidth}%
	\abovedisplayskip=-1em%
\begin{align*}
	& \annotate{
	\bigoplus_{n\in\Nats}~\E[(1+n)^2]\odot 0.5^{n+1}-
	\big(\bigoplus_{n\in\Nats}~\E[1+n]\odot 0.5^{n+1}\big)^2}   \\
	& \ASSIGN{x}{1}\\
	& \annotate{\bigoplus_{n\in\Nats}~\E[(x+n)^2]\odot 0.5^{n+1}-
	\big(\bigoplus_{n\in\Nats}~\E[x+n]\odot 0.5^{n+1}\big)^2} \\
	& \LOOP{(\ASSIGN{x}{x+1})}{\frac 1 2}{\frac 1 2}\\
	& \annotate{\E[x^2]-\E[x]^2}\\ 
	& \annotate{\cov[x,x]}
\end{align*}%
	\captionof{figure}{Computing the variance of a random variable}
	\label{fig:ex:variance}
\end{minipage}%
\end{center}

\subsubsection{Variance}
\label{ex:variance}

We show how to compute the variance of a random variable using $\whpsymbol$. Let's consider the following gaming scenario: a player flips a fair coin continuously until a head appears. To assess the variance in the number of flips required to conclude the game, we model this scenario with the program in~\Cref{fig:ex:variance}.
%
%
%
We leverage~\Cref{ex:cov} to compute $\whp{\LOOP{\ASSIGN{x}{x+1}}{\frac 1 2}{\frac 1 2}}{\cov[x,x]}$ compositionally, by computing $\whp{\LOOP{\ASSIGN{x}{x+1}}{\frac 1 2}{\frac 1 2}}{\E[x^2]}$ and $\whp{\LOOP{\ASSIGN{x}{x+1}}{\frac 1 2}{\frac 1 2}}{\E[x]}$ individually (full calculations of the Kleene's iterates are in~\Cref{app:examples}), obtaining:%
\begin{align*}
	\whp{\LOOP{\ASSIGN{x}{x+1}}{\frac 1 2}{\frac 1 2}}{\E[x^2]-\E[x]^2} &\eeq \bigoplus_{n\in\Nats}~W_{0.5}^n (\E[x^2]\odot 0.5)-
	\big(\bigoplus_{n\in\Nats}~W_{0.5}^n (\E[x]\odot 0.5)\big)^2
	\\
	&\eeq \bigoplus_{n\in\Nats}~\E[(x+n)^2]\odot 0.5^{n+1}-
	\big(\bigoplus_{n\in\Nats}~\E[x+n]\odot 0.5^{n+1}\big)^2
\end{align*}%
Finally, we take as input any probability distribution $\mu$ and compute the variance via:%
\begin{align*}
	\whp{C}{\cov[x,x]}(\mu)
	&\eeq \big(\bigoplus_{n\in\Nats}~\E[(1+n)^2]\odot 0.5^{n+1}-
		\big(\bigoplus_{n\in\Nats}~\E[1+n]\odot 0.5^{n+1}\big)^2\big)(\mu)\\ 
	&\eeq \sum (1+n)^2\cdot 0.5^{n+1}- (\sum (1+n)\cdot 0.5^{n+1})^2 \eeq 6-4=2~.
\end{align*}%
We contend that employing $\whpsymbol$ offers the advantage of mechanization and compositional computation without necessitating specialized knowledge of probability theory.

\section{Related Work}

\subsubsection*{Relational program logics}
Relational Hoare Logics were initially introduced by~\citet{benton2004simple}. Subsequently, several extensions emerged, including to reason about probabilistic programs via couplings \cite{barthe2009formal}.
Later, \citet{Maillard19}, proposed a general framework for developing relational program logics with effects based on Dijkstra Monads \cite{maillard2019dijkstra}. While effective, this framework is limited to 2-properties and thus does not apply to, e.g., monotonicity and transitivity, which are properties of \emph{more than} two executions.

\citet{Sousa16,Osualdo22} introduced logics for $k$-safety properties, but they cannot prove liveness. \citet{Dickerson22} introduced the first logic tailored for $\forall^*\exists^*$-hyperproperties, enabling, among others, proof and disproof of $k$-safety properties. Nonetheless, it has limited under-approximation capabilities: e.g., it does not suport incorrectness \`a la \citet{OHearn19}, and cannot disprove triples within the same logic. For instance, it cannot disprove GNI, a task which can only be completed by---to the best of our knowledge---\HHL, \OL, and our framework.

\subsubsection*{Unified Program Logics}

Similar to Outcome Logic (\OL) \cite{Zilberstein2023,zilberstein2024outcome} and Weighted Programming \cite{batz2022weighted}, our calculus utilizes semirings to capture branch weights. This approach enables the development of a weakest-pre style calculus for Outcome Logic. While \OL is relatively complete \cite{zilberstein2024relatively}, the derivations are not always straightforward.
Weakest Hyper-pre can be used to \emph{mechanically} derive \OL triples with the weakest precondition for a given postcondition. 
Weakest Hyper-pre also subsumes Hyper Hoare Logic \cite{Dardinier2023hyper}, which is similar to \OL, but specialized to nondeterministic programs.

Our approach surpasses Weighted Programming by facilitating reasoning about multiple outcomes. Our calculus also supports quantitative reasoning, demonstrating its versatility by encompassing various existing quantitative $\wpsymbol$ instances through the adaptation of hyperquantities.

\subsubsection*{Predicate Transformers} These were first introduced by \citet{Dijkstra76,Dijkstra1990}, who created propositional weakest pre- and strongest postcondition calculi. \citet{DBLP:journals/jcss/Kozen85,McIverM05} lifted these to a quantitative setting, introducing Probabilistic Propositional Dynamic Logic and weakest preexpectations for computing expected values over probabilistic programs. Many variants of weakest preexpectation now exist \cite{benni_diss,quantitative_sl}.
We build on this line of work by extending these predicate transformers to hyperproperties. This gives us the flexibility to express a broader range of quantitative properties, as shown in \Cref{sec:case-studies}.%

\section{Conclusion}

Recent years have seen a focus on logics for proving properties other than classical partial correctness. E.g., program \emph{security} is a \emph{hyperproperty}, and \emph{incorrectness} must \emph{witness a faulty execution}.

Recent work on Outcome Logic \cite{Zilberstein2023,zilberstein2024relatively,zilberstein2024outcome} and Hyper Hoare Logic \cite{Dardinier2023hyper} has shown that all of these properties can be captured via a single proof system. In this paper, we build upon those logics, but approach the problem using quantitative predicate transformers. 
This has allowed us to create a single calculus that can be used to \emph{prove}, but also \emph{disprove}, a variety of correctness properties. 
In addition, it can be used to derive advanced quantitative properties for programs too, such as variance in probabilistic programs.

The predicate transformer approach has two key benefits. First, it provides a calculus to mechanically derive specifications. Second, it finds the \emph{most precise} pre, so as to remove guesswork around obtaining a precondition in the aforementioned logics. As we have demonstrated, this brings about new ways of proving---and disproving---hyperproperties for a variety of program types.

\bibliography{literature}
\pagebreak
\appendix
\section{Quantitative Strongest Post and Weakest Pre}

\subsection{Proof of Soundness for \textnormal{$\spsymbol$}, Thereom~\ref{thm:sp-soundness}}
\label{proof:sp-soundness}

\spsoundness*
\begin{proof}
	We prove Theorem~\ref{thm:sp-soundness} by induction on the structure of $C$.
	For the induction base, we have the atomic statements:%
	\paragraph{The assignment $\ASSIGN{x}{e}$:} We have%
	\begin{align*}
		\sp{\ASSIGN{x}{\ee}}{f}(\tau)
		\eeq & \big(\bigoplus_\alpha~  f\subst{x}{\alpha}\odot \iverson{x = e\subst{x}{\alpha}}\big)(\tau)\\
		\eeq & \bigoplus_{\alpha\colon\tau(x)=\tau(e\subst{x}{\alpha})}~ f\subst{x}{\alpha}(\tau)\\
		\eeq & \bigoplus_{\alpha\colon\tau(x)=\tau(e\subst{x}{\alpha})}~ f(\tau\subst{x}{\alpha})\\
		\eeq & \bigoplus_{\alpha\colon \tau\subst{x}{\alpha}\subst{x}{\tau(\ee\subst{x}{\alpha})}=\tau}~ f(\tau\subst{x}{\alpha})\\
		\eeq & \bigoplus_{\alpha\colon\tau\subst{x}{\alpha}\subst{x}{\tau\subst{x}{\alpha}(\ee)}=\tau}~ f(\tau\subst{x}{\alpha})\\
		\eeq &	\bigoplus_{\sigma\in\States,\sigma\subst{x}{\sigma(\ee)}=\tau}~ f(\sigma) \tag{by taking $\sigma=\tau\subst{x}{\alpha}$}\\
		\eeq &	\bigoplus_{\sigma\in\States}~ f(\sigma)\odot \iverson{\sigma\subst{x}{\sigma(\ee)}=\tau} \\		%
		\eeq &	
		\bigoplus_{\sigma\in\States}~f(\sigma)\odot \eval{\ASSIGN{x}{\ee}}(\sigma,\tau)~.
	\end{align*}%
	\paragraph{The nondeterministic assignment $\ASSIGNNONDET{x}$:} We have%
	\begin{align*}
		\sp{\ASSIGNNONDET{x}}{f}(\tau)
		\eeq & \big(\bigoplus_\alpha~  f\subst{x}{\alpha}\big)(\tau)\\
		\eeq & \bigoplus_{\alpha}~ f(\tau\subst{x}{\alpha})\\
		\eeq &	\bigoplus_{\sigma\in\States,\exists\alpha\mydot \tau\subst{x}{\alpha}=\sigma}~ f(\sigma) \tag{by taking $\sigma=\tau\subst{x}{\alpha}$}\\
		\eeq &	\bigoplus_{\sigma\in\States}~ f(\sigma)\odot \bigoplus_{\alpha\in\Nats}   \iverson{\sigma\subst{x}{\alpha}=\tau} \\		%
		\eeq &	
		\bigoplus_{\sigma\in\States}~f(\sigma)\odot \eval{\ASSIGNNONDET{x}}(\sigma,\tau)~.
	\end{align*}%

	\paragraph{The weighting $\WEIGHT{w}$:} We have%
	\begin{align*}
		\sp{\WEIGHT{w}}{f}(\tau) 
		\eeq &  (f\odot w)(\tau)\\
		\eeq &  f(\tau) \odot w(\tau)\\
		\eeq & \bigoplus_{\sigma\in\States} 
		~f(\sigma)\odot w(\tau)\odot \iverson{\sigma=\tau} \\
		\eeq & \bigoplus_{\sigma\in\States} 
		~f(\sigma)\odot w(\sigma)\odot \iverson{\sigma=\tau} \\
		\eeq & \bigoplus_{\sigma\in\States}~f(\sigma)\odot \eval{\WEIGHT{w}}(\sigma,\tau)~.
	\end{align*}%
	\noindent{}%

	%
	
	%
	This concludes the proof for the atomic statement.%
	\paragraph{Induction Hypothesis:} 
	For arbitrary but fixed programs $C$, $C_1$, $C_2$, we proceed with the inductive step on the composite statements.%
	\paragraph{The sequential composition $\COMPOSE{C_1}{C_2}$:} We have%
	\begin{align*}
		\sp{\COMPOSE{C_1}{C_2}}{f}(\tau) 
		\eeq & \sp{C_2}{\sp{C_1}{f}}(\tau) \\
		\eeq & \bigoplus_{\sigma'\in\States} \sp{C_1}{f}(\sigma')\odot\eval{C_2} (\sigma',\tau)
		\tag{by I.H.~on $C_2$}\\
		\eeq & \bigoplus_{\sigma'\in\States} 
		\bigoplus_{\sigma\in\States}
		~f(\sigma)\odot \eval{C_1}(\sigma,\sigma')
		\odot\eval{C_2} (\sigma',\tau)
		\tag{by I.H.~on $C_1$}\\
		\eeq & \bigoplus_{\sigma\in\States} 
		\bigoplus_{\sigma'\in\States}
		~f(\sigma)\odot \eval{C_1}(\sigma,\sigma')
		\odot\eval{C_2} (\sigma',\tau) \tag{by commutativity of $\oplus$}\\	
		\eeq & \bigoplus_{\sigma\in\States} 
		~f(\sigma)\odot
		\bigoplus_{\sigma'\in\States}
		~ \eval{C_1}(\sigma,\sigma')
		\odot\eval{C_2} (\sigma',\tau) \tag{by distributivity of $\odot$}\\
		\eeq & \bigoplus_{\sigma\in\States}~f(\sigma)\odot \eval{\COMPOSE{C_1}{C_2}}(\sigma,\tau)~.
	\end{align*}%
	\noindent{}%
	\paragraph{The nondeterministic choice $\NDCHOICE{C_1}{C_2}$:} We have%
	\begin{align*}
		\sp{\NDCHOICE{C_1}{C_2}}{f}(\tau) 
		\eeq &  \sp{C_1}{f} \oplus \sp{C_2}{f} \\
		\eeq & \bigoplus_{\sigma\in\States}f(\sigma)\odot\sem{C_1}{\sigma,\tau}
		\oplus \bigoplus_{\sigma\in\States}f(\sigma)\odot\sem{C_2}{\sigma,\tau}
		\tag{by I.H.~on $C_1, C_2$}\\
		\eeq & \bigoplus_{\sigma\in\States} 
		~f(\sigma)
		\odot (\sem{C_1}{\sigma,\tau}\oplus \sem{C_2}{\sigma,\tau})\tag{by distributivity of $\odot$}\\
		\eeq & \bigoplus_{\sigma\in\States}~f(\sigma)\odot \eval{\NDCHOICE{C_1}{C_2}}(\sigma,\tau)~.
	\end{align*}%
	\noindent{}%

	\paragraph{The Iteration $\LOOP{C}{e}{e'}$:} 
	Let%
	\begin{align*}
		\Psi_{f}(X) &\eeq 
		f \oplus  \sp{C}{X\odot \eval{e}}~,
	\end{align*}%
	be the $\spsymbol$-characteristic function of the iteration $\LOOP{C}{e}{e'}$ with respect to any preanticipation $f$ and
	\begin{align*}
		F(X)(\sigma,\tau) &\eeq 
		\sigma(e)\odot\left(\bigoplus_{\sigma'\in\States} 
		\eval{C}(\sigma,\sigma')\odot
		 X (\sigma', \tau)
		\right) \oplus \sigma(e')\odot\iverson{\sigma=\tau}~,
	\end{align*}
	be the denotational semantics characteristic function of the loop $\LOOP{C}{e}{e'}$ for any input $
		\sigma,\tau\in\States$.
	We first prove by induction on $m$ that, for all $\tau\in\States,f\in\A$  we have: 
		\begin{equation}
			\label{eq:sp-loop-composition}
		\bigoplus_{\sigma\in\States}
		\Psi_{f}^m(\bzero)(\sigma) \odot \sigma (e)\odot  \eval{C}(\sigma,\tau) 
		\eeq  
		\bigoplus_{\sigma\in\States}\Psi_{
			\lambda \sigma'.f(\sigma)\odot\sigma (e)\odot  \eval{C}(\sigma,\sigma')}^m(\bzero)(\tau)~.
		\end{equation}
	For the induction base $m = 0$, consider the following:%
		\begin{align*}
			\bigoplus_{\sigma\in\States}
			\Psi_{f}^0(\bzero)(\sigma)\odot\sigma (e) \odot \eval{C}(\sigma,\tau) 
			& \eeq  \bzero \\
			& \eeq \bigoplus_{\sigma\in\States}\Psi_{
				\lambda \sigma'.f(\sigma)\odot\sigma (e)\odot \eval{C}(\sigma,\sigma')}^0(\bzero)(\tau)~.
		\end{align*}%
		As induction hypothesis, we have for arbitrary but fixed $m$ and all $\tau\in\States,f\in\A$%
		\begin{align*}
			\bigoplus_{\sigma\in\States}   \Psi_{f}^m(\bzero)(\sigma)\odot \sigma (e)\odot\eval{C}(\sigma,\tau) 
			\eeq  
			\bigoplus_{\sigma\in\States}\Psi_{
				\lambda \sigma'.f(\sigma)\odot\sigma (e)\odot \eval{C}(\sigma,\sigma')}^m(\bzero)(\tau)~.
		\end{align*}%
		For the induction step $m \longrightarrow m + 1$, consider the following:%
	\begin{align*}
		&\bigoplus_{\sigma\in\States} \Psi_{f} ^{m+1}(\bzero)(\sigma)\odot\sigma (e)\odot \eval{C}(\sigma,\tau) \\
		&\eeq 
		\bigoplus_{\sigma\in\States}
		\big(f \oplus \sp{C}{\Psi_{f}^m(\bzero)\odot \eval{e}}\big)(\sigma)
		\odot \sigma (e)\odot \eval{C}(\sigma,\tau) \\
		&\eeq 
		\bigoplus_{\sigma\in\States}
			f(\sigma)\odot\sigma (e)\odot \eval{C}(\sigma,\tau)
			\oplus
			 \sp{C}{\Psi_{f}^m(\bzero)\odot \eval{e}}(\sigma)\odot \sigma(e)\odot\eval{C}(\sigma,\tau)\tag{by distributivity of $\odot$}\\
		&\eeq 
	\bigoplus_{\sigma\in\States}
		f(\sigma)\odot\sigma(e)\odot \eval{C}(\sigma,\tau)
		\oplus
		\big(\bigoplus_{\sigma'\in\States}
			\Psi_{f} ^m(\bzero)(\sigma')\odot\sigma'(e)\odot\eval{C}(\sigma',\sigma)
			\big)\odot\sigma(e)
			\odot \eval{C}(\sigma,\tau)
		\tag{by I.H.\ on $C$}\\
		&\eeq 
		\bigoplus_{\sigma\in\States}
		f(\sigma)\odot\sigma (e)\odot  \eval{C}(\sigma,\tau)
			\oplus
			\big(\bigoplus_{\sigma'\in\States}
			\Psi_{
			\lambda \sigma''.f(\sigma')\odot\sigma'(e)\odot\eval{C}(\sigma',\sigma'')} ^m(\bzero)(\sigma)\big)
			\odot \sigma (e)\odot\eval{C}(\sigma,\tau)\tag{by I.H.\ on $m$}\\
			&\eeq 
			\bigoplus_{\sigma\in\States}
			 f(\sigma)\odot \sigma (e)\odot\eval{C}(\sigma,\tau)
			\\
			&\qquad\oplus\bigoplus_{\sigma\in\States}
			\big(\bigoplus_{\sigma'\in\States}
			\Psi_{
			\lambda \sigma''.f(\sigma')\odot\sigma'(e)\odot\eval{C}(\sigma',\sigma'')} ^m(\bzero)(\sigma)\big)
			\odot\sigma (e)\odot \eval{C}(\sigma,\tau)\tag{by associativity of $\oplus$}\\
			&\eeq 
			\bigoplus_{\sigma\in\States}
			f(\sigma)\odot\sigma (e)\odot  \eval{C}(\sigma,\tau)
			\\
			&\qquad\oplus\bigoplus_{\sigma\in\States}
			\bigoplus_{\sigma'\in\States}
			\Psi_{
			\lambda \sigma''.f(\sigma')\odot\sigma'(e)\odot\eval{C}(\sigma',\sigma'')} ^m(\bzero)(\sigma)
			\odot\sigma (e)\odot \eval{C}(\sigma,\tau)\tag{by distributivity of $\odot$}\\
			& \eeq  
			\bigoplus_{\sigma\in\States}
				f(\sigma)\odot \sigma (e)\odot\eval{C}(\sigma,\tau)
				\quad\oplus\quad
				\bigoplus_{\sigma\in\States}
				\bigoplus_{\sigma'\in\States}
				\Psi_{
					\lambda \sigma''.f(\sigma)\odot\sigma(e)\odot\eval{C}(\sigma,\sigma'')} ^m(\bzero)(\sigma')
					\odot \sigma'(e)\odot\eval{C}(\sigma',\tau) \tag{by commutativity of $\oplus$}\\
		& \eeq  
	\bigoplus_{\sigma\in\States}
	 f(\sigma)\odot \sigma (e)\odot\eval{C}(\sigma,\tau)
		\quad \oplus \quad
\sp{C}{\Psi_{
			\lambda \sigma''.f(\sigma)\odot\sigma(e)\odot \eval{C}(\sigma,\sigma'')
			} ^m(\bzero)\odot \eval{e}}(\tau)
		\tag{by I.H.\ on $C$ and associativity of $\oplus$}
		\\
		& \eeq 
		\bigoplus_{\sigma\in\States} 
		\Psi_{
			\lambda \sigma''.f(\sigma)\odot \sigma(e)\odot\eval{C}(\sigma,\sigma'')
			} ^{m+1}(\bzero)(\tau)
			\\
			& \eeq 
		\bigoplus_{\sigma\in\States} 
		\Psi_{
			\lambda \sigma'.f(\sigma)\odot \sigma(e)\odot\eval{C}(\sigma,\sigma')
			} ^{m+1}(\bzero)(\tau)
	\end{align*}
	This concludes the induction on $m$. We now prove by induction on $n$ that, for all $\tau\in\States,f\in\A$
	\begin{equation}
		\label{eq:sp-soundness-induction}
		 \Psi_{f}^{n}(\bzero)(\tau)\odot\tau(e')
		\eeq
		\bigoplus_{\sigma\in\States}
		f (\sigma)\odot F^n(\bzero)(\sigma,\tau)~.
	\end{equation}%
	For the induction base $n = 0$, consider the following:%
	\begin{align*}
		\Psi_{f}^{\bzero}(\bzero )(\tau)\odot\tau(e')
		& \eeq  \bzero \\
		& \eeq \bigoplus_{\sigma\in\States} f(\sigma)\odot F^0(\bzero)(\sigma,\tau)~.
	\end{align*}%
	As induction hypothesis, we have for arbitrary but fixed $n$ and all $\tau\in\States,f\in\A$%
	\begin{align*}
		\Psi_{f}^{n}(\bzero)(\tau)\odot\tau(e')
		\eeq
		\bigoplus_{\sigma\in\States}
		f (\sigma)\odot F^n(\bzero)(\sigma,\tau)~.
	\end{align*}%
	For the induction step $n \longrightarrow n + 1$, consider the following:%
	\begin{align*}
		&	\Psi_{f}^{n+1}(\bzero )(\tau)\odot \tau(e') \\
		& \eeq \left(f \oplus \sp{C}{\Psi_{f}^n(\bzero)\odot \eval{e}} \right)(\tau)\odot \tau(e') \\
		& \eeq  f(\tau) \odot \tau(e') \oplus  \sp{C}{\Psi_{f}^n(\bzero)\odot \eval{e}}(\tau)\odot \tau(e') \\
		& \eeq  f(\tau)\odot \tau(e')  \oplus
		\bigoplus_{\sigma\in\States}
		 	\Psi_{f}^n(\bzero)(\sigma)\odot \sigma(e)\odot\eval{C}(\sigma,\tau) \odot \tau(e') 
		\tag{by I.H.\ on $C$}\\
		& \eeq f(\tau) \odot \tau(e') \oplus
		\bigoplus_{\sigma\in\States}
		 	\Psi_{
				\lambda \sigma'.f(\sigma)\odot\sigma(e)\odot \eval{C}(\sigma,\sigma')}^n(\bzero)(\tau)\odot\tau(e')
				\tag{by Equation~\ref{eq:sp-loop-composition}}\\
		& \eeq
		f(\tau) \odot \tau(e') \oplus
		\bigoplus_{\sigma\in\States}
		\bigoplus_{\sigma'\in\States}
		f (\sigma)\odot \sigma(e)\odot
		\eval{C}(\sigma,\sigma')\odot F^n(\bzero) (\sigma', \tau)
		\tag{by I.H.\ on $n$} 
		\\
		& \eeq
		f(\tau) \odot \tau(e') \oplus
		\bigoplus_{\sigma\in\States}
		f (\sigma)\odot 
		\sigma(e)\odot\bigoplus_{\sigma'\in\States}\eval{C}(\sigma,\sigma')\odot F^n(\bzero) (\sigma', \tau) 
		\tag{by distributivity of $\odot$}
		\\
		& \eeq
		\big(\bigoplus_{\sigma\in\States}
		f(\sigma)\odot
		\sigma(e')\odot
		\iverson{\sigma=\tau}
		\big)
		\oplus
		\big(
		\bigoplus_{\sigma\in\States}
		f (\sigma)\odot 
		\sigma(e)\odot\bigoplus_{\sigma'\in\States} \eval{C}(\sigma,\sigma')\odot F^n(\bzero) (\sigma', \tau)\big)
		\\
		& \eeq
		\big(
		\bigoplus_{\sigma\in\States}
		f (\sigma)\odot 
		\sigma(e)\odot
		\bigoplus_{\sigma'\in\States} \eval{C}(\sigma,\sigma')\odot F^n(\bzero) (\sigma', \tau)\big)
		 \oplus
		 \big(\bigoplus_{\sigma\in\States}
		f(\sigma)\odot
		\sigma(e')\odot
		\iverson{\sigma=\tau}
		\big)
		\tag{by commutativity of $\oplus$}
		\\
		& \eeq
		\bigoplus_{\sigma\in\States}
		f (\sigma)\odot 
		\left(
			\big( \sigma(e)\odot\bigoplus_{\sigma'\in\States} \eval{C}(\sigma,\sigma')\odot F^n(\bzero) (\sigma', \tau)\big) \oplus 
		\sigma(e')\odot
		\iverson{\sigma=\tau}
		\right)
		\tag{by associativity of $\oplus$ and distributivity of $\odot$}
		\\
		& \eeq
		\bigoplus_{\sigma\in\States}
		f (\sigma)\odot F^{n+1}(\bzero)(\sigma,\tau)~.
		\end{align*}
	
	This concludes the induction on $n$. Now we have:

	\begin{align*}
		\sp{\LOOP{C}{e}{e'}}{f}(\tau) 
		\eeq & \big(\lfp  X\mydot  f \oplus \sp{C}{X\odot \eval{e}}\big)(\tau) \odot \eval{e'}(\tau)\\
		\eeq & \big(\sup_{n\in\mathbb{N}}~\Psi_{f}^{n}(0 ) (\tau)\big) \odot \tau(e')\tag{by Kleene's fixpoint theorem}\\
		\eeq & \sup_{n\in\mathbb{N}}~\Psi_{f}^{n}(0 ) (\tau)\odot \tau(e')\\
		\eeq&	\sup_{n\in\mathbb{N}}~
		\bigoplus_{\sigma\in\States}
		f (\sigma)\odot F^n(\bzero)(\sigma,\tau)
		\tag{by Equation~\ref{eq:sp-soundness-induction}}\\
		\eeq&
		\bigoplus_{\sigma\in\States}
		f (\sigma)\odot \sup_{n\in\mathbb{N}}~ F^n(\bzero)(\sigma,\tau)
		\tag{by continuity of $\lambda X\mydot\bigoplus_\sigma f(\sigma)\odot X(\sigma,\tau)$}
		\\
		\eeq&\bigoplus_{\sigma\in\States}~f(\sigma)\odot \eval{\LOOP{C}{e}{e'}}(\sigma,\tau)
		\tag{by Kleene's fixpoint theorem}~.
	\end{align*}

	and this concludes the proof.
\end{proof}

\subsection{A Weakest Pre Calculus for \wrcl}
\label{se:weighted-wp}
\begin{table}[t]
	\renewcommand{\arraystretch}{1.5}
	\begin{tabular}{@{\hspace{.5em}}l@{\hspace{2em}}l@{\hspace{.5em}}}
		\hline
		$\boldsymbol{C}$			& $\boldwp{C}{f}$ \\
		\hline
		$\ASSIGN{x}{\ee}$			& $f\subst{x}{\ee}$ 	\\
		$\ASSIGNNONDET{x}$			& $\bigoplus_{\alpha} f\subst{x}{\alpha}$ 	\\
		$\WEIGHT{w}$		& 	$w \odot f$\\
		$\COMPOSE{C_1}{C_2}$		& $\wp{C_1}{\vphantom{\big(}\wp{C_2}{f}}$\\
		$\NDCHOICE{C_1}{C_2}$		& $\wp{C_1}{f}\oplus \wp{C_2}{f}$\\
		$\LOOP{C}{e}{e'}$		& $\lfp X\mydot \eval{e'}\odot f\oplus \eval{e}\odot\wp{C}{X}$ 	\\
		\hline
	\end{tabular}%
	\renewcommand{\arraystretch}{1}
	\caption{\small Rules for the weakest pre transformer.}
	\label{table:wp}
\end{table}%
\wpsoundness*
\begin{proof}
	We define our weighted $\wpsymbol$ in~\Cref{table:wp}.
	We prove Theorem~\ref{thm:wp-soundness} by induction on the structure of $C$.
	For the induction base, we have the atomic statements:%
	\paragraph{The assignment $\ASSIGN{x}{e}$:} We have%
	\begin{align*}
		\wp{\ASSIGN{x}{\ee}}{f}(\sigma)
		\eeq & f\subst{x}{\ee} (\sigma) \\
		\eeq & f(\sigma\subst{x}{\sigma(\ee)}) \\
		\eeq &	\bigoplus_{\tau\in\States} \iverson{\sigma\subst{x}{\sigma(e)}=\tau}\odot  f (\tau)\\
		\eeq &	\bigoplus_{\tau\in\States} \eval{\ASSIGN{x}{\ee}}(\sigma, \tau)\odot f (\tau)~.
	\end{align*}%
	\paragraph{The nondeterministic assignment $\ASSIGNNONDET{x}$:} We have%
	\begin{align*}
		\wp{\ASSIGNNONDET{x}}{f}(\sigma)
		\eeq & \big(\bigoplus_\alpha~  f\subst{x}{\alpha}\big)(\sigma)\\
		\eeq & \bigoplus_{\alpha}~ f(\sigma\subst{x}{\alpha})\\
		\eeq &	\bigoplus_{\tau\in\States,\exists\alpha\mydot \sigma\subst{x}{\alpha}=\tau}~ f(\tau) \tag{by taking $\tau=\sigma\subst{x}{\alpha}$}\\
		\eeq &	\bigoplus_{\tau\in\States}~ \bigoplus_{\alpha\in\Nats}   \iverson{\sigma\subst{x}{\alpha}=\tau}\odot  f(\tau) \\		%
		\eeq &	
		\bigoplus_{\tau\in\States}~\eval{\ASSIGNNONDET{x}}(\sigma,\tau)\odot f(\tau) ~.
	\end{align*}%
	\paragraph{The weighting $\WEIGHT{w}$:} We have%
	\begin{align*}
		\wp{\WEIGHT{w}}{f}(\sigma) 
		\eeq &  (w\odot f)(\sigma)\\
		\eeq &  w(\sigma) \odot f(\sigma)\\
		\eeq & \bigoplus_{\tau\in\States} 
		~ w(\sigma)\odot \iverson{\sigma=\tau}  \odot f(\sigma)\\
		\eeq & \bigoplus_{\tau\in\States} 
		~w(\sigma)\odot \iverson{\sigma=\tau}\odot f(\tau) \\
		\eeq & \bigoplus_{\tau\in\States}~
		\eval{\WEIGHT{w}}(\sigma,\tau)\odot f(\tau) ~.
	\end{align*}%
	\noindent{}%
	This concludes the proof for the atomic statement.%
	\paragraph{Induction Hypothesis:} 
	For arbitrary but fixed programs $C$, $C_1$, $C_2$, we proceed with the inductive step on the composite statements.%
	\paragraph{The sequential composition $\COMPOSE{C_1}{C_2}$:} We have%
	\begin{align*}
		\wp{\COMPOSE{C_1}{C_2}}{f}(\sigma) 
		\eeq & \wp{C_1}{\wp{C_2}{f}}(\sigma) \\
		\eeq & \bigoplus_{\sigma'\in\States} ~\eval{C_1} (\sigma,\sigma') \odot \wp{C_2}{f}(\sigma')
		\tag{by I.H.~on $C_1$}\\
		\eeq & \bigoplus_{\sigma'\in\States} ~
		\eval{C_1}(\sigma,\sigma') \odot 
		\bigoplus_{\tau\in\States}
		~
		\eval{C_2} (\sigma',\tau)\odot f(\tau)
		\tag{by I.H.~on $C_2$}\\
		\eeq & \bigoplus_{\sigma'\in\States} \bigoplus_{\tau\in\States}~
		\eval{C_1}(\sigma,\sigma') \odot 
		~
		\eval{C_2} (\sigma',\tau)\odot f(\tau)\tag{by distributivity of $\odot$}\\	
		\eeq & \bigoplus_{\tau\in\States} 
		~ 
		\big( \bigoplus_{\sigma'\in\States}
		~ \eval{C_1}(\sigma,\sigma')
		\odot\eval{C_2} (\sigma',\tau) \big)\odot  f(\tau)  \tag{by commutativity of $\oplus$}\\
		\eeq & \bigoplus_{\tau\in\States}~ \eval{\COMPOSE{C_1}{C_2}}(\sigma,\tau) \odot f(\tau)~.
	\end{align*}%
	\noindent{}%
	\paragraph{The nondeterministic choice $\NDCHOICE{C_1}{C_2}$:} We have%
	\begin{align*}
		\wp{\NDCHOICE{C_1}{C_2}}{f}(\sigma) 
		\eeq &  \wp{C_1}{f} \oplus \wp{C_2}{f} \\
		\eeq & \bigoplus_{\tau\in\States}\sem{C_1}{\sigma,\tau}\odot f(\tau)
		\oplus \bigoplus_{\tau\in\States}\sem{C_2}{\sigma,\tau}\odot f(\tau)
		\tag{by I.H.~on $C_1, C_2$}\\
		\eeq & \bigoplus_{\tau\in\States} 
		~(\sem{C_1}{\sigma,\tau}\oplus \sem{C_2}{\sigma,\tau})
		\odot f(\tau) \tag{by distributivity of $\odot$}\\
		\eeq & \bigoplus_{\tau\in\States}~ \eval{\NDCHOICE{C_1}{C_2}}(\sigma,\tau)\odot f(\sigma) ~.
	\end{align*}%
	\noindent{}%

	\paragraph{The Iteration $\LOOP{C}{e}{e'}$:} 
	Let%
	\begin{align*}
		\Phi_{f}(X) &\eeq 
		\eval{e'}\odot f \oplus  \eval{e} \odot  \wp{C}{X}~,
	\end{align*}%
	be the $\wpsymbol$-characteristic function of the iteration $\LOOP{C}{e}{e'}$ with respect to any preanticipation $f$ and
	\begin{align*}
		F(X)(\sigma,\tau) &\eeq 
		\sigma(e)\odot\left(\bigoplus_{\sigma'\in\States} 
		\eval{C}(\sigma,\sigma')\odot
		 X (\sigma', \tau)
		\right) \oplus \sigma(e')\odot\iverson{\sigma=\tau}~,
	\end{align*}
	be the denotational semantics characteristic function of the loop $\LOOP{C}{e}{e'}$ for any input $
		\sigma,\tau\in\States$.
	We first prove by induction on $n$ that, for all $\sigma\in\States,f\in\A$
	\begin{equation}
		\label{eq:wp-soundness-induction}
		 \Phi_{f}^{n}(\bzero)(\sigma)
		\eeq
		\bigoplus_{\tau\in\States}~
		F^n(\bzero)(\sigma,\tau) \odot 
		f (\tau) ~.
	\end{equation}%
	For the induction base $n = 0$, consider the following:%
	\begin{align*}
		\Phi_{f}^{n}(\bzero)(\sigma)
		& \eeq  \bzero \\
		& \eeq\bigoplus_{\tau\in\States}~
		F^0(\bzero)(\sigma,\tau) \odot 
		f (\tau)~.
	\end{align*}%
	As induction hypothesis, we have for arbitrary but fixed $n$ and all $\tau\in\States,f\in\A$%
	\begin{align*}
		\Phi_{f}^{n}(\bzero)(\sigma)
		\eeq
		\bigoplus_{\tau\in\States}~
		F^n(\bzero)(\sigma,\tau) \odot 
		f (\tau)~.
	\end{align*}%
	For the induction step $n \longrightarrow n + 1$, consider the following:%
	\begin{align*}
		&	\Phi_{f}^{n+1}(\bzero )(\sigma) \\
		& \eeq \left(\eval{e'}\odot f \oplus \eval{e}\odot\wp{C}{\Phi_{f}^n(\bzero)} \right)(\sigma) \\
		& \eeq \eval{e'}(\sigma)\odot f(\sigma) \oplus \eval{e}(\sigma)\odot\wp{C}{\Phi_{f}^n(\bzero)}(\sigma)\\
		& \eeq  \sigma(e')\odot f(\sigma)   \oplus
		\sigma(e)\odot 
		\bigoplus_{\sigma'\in\States}~
		\eval{C}(\sigma,\sigma') 
		\odot\Phi_{f}^n(\bzero)(\sigma') 
		\tag{by I.H.\ on $C$}\\
		& \eeq  \sigma(e')\odot f(\sigma)   \oplus
		\sigma(e)\odot 
		\bigoplus_{\sigma'\in\States}~
		\eval{C}(\sigma,\sigma') 
		\odot
		\bigoplus_{\tau\in\States}~
		F^n(\bzero)(\sigma',\tau) \odot 
		f (\tau)
		\tag{by I.H.\ on $n$} \\
		& \eeq
		\sigma(e')\odot f(\sigma)  \oplus
		\bigoplus_{\tau\in\States}~
		\big(
		\sigma(e)\odot\bigoplus_{\sigma'\in\States} \eval{C}(\sigma,\sigma')\odot F^n(\bzero) (\sigma', \tau)\big)\odot f(\tau)
		\tag{by distributivity of $\odot$, commutativity and associativity of $\oplus$}
		\\
		& \eeq
		\big(\bigoplus_{\tau\in\States}
		\sigma(e') \odot
		\iverson{\sigma=\tau}\odot
		f(\tau)
		\big)
		\oplus
		\bigoplus_{\tau\in\States}~
		\big(
		\sigma(e)\odot\bigoplus_{\sigma'\in\States} \eval{C}(\sigma,\sigma')\odot F^n(\bzero) (\sigma', \tau)\big)\odot f(\tau)
		\\
		& \eeq
		\bigoplus_{\tau\in\States}~
		\big(
		\sigma(e)\odot
		\bigoplus_{\sigma'\in\States} ~
		\eval{C}(\sigma,\sigma')\odot
		F^n(\bzero) (\sigma', \tau)\big) \odot f (\tau) 
		 \oplus
		 \big(\bigoplus_{\tau\in\States}~
		 \sigma(e')\odot 
		\iverson{\sigma=\tau}\odot  f(\tau)
		\big)
		\tag{by commutativity of $\oplus$}
		\\
		& \eeq
		\bigoplus_{\tau\in\States}
		\big(
			 \sigma(e)\odot\bigoplus_{\sigma'\in\States} \eval{C}(\sigma,\sigma')\odot F^n(\bzero) (\sigma', \tau) \oplus 
		\sigma(e')\odot
		\iverson{\sigma=\tau}
		\big)
		\odot 
		f (\tau)
		\tag{by associativity of $\oplus$ and distributivity of $\odot$}
		\\
		& \eeq
		\bigoplus_{\tau\in\States}
		F^{n+1}(\bzero)(\sigma,\tau)
		\odot f (\tau) ~.
		\end{align*}
	
	This concludes the induction on $n$. Now we have:

	\begin{align*}
		\wp{\LOOP{C}{e}{e'}}{f}(\sigma) 
		\eeq & \big(\lfp  X\mydot  \eval{e'}\odot f \oplus \eval{e}\odot\wp{C}{X}\big)(\sigma)\\
		\eeq &\sup_{n\in\mathbb{N}}~\Phi_{f}^{n}(0 ) (\sigma) \tag{by Kleene's fixpoint theorem}\\
		\eeq&	\sup_{n\in\mathbb{N}}~
		\bigoplus_{\tau\in\States}~
		F^n(\bzero)(\sigma,\tau) \odot 
		f (\tau) 
		\tag{by Equation~\ref{eq:wp-soundness-induction}}\\
		\eeq&
		\bigoplus_{\tau\in\States}
		 \sup_{n\in\mathbb{N}}~ F^n(\bzero)(\sigma,\tau)\odot f (\tau)
		\tag{by continuity of $\lambda X\mydot\bigoplus_\tau X(\sigma,\tau)\odot f(\tau)$}
		\\
		\eeq&\bigoplus_{\tau\in\States}~\eval{\LOOP{C}{e}{e'}}(\sigma,\tau) \odot f(\tau)
		\tag{by Kleene's fixpoint theorem}~.
	\end{align*}

	and this concludes the proof.
\end{proof}

\subsection{Proof of $\spsymbol$-$\wpsymbol$ Duality for probabilistic programs, Thereom~\ref{thm:forwardbackward}}
\forwardbackward*
\begin{proof}
	\begin{align*}
	  \bigoplus_{\tau \in \States}~ \sp{C}{\mu}(\tau) \odot g(\tau) \eeq& 
	  \bigoplus_{\tau \in \States}~ \bigoplus_{\sigma\in\States}
	  \mu(\sigma) \odot \eval{C}(\sigma,\tau)
		\odot g(\tau) 
		\tag{by Theorem~\ref{thm:sp-soundness}}
		\\
	  \eeq&
	  \bigoplus_{\sigma\in\States}
	  \bigoplus_{\tau \in \States}
	  \mu(\sigma) \odot \eval{C}(\sigma,\tau)
	  \odot g(\tau) 
	  \\
	  \eeq&
	  \bigoplus_{\sigma\in\States}
	  \mu(\sigma) \odot
	  \bigoplus_{\tau \in \States} \eval{C}(\sigma,\tau)
	  \odot g(\tau) 
	  \\
	  \eeq&
	  \bigoplus_{\sigma \in \States}~ \mu(\sigma) \odot \wp{C}{g}(\sigma)\tag{by Theorem~\ref{thm:wp-soundness}}~.
	\end{align*}
\end{proof}%
\section{Quantitative Weakest Hyper Pre}

\whpsoundness*
\begin{proof}
	We prove Theorem~\ref{thm:whp-soundness} by induction on the structure of $C$.
	For the induction base, we have the atomic statement:%
	\paragraph{The assignment $\ASSIGN{x}{e}$:} We have%
	\begin{align*}
		\whp{\ASSIGN{x}{\ee}}{\hyperf}(\mu)
		\eeq & \hyperf(\bigoplus_{\alpha} \iverson{x= \ee\subst{x}{\alpha}} \odot \mu\subst{x}{\alpha})\\
		\eeq &	\hyperf(\sp{\ASSIGN{x}{\ee}}{\mu}) ~.
	\end{align*}%
	\paragraph{The nondeterministic assignment $\ASSIGNNONDET{x}$:} We have%
	\begin{align*}
		\whp{\ASSIGNNONDET{x}}{\hyperf}(\mu) 
		\eeq & \hyperf(\bigoplus_{\alpha}  \mu\subst{x}{\alpha})\\
		\eeq &	\hyperf(\sp{\ASSIGNNONDET{x}}{\mu}) ~.
	\end{align*}%
	\paragraph{The weighting $\WEIGHT{w}$:} We have%
	\begin{align*}
		\whp{\WEIGHT{w}}{\hyperf}(\mu) 
		\eeq & (\hyperf \odot w)(\mu)\\
		\eeq&	\hyperf(\mu \odot w)\\
		\eeq&	\hyperf(\sp{\WEIGHT{w}}{\mu})~.
	\end{align*}%

%
This concludes the proof for the atomic statements.%
\paragraph{Induction Hypothesis:} 
For arbitrary but fixed programs $C$, $C_1$, $C_2$, we proceed with the inductive step on the composite statements.%
\paragraph{The sequential composition $\COMPOSE{C_1}{C_2}$:} We have%
\begin{align*}
	\whp{\COMPOSE{C_1}{C_2}}{\hyperf}(\mu) 
	\eeq & \whp{C_1}{\whp{C_2}{\hyperf}}(\mu) \\
	\eeq & \whp{C_2}{\hyperf}(\sp{C_1}{\mu}) \tag{by I.H.~on $C_1$}\\
	\eeq & \hyperf(\sp{C_2}{\sp{C_1}{\mu}}) \tag{by I.H.~on $C_2$}\\
	\eeq &	\hyperf(\sp{\COMPOSE{C_1}{C_2}}{\mu})\\
\end{align*}%

\paragraph{The nondeterministic choice $\NDCHOICE{C_1}{C_2}$:} We have%
\begin{align*}
	\whp{\NDCHOICE{C_1}{C_2}}{\hyperf}(\mu)
	\eeq & \bigoplus_{\nu_1,\nu_2}~\hyperf(\nu_1\oplus\nu_2)\odot \whp{C_1}{\iverson{\nu_1}}(\mu) \odot \whp{C_2}{\iverson{\nu_2}}(\mu) \\
	\eeq &	\bigoplus_{\nu_1,\nu_2}~\hyperf(\nu_1\oplus\nu_2)\odot \iverson{\nu_1}(\sp{C_1}{\mu})\odot \iverson{\nu_2}(\sp{C_2}{\mu}) \tag{by I.H.~on $C_1,C_2$}\\
	\eeq&	\hyperf(\sp{C_1}{\mu}\oplus\sp{C_2}{\mu})\\
	\eeq&	\hyperf(\sp{\NDCHOICE{C_1}{C_2}}{\mu})~.
\end{align*}%

	\paragraph{The Iteration $\LOOP{C}{e}{e'}$:} 
	\begin{align*}
		\whp{\LOOP{C}{e}{e'}}{\hyperf}(\mu) 
		\eeq & \hyperf\big(\big(\lfp  X\mydot  \mu \oplus \sp{C}{X\odot \eval{e}}\big) \odot \eval{e'}\big) \\
		\eeq&	\hyperf(\sp{\LOOP{C}{e}{e'}}{\mu})~.
	\end{align*}%
	and this concludes the proof.
\end{proof}

\subsection{Proof of Consistency of iteration rule, ~\Cref{thm:forwardbackward}}
\propfixpointuniqueness*
\begin{proof}
		\begin{align*}
			&\Phi(  \lambda \hyperf\,\lambda \mu\mydot \hyperf (\sp{\LOOP{C}{e}{e'}}{\mu}))\\
			\eeq & \lambda \hyperh\,\lambda f\mydot 
			\bigoplus_{\nu}~\hyperh(\nu \oplus f\odot\eval{e'}) \odot \whp{C}{ \lambda \mu\mydot \iverson{\nu}(\sp{\LOOP{C}{e}{e'}}{\mu})}(f\odot\eval{e})\\
			\eeq & \lambda \hyperh\,\lambda f\mydot \bigoplus_{\nu}~\hyperh(\nu \oplus f\odot\eval{e'})\odot \iverson{\nu}(\sp{\LOOP{C}{e}{e'}}{\sp{C}{f\odot\eval{e}}})\tag{by I.H.~on $C$} \\
			\eeq & \lambda \hyperh\,\lambda f\mydot ~\hyperh(\sp{\LOOP{C}{e}{e'}}{\sp{C}{f\odot\eval{e}}} \oplus f\odot\eval{e'}) \\
			\eeq & \lambda \hyperh\,\lambda f\mydot ~\hyperh(\sp{\LOOP{C}{e}{e'}}{f}) \tag{$\spC{\LOOP{C}{e}{e'}}$ is a fixpoint of $\Psi(X)=\lambda f\mydot X(\sp{C}{f\odot\eval{e}}\oplus f \odot \eval{e'})$} \\
			\eeq&	  \lambda \hyperf\,\lambda \mu\mydot \hyperf (\sp{\LOOP{C}{e}{e'}}{\mu})~.
		\end{align*}%
\end{proof}
\subsection{Properties}
\subhyperhoare*
\begin{proof}
	\begin{align*}
		\hhl{\hyperpsi}{C}{\hyperphi}
		&\qiff \forall S\in\powerset{\States}\mydot S\in \hyperpsi \implies \sem{C}{S}\in \hyperphi\\
		&\qiff \forall S\in\powerset{\States}\mydot S\in \hyperpsi \implies \supp{\sp{C}{\iverson{S}}}\in \hyperphi\\
		&\qiff \forall S\in\powerset{\States}\mydot \iverson{\hyperpsi}(\iverson{S}) \leq \iverson{\hyperphi}(\sp{C}{\iverson{S}})\\
		&\qiff \forall S\in\powerset{\States}\mydot \iverson{\hyperpsi}(\iverson{S}) \leq \whp{C}{\iverson{\hyperphi}}(\iverson{S})\\
		&\qiff \forall \mu\in\A\mydot \iverson{\hyperpsi}(\mu) \leq  \whp{C}{\iverson{\hyperphi}}(\mu)\\
		&\qiff\supp{\indicator{\hyperpsi}} \implies  \supp{\whp{C}{\indicator{\hyperphi}}}
	\end{align*}
\end{proof}
\disprovingTriples*
\begin{proof}
	First, let us observe that
	\begin{align*}
		A\subseteq B &\qiff \forall x\in A \mydot \{x\}\cap B  \neq \emptyset 
	\end{align*}
	Now, we have:
	\begin{enumerate}
		\item
		\begin{align*}
			\pcl{P}{C}{Q}
			&\qiff P\subseteq \wlp{C}{Q}\\
			&\qiff \forall \sigma\in P\mydot \{\sigma\}\cap \wlp{C}{Q} \neq \emptyset  \\
			&\qiff \forall \sigma\in P\mydot  \not\tcl{\{\sigma\}}{C}{\neg Q}
		\end{align*}
		\item
		\begin{align*}
			\tcl{P}{C}{Q}
			&\qiff P\subseteq \wp{C}{Q}\\
			&\qiff \forall \sigma\in P\mydot \{\sigma\}\cap \wp{C}{Q} \neq \emptyset \\
			&\qiff \forall \sigma\in P\mydot\not\pcl{\{\sigma\}}{C}{\neg Q}
		\end{align*}
		\item
		\begin{align*}
			\pil{P}{C}{Q}
			&\qiff Q\subseteq \slp{C}{P}\\
			&\qiff \forall \sigma\in Q\mydot\{\sigma\}\cap\slp{C}{P} \neq \emptyset  \\
			&\qiff \forall \sigma\in Q\not\til{\neg P}{C}{\{\sigma\}}
		\end{align*}
		\item
		\begin{align*}
			\til{P}{C}{Q}
			&\qiff Q\subseteq \sp{C}{P}\\
			&\qiff \forall \sigma\in Q\mydot\{\sigma\}\cap\sp{C}{P} \neq \emptyset  \\
			&\qiff \forall \sigma\in Q\not\pil{\neg P}{C}{\{\sigma\}}
		\end{align*}
	\end{enumerate}
\end{proof}

\subwpwlp*
\begin{proof}
	\begin{align*}
		\whp{C}{\bigcurlyvee [f]}(\iversone{\sigma}) 
		& \eeq \bigcurlyvee [f] (\sp{C}{\iversone{\sigma}})\\ 
		& \eeq \bigcurlyvee_{\tau\colon\sp{C}{\iversone{\sigma}}(\tau)>0} f(\tau)\\ 
		& \eeq\wp{C}{f}(\sigma)
	\end{align*}
	\begin{align*}
		\whp{C}{\bigcurlywedge [f]}(\iversone{\sigma}) 
		& \eeq \bigcurlywedge [f] (\sp{C}{\iversone{\sigma}})\\ 
		& \eeq \bigcurlywedge_{\tau\colon\sp{C}{\iversone{\sigma}}(\tau)>0} f(\tau)\\ 
		& \eeq\wlp{C}{f}(\sigma)
	\end{align*}
\end{proof}

\subwp*
\begin{proof}
	\begin{align*}
		\whp{C}{\E [f]}(\iversone{\sigma}) & \eeq \E [f] (\sp{C}{\iversone{\sigma}}) \\
		& \eeq \wp{C}{f}(\sigma)
	\end{align*}
	\begin{align*}
		\whp{C}{\E [f]+1-\E [1]}(\iversone{\sigma}) & \eeq (\E [f]+1-\E [1])(\sp{C}{\iversone{\sigma}}) \\
		& \eeq \wp{C}{f}(\sigma)+1- \wp{C}{1}(\sigma)\\
		& \eeq \wlp{C}{f}(\sigma)
	\end{align*}
\end{proof}

\section{Proofs of Section~\ref{se:healthiness}}
\subsection{Proof of Healthiness Properties of Quantitative Transformers, \Cref{thm:whphealthiness}}
Each of the properties is proven individually below.
\begin{itemize}
	\item Quantitative universal conjunctiveness: \Cref{thm:whpconjunctive};
	\item Quantitative universal disjunctiveness: \Cref{thm:whpdisjunctive};
	\item Strictness: Corollary~\ref{thm:whpstrict};
	\item Costrictness: Corollary~\ref{thm:whpcostrict};
	\item Monotonicity: Corollary~\ref{thm:mono}.
\end{itemize}

\begin{restatable}[Quantitative universal conjunctiveness of \textnormal{$\whpsymbol$}]{theorem}{whpconjunctive}
	\label{thm:whpconjunctive}
	For any set of quantities \mbox{$S \subseteq \hyperA$},%
	\begin{align*}
		\whp{C}{\prod S} \eeq \prod_{\hyperf \in S}~ \whp{C}{\hyperf}~.
	\end{align*}%
\end{restatable}%
\begin{proof}
	\begin{align*}
		\whp{C}{\prod S}
		\eeq & \lambda \mu\mydot (\prod S) (\sp{C}{\mu}) \tag{by \Cref{thm:whp-soundness}}\\
		\eeq & \lambda \mu\mydot \prod_{\hyperf \in S} \hyperf(\sp{C}{\mu})  \\
		\eeq &  \prod_{\hyperf \in S}~ \whp{C}{\hyperf} \tag{by \Cref{thm:whp-soundness}} ~.
	\end{align*}%
\end{proof}
\begin{restatable}[Quantitative universal disjunctiveness of \textnormal{$\whpsymbol$}]{theorem}{whpdisjunctive}
	\label{thm:whpdisjunctive}
	For any set of quantities \mbox{$S \subseteq \hyperA$},%
	\begin{align*}
		\whp{C}{\sum S} \eeq \sum_{\hyperf \in S}~ \whp{C}{\hyperf}~.
	\end{align*}%
\end{restatable}%
\begin{proof}
	\begin{align*}
		\whp{C}{\sum S}
		\eeq & \lambda \mu\mydot (\sum S) (\sp{C}{\mu}) \tag{by \Cref{thm:whp-soundness}} \\
		\eeq & \lambda \mu\mydot \sum_{\hyperf \in S} \hyperf(\sp{C}{\mu}) \\
		\eeq &  \sum_{\hyperf \in S}~ \whp{C}{\hyperf} \tag{by \Cref{thm:whp-soundness}}~.
	\end{align*}%
\end{proof}
\begin{restatable}[Strictness of \textnormal{$\whpsymbol$}]{corollary}{wpstrictness}%
	\label{thm:whpstrict}%
	For all programs $C$, $\whpC{C}$ is strict, i.e.\ 
	\begin{align*}
		\whp{C}{0} \eeq 0~.
	\end{align*}%
\end{restatable}%
\begin{proof}
	\begin{align*}
		\whp{C}{0} \eeq& \lambda \mu\mydot (0) (\sp{C}{\mu}) \tag{by \Cref{thm:whp-soundness}} \\
		\eeq & 0~.
	\end{align*}
\end{proof}
\begin{restatable}[Co-strictness of \textnormal{$\whpsymbol$}]{corollary}{whpcostrictness}%
	\label{thm:whpcostrict}%
	For all programs $C$, $\wpC{C}$ is co-strict, i.e.\ 
	\begin{align*}
		\whp{C}{\pinfty} \eeq \pinfty~.
	\end{align*}%
\end{restatable}%
\begin{proof}
	\begin{align*}
		\whp{C}{\pinfty} \eeq& \lambda \mu\mydot (\pinfty) (\sp{C}{\mu}) \tag{by \Cref{thm:whp-soundness}} \\
		\eeq & \pinfty~.
	\end{align*}
\end{proof}

\begin{corollary}[Monotonicity of Quantitative Transformers]
	\label{thm:mono}
	For all programs $C$, $\hyperf, \hyperg \in \AA$, we have%
	\begin{align*}
		\hyperf \ppreceq \hyperg \qqimplies \whp{C}{\hyperf} \ppreceq\whp{C}{\hyperg}
	\end{align*}
\end{corollary}%
\begin{proof}
	\begin{align*}
		\whp{C}{\hyperf} 
		&\eeq \lambda \mu  \mydot \hyperf(\sp{C}{\mu}) \tag{by \Cref{thm:whp-soundness}} \\
		&\ppreceq \lambda \mu  \mydot \hyperg(\sp{C}{\mu}) \tag{$\hyperf \ppreceq \hyperg$}\\
		&\eeq  \whp{C}{\hyperg} \tag{by \Cref{thm:whp-soundness}} 
	\end{align*}
\end{proof}

\subsection{Proof of Linearity, \Cref{thm:whp:linearity}}
\whplinearity*
\begin{proof}
	\begin{align*}
	& \whp{C}{r\cdot \hyperf +  \hyperg} \\
	&\eeq \lambda \mu\mydot (r\cdot \hyperf +  \hyperg)(\sp{C}{\mu}) \tag{by \Cref{thm:whp-soundness}}
	\\
	&\eeq
	\lambda \mu\mydot (r\cdot \hyperf)(\sp{C}{\mu}) +  \hyperg(\sp{C}{\mu})\\
	&\eeq
	\lambda \mu\mydot r\cdot \hyperf(\sp{C}{\mu}) +  \hyperg(\sp{C}{\mu})\\
	&\eeq  r\cdot\whp{C}{\hyperf} + \whp{C}{\hyperg} \tag{by \Cref{thm:whp-soundness}}~.
	\end{align*}
\end{proof}
\subsection{Proof of Multiplicativity, \Cref{thm:whp:multiplicativity}}
\whpmultiplicativity*
\begin{proof}
	\begin{align*}
		& \whp{C}{r\cdot \hyperf\cdot  \hyperg} \\
		&\eeq \lambda \mu\mydot (r\cdot \hyperf \cdot  \hyperg)(\sp{C}{\mu}) \tag{by \Cref{thm:whp-soundness}}
		\\
		&\eeq
		\lambda \mu\mydot r\cdot\hyperf(\sp{C}{\mu}) \cdot  \hyperg(\sp{C}{\mu})\\
		&\eeq  r\cdot\whp{C}{\hyperf} \cdot \whp{C}{\hyperg} \tag{by \Cref{thm:whp-soundness}}~.
	\end{align*}
\end{proof}

\subsection{Proof of Liberal-Non-liberal Duality, \Cref{thm:whp:duality}}

\whpduality*
\begin{proof}
	\begin{align*}
		\whp{C}{\hyperf}\eeq &
		\lambda \mu\mydot \hyperf (\sp{C}{\mu})
		f(\tau) \tag{by \Cref{thm:whp-soundness}}\\
		\eeq &
		k-\lambda \mu\mydot k-\hyperf (\sp{C}{\mu})\\
		\eeq &	k-\whp{C}{k-\hyperf}~.
	\end{align*}
\end{proof}

\subsection*{Proof of rules for linear hyperquantities, \Cref{thm:whp-soundness-linear}}
\hyperlinear*
\begin{proof}
	We prove \Cref{thm:whp-soundness} by induction on the structure of $C$.
	For the induction base, we have the atomic statement:%
	\paragraph{The assignment $\ASSIGN{x}{e}$:} We have%
	\begin{align*}
		\whp{\ASSIGN{x}{\ee}}{\hyperf}(\mu)
		\eeq & \bigoplus_{\alpha} \hyperf(\iverson{x= \ee\subst{x}{\alpha}} \odot \mu\subst{x}{\alpha})\\
		\eeq & \hyperf(\bigoplus_{\alpha} \iverson{x= \ee\subst{x}{\alpha}} \odot \mu\subst{x}{\alpha})\\
		\eeq &	\hyperf(\sp{\ASSIGN{x}{\ee}}{\mu}) ~.
	\end{align*}%
	\paragraph{The nondeterministic assignment $\ASSIGNNONDET{x}$:} We have%
	\begin{align*}
		\whp{\ASSIGNNONDET{x}}{\hyperf}(\mu) 
		\eeq & \hyperf(\bigoplus_{\alpha}  \mu\subst{x}{\alpha})\\
		\eeq &	\hyperf(\sp{\ASSIGNNONDET{x}}{\mu}) ~.
	\end{align*}%

		%
	\paragraph{The weighting $\AWEIGHT$:} We have%
	\begin{align*}
		\whp{\WEIGHT{w}}{\hyperf}(\mu) 
		\eeq & (\hyperf \odot w)(\mu)\\
		\eeq&	\hyperf(\mu \odot w)\\
		\eeq&	\hyperf(\sp{\WEIGHT{w}}{\mu})~.
	\end{align*}%
	This concludes the proof for the atomic statements.%
	\paragraph{Induction Hypothesis:} 
	For arbitrary but fixed programs $C$, $C_1$, $C_2$, we proceed with the inductive step on the composite statements.%
	\paragraph{The sequential composition $\COMPOSE{C_1}{C_2}$:} We have%
\begin{align*}
\whp{\COMPOSE{C_1}{C_2}}{\hyperf}(\mu) 
\eeq & \whp{C_1}{\whp{C_2}{\hyperf}}(\mu) \\
\eeq & \whp{C_2}{\hyperf}(\sp{C_1}{\mu}) \tag{by I.H.~on $C_1$}\\
\eeq & \hyperf(\sp{C_2}{\sp{C_1}{\mu}}) \tag{by I.H.~on $C_2$}\\
\eeq &	\hyperf(\sp{\COMPOSE{C_1}{C_2}}{\mu})\\
\end{align*}%

\paragraph{The nondeterministic choice $\NDCHOICE{C_1}{C_2}$:} We have%
\begin{align*}
\whp{\NDCHOICE{C_1}{C_2}}{\hyperf}(\mu)
\eeq & \whp{C_1}{\hyperf}(\mu)\oplus\whp{C_2}{\hyperf}(\mu) \\
\eeq & \hyperf(\sp{C_1}{\mu})\oplus\hyperf(\sp{C_2}{\mu})\tag{by I.H.~on $C_1,C_2$}\\
\eeq & \hyperf(\sp{C_1}{\mu}\oplus\sp{C_2}{\mu})\tag{by~\Cref{def:quantities:linear}}\\
\eeq &	\bigoplus_{\nu_1,\nu_2}~\hyperf(\nu_1\oplus\nu_2)\odot \iverson{\nu_1}(\sp{C_1}{\mu})\odot \iverson{\nu_2}(\sp{C_2}{\mu}) \\
\eeq&	\hyperf(\sp{C_1}{\mu}\oplus\sp{C_2}{\mu})\\
\eeq&	\hyperf(\sp{\NDCHOICE{C_1}{C_2}}{\mu})~.
\end{align*}%

%

%

\paragraph{The Iteration $\LOOP{C}{e}{e'}$:} 
Let $W_{e}(X)=\whp{C}{X}\odot \eval{e}$ and $S(X)=\sp{C}{X\odot\eval{e}}$.
We first prove by induction on $n$ that:
\begin{align*}
	W_{e}^n(\hyperf \odot \eval{e'})(\mu)\eeq \hyperf (S^n(\mu) \odot \eval{e'})
\end{align*}
For the induction base $n = 0$, consider the following:%
	\begin{align*}
		W_{e}^n(\hyperf \odot \eval{e'})(\mu)
		& \eeq  (\hyperf \odot \eval{e'}) (\mu) \\
		& \eeq  \hyperf (\mu\odot\eval{e'}) \\
		& \eeq \hyperf (S^n(\mu) \odot \eval{e'})~.
	\end{align*}%
	As induction hypothesis, we have for arbitrary but fixed $n$ and all $\mu$%
	\begin{align*}
		W_{e}^n(\hyperf \odot \eval{e'})(\mu)\eeq \hyperf (S^n(\mu) \odot \eval{e'})
	\end{align*}%
	For the induction step $n \longrightarrow n + 1$, consider the following:%
	\begin{align*}
		W_{e}^{n+1}(\hyperf \odot \eval{e'})(\mu)
		&\eeq \big(W_{e}(W_{e}^n(\hyperf \odot \eval{e'}))\big)(\mu)\\
		&\eeq \big(\whp{C}{W^n(\hyperf \odot \eval{e'})}\odot\eval{e}\big)(\mu) \\
		&\eeq \big(\whp{C}{W^n(\hyperf \odot \eval{e'})}\big)(\mu\odot\eval{e}) \\
		&\eeq W_{e}^n(\hyperf \odot \eval{e'})(\sp{C}{\mu\odot\eval{e}})  \tag{by I.H.~on $C$}\\
		&\eeq \hyperf (S^n(\sp{C}{\mu\odot\eval{e}}) \odot \eval{e'})\tag{by I.H.~on $n$}\\
		&\eeq \hyperf (S^n(S(\mu)) \odot \eval{e'})\\
		&\eeq \hyperf (S^{n+1}(\mu) \odot \eval{e'})
	\end{align*}%
This concludes the induction on $n$. Now we have:
\begin{align*}
	\whp{\LOOP{C}{e}{e'}}{\hyperf}(\mu) 
	\eeq & \bigoplus_{n\in\Nats} W_{e}^n(\hyperf \odot \eval{e'})(\mu)\\
	\eeq & 
		\bigoplus_{n\in\Nats}
		 \hyperf (S^n(\mu) \odot \eval{e'})
		 \\
	\eeq & \hyperf\big(
		\big(\bigoplus_{n\in\Nats} S^n(\mu) \big)
		\odot \eval{e'}\big) \tag{by~\Cref{def:quantities:linear}}\\
	\eeq& \hyperf\big(\big(\lfp  X\mydot  \mu \oplus \sp{C}{X\odot \eval{e}}\big) \odot \eval{e'}\big)\\
	\eeq&	\hyperf(\sp{\LOOP{C}{e}{e'}}{\mu})~.
	\end{align*}%
\end{proof}
\section{Well-definedness of the semantics}
\label{app:wd-semantics}

In this section we prove that the denotational semantics of~\Cref{se:semantics} is a total function.

\subsection{Additional definitions omitted from the main text}
We assume that the operations $\oplus$, $\odot$ belong to a complete, Scott continuous, naturally ordered, partial semiring with a top element.

\begin{definition}[Complete semirings~\cite{Golan03}]
	A (partial) semiring $\langle U, \oplus, \odot, \bzero, \bone\rangle$ is complete if there is a sum operator $\bigoplus _{i\in I}$ with the following properties:
	\begin{enumerate}
		\item If $I = \{i_1, \dots, i_n\}$ is finite, then $\bigoplus _{i\in I} u_i = u_{i_1} + \dots + u_{i_n}$.
		\item If $\bigoplus _{i\in I} x_i$ is defined, then $v \odot \bigoplus _{i\in I} u_i = \bigoplus _{i\in I} v\odot u_i $ and $(\bigoplus _{i\in I} u_i )\odot v = \bigoplus _{i\in I} u_i\odot v$.
		\item Let $(J_k)_{k \in K}$ be a family of nonempty disjoint subsets of $I$ ($I = \bigcup_{k \in K} J_k$ and $J_k \cap J_l = \emptyset$ if $k \neq l$), then $\bigoplus_{k\in K}\bigoplus_{j\in J_k}u_j=\bigoplus_{i\in I}u_i$.
	\end{enumerate}
	\end{definition}
	
\begin{definition}[Scott Continuity~\cite{Karner04}]
A (partial) semiring with order $\leq$ is Scott Continuous if for any directed set $D \subseteq X$ (where all pairs of elements in $D$ have a supremum), the following hold:
\begin{align*}
	\sup_{x \in D}(x \oplus y) &= (\sup D) \oplus y \\
	\sup_{x \in D}(x \odot y) &= (\sup D) \odot y \\
	\sup_{x \in D}(y \odot x) &= y \odot \sup D 
\end{align*}
\end{definition}

\subsection{Fixed point existence}

\begin{proposition}
	\label{prop:fixpoint-esistence-semantics}
	Let $\Phi_{C,e,e'}(X)(\sigma,\tau)=\sem{e}{\sigma}\odot\left(\bigoplus_{\iota\colon \sem{C}{\sigma,\iota}\neq\bzero}~ 
	\eval{C}(\sigma,\iota)\odot
	X (\iota, \tau)
   \right) \ooplus \sem{e'}{\sigma}\odot\iverson{\sigma=\tau}$.
	If $\Phi_{C,e,e'}$ is a total function, the semantics of loops:
	\begin{align*}
	\eval{\LOOP{C}{e}{e'}}(\sigma,\tau) = 
	(\lfp X\mydot  \Phi_{C,e,e'}(X))(\sigma, \tau)
	\end{align*}
	is well-defined, i.e., the least fixed point of $\Phi_{C,e,e'}$ exists.
\end{proposition}
\begin{proof}
	It is sufficient to show that $\Phi_{C,e,e'}$ is Scott-continuous and rely on Kleene's fixpoint theorem to conclude that the fixpoint exists. For all directed sets $D\subseteq (\States\times\States\to W(\States))$ we have:
	\begin{align*}
		&\sup _{f\in D} ~\Phi_{C,e,e'}(f)(
			\sigma,
			\tau
		)\\
		& \eeq 
		\sup _{f\in D}~ \sem{e}{\sigma}\odot\big(\bigoplus_{\iota\in\States}~ 
		\eval{C}(\sigma,\iota)\odot
		f (\iota, \tau)
	   \big) \ooplus \sem{e'}{\sigma}\odot\iverson{\sigma=\tau}\\
		& \eeq 
		 \sem{e}{\sigma}\odot	\big(\sup _{f\in D}~\bigoplus_{\iota\in\States}~ 
									\eval{C}(\sigma,\iota)\odot
		f (\iota, \tau)
	   \big)
	    \ooplus \sem{e'}{\sigma}\odot\iverson{\sigma=\tau}\tag{by continuity of $\oplus$ and $\odot$}\\
		& \eeq 
		\sem{e}{\sigma}\odot	\big(
		\bigoplus_{\iota\in\States}~ 
		\eval{C}(\sigma,\iota)\odot
		\sup ~D (\iota, \tau)
	  \big)
	   \ooplus \sem{e'}{\sigma}\odot\iverson{\sigma=\tau} \tag{by~\cite[Lemma A.4]{zilberstein2024relatively} with $f_\iota(X) = \sem{C}{\sigma,\iota}\odot X(\iota,\tau)$ for $\iota\in\States$}\\
	   & \eeq 
	   \Phi_{C,e,e'}(\sup ~D)(
			\sigma,
			\tau
		)
	\end{align*}
	And hence we conclude by Kleene's fixpoint theorem.
\end{proof}

\subsection{Syntactic restrictions for partial semirings}
\Cref{prop:fixpoint-esistence-semantics} ensures the well-definedness of the iteration rule, provided that $\Phi_{C,e,e'}$ is total. In this section, we investigate syntactic constraints to ensure the totality of $\Phi_{C,e,e'}$ (and all other statements). Notably, challenges arise in partial semirings only, where $\oplus$ might be undefined. The constraints and results above are adapted from~\cite[Appendix A.3]{zilberstein2024relatively} to our framework.

\begin{definition}[Compatibility~\cite{zilberstein2024relatively}]
	The expressions $e_1$ and $e_2$ are compatible in semiring $A = \langle U, \oplus, \odot, \bzero, \bone \rangle$ if $\sem{e_1}{ \sigma} \oplus \sem{e_2}{ \sigma}$ is defined for any $\sigma \in \States$.
\end{definition}

\begin{proposition}
	If $e_1,e_2$ are compatible and $\eval{C_1},\eval{C_2}$ are total functions, then
	\[\eval{\NDCHOICE{\COMPOSE{\WEIGHT{e_1}}{C_1}}{\COMPOSE{\WEIGHT{e_2}}{C_2}}}\] is a total function.
\end{proposition}
\begin{proof}
	\begin{align*}
		& \sem{\NDCHOICE{\COMPOSE{\WEIGHT{e_1}}{C_1}}{\COMPOSE{\WEIGHT{e_2}}{C_2}}}{\sigma} \\ 
		& \eeq
		\sem{\COMPOSE{\WEIGHT{e_1}}{C_1}}{\sigma,\tau} \ooplus \sem{\COMPOSE{\WEIGHT{e_2}}{C_2}}{\sigma,\tau} \\ 
		& \eeq
		\bigoplus_{\iota\colon \sem{\WEIGHT{e_1}}{\sigma,\iota}\neq \bzero}~ \sem{\WEIGHT{e_1}}{\sigma,\iota}\odot \eval{C_1}(\iota,\tau)\\
		&\qquad\oplus
		\bigoplus_{\iota\colon \sem{\WEIGHT{e_2}}{\sigma,\iota}\neq \bzero}~ \sem{\WEIGHT{e_2}}{\sigma,\iota}\odot \eval{C_2}(\iota,\tau)
		\\
		& \eeq
		\bigoplus_{\iota\colon\eval{e_1}(\sigma)\odot \iverson{\sigma=\iota}\neq \bzero}~ \eval{e_1}(\sigma)\odot \iverson{\sigma=\iota}\odot \eval{C_1}(\iota,\tau)\\
		&\qquad\oplus
		\bigoplus_{\iota\colon\eval{e_2}(\sigma)\odot \iverson{\sigma=\iota}\neq \bzero}~\eval{e_2}(\sigma)\odot \iverson{\sigma=\iota}\odot \eval{C_2}(\iota,\tau)
		\\
		& \eeq
		\eval{e_1}(\sigma)\odot\eval{C_1}(\sigma,\tau)\oplus
		\eval{e_2}(\sigma)\odot \eval{C_2}(\sigma,\tau)
	\end{align*}
	which is well-defined by~\cite[Lemma A.5]{zilberstein2024relatively} (since $\sem{e_1}{ \sigma} \oplus \sem{e_2}{ \sigma}$ is well-defined).
\end{proof}
\begin{proposition}[Well-definedness of $\LOOP{C}{e}{e'}$]
	If $e,e'$ are compatible and $\eval{C}$ is a total function, then $\eval{\LOOP{C}{e}{e'}}$ is a total function.
\end{proposition}
\begin{proof}
	Let $\Phi_{C,e,e'}(X)(\sigma,\tau)=\sem{e}{\sigma}\odot\left(\bigoplus_{\iota\in\States}~ 
	\eval{C}(\sigma,\iota)\odot
	X (\iota, \tau)
   \right) \ooplus \sem{e'}{\sigma}\odot\iverson{\sigma=\tau}$. By~\cite[Lemma A.5]{zilberstein2024relatively}, $\Phi_{C,e,e'}(X)(\sigma,\tau)$ is well-defined, ensuring the well-definedness of $\eval{\LOOP{C}{e}{e'}}$ as well (as per~\Cref{prop:fixpoint-esistence-semantics}).
\end{proof}
\section{Nontermination and Unreachability}
\label{app:nontermination}

However, we can represent these situations using "angelic partial correctness" and "demonic total correctness" triples, respectively.
\begin{table}[h]
	\small
    \centering
	\begin{tabularx}{\textwidth}{|X|c|c|}       
		\hline
        \textbf{Triple}  &  \textbf{Property}\\  
		\hline
		$\apcl{P}{C}{\false}$ &  May-Nontermination\\
		\hline
        $\dtcl{P}{C}{\true}$ & Must-Termination\\
		\hline
		$\not\apcl{P}{C}{\false}$ & Must-Termination\\
		\hline
        $\not\dtcl{P}{C}{\true}$ &   May-Nontermination\\
		\hline
    \end{tabularx}
    \caption{Nontermination and unreachability.}
    \label{tab:other-non-termination}
\end{table}

for a reasonable definition of $\sem{\STAR{C}}{\sigma} \text{ may diverge}$ which we omit as this is not the main focus of the paper.

As angelic total correctness triples can be expressed by $\whpsymbol$, our calculus also subsume nontermination proving, i.e., the following holds:
\begin{align*}
	(\lambda \rho.~ P\cap \rho \neq \emptyset) \subseteq \whp{C}{\lambda \rho.~ P\cap \rho \neq \emptyset} & \implies \forall \sigma\in P\mydot \sem{\STAR{C}}{\sigma} \text{ may diverge}
\end{align*}

Whilst~\cite[Section 1, "Formal Interpretation of Divergent Triples"]{Raad24} focuses on a stronger interpretation of triples where $\tcl{P}{C}{\infty}$ means \emph{every} state $\sigma\in P$ have \emph{at least} a diverging trace, our framework allows to express \emph{three} novel interpretation as well. We start with the weaker interpretation that mandates the existence of at least one state in the precondition that may diverge.
\begin{align*}
	\{ P\} \subseteq \whp{C}{\lambda \rho.P \cap \rho \neq \emptyset} & \implies \exists \sigma\in P\mydot \sem{\STAR{C}}{\sigma} \text{ may diverge}
\end{align*}
which can be rewritten as a program logics, using~\Cref*{tab:disprove-logics}
\begin{align*}
	\infer[]{
		\exists \sigma\in P\mydot \sem{\STAR{C}}{\sigma} \text{ may diverge}
			}{
		\not\pcl{P}{C}{\neg P}
			}
\end{align*}
It's not surprising that the premise involves the falsification of a triple since the objective is to establish an $\exists$ property. It's worth noting that we can always convert it back to a valid triple in some other logics through ~\Cref*{cor:falsify-triples}. However, we choose not to do so, as it would introduce an additional quantifier.

For the remaining two interpretations, we will focus on what we term \emph{must divergence}. Unlike \emph{may divergence}, \emph{must divergence} asserts that all traces originating from a given initial state must diverge. We highlight the inadequacy of $\STAR{C}$ due to its semantics implicitly assuming that divergence should never be necessary. Consequently, our subsequent exploration will revolve around $\WHILEDO{\guard}{C}$, and we will present rules for all four interpretations.

First all, we show the nontermination rules for $\WHILEDO{\guard}{C}$ via $\whpsymbol$.
\begin{align*}
	P\subseteq \guard \qand (\lambda \rho.~ P\cap \rho \neq \emptyset) \subseteq \whp{C}{\lambda \rho.~ P\cap \rho \neq \emptyset} & \implies \forall \sigma\in P\mydot \sem{\WHILEDO{\guard}{C}}{\sigma} \text{ may diverge}\\
	P\subseteq \guard \qand \{ P\} \subseteq \whp{C}{\lambda \rho.P \cap \rho \neq \emptyset} & \implies \exists \sigma\in P\mydot \sem{\WHILEDO{\guard}{C}}{\sigma} \text{ may diverge}\\
	P\subseteq \guard \qand \{P\} \subseteq \whp{C}{\lambda \rho.~ \rho\subseteq P}& \implies \forall \sigma\in P\mydot \sem{\WHILEDO{\guard}{C}}{\sigma} \text{ must diverge}\\
	P\subseteq \guard \qand \exists \sigma \in P \mydot  \{\{\sigma\}\}\subseteq \whp{C}{\lambda \rho.~ \rho\subseteq P} & \implies \exists \sigma\in P\mydot \sem{\WHILEDO{\guard}{C}}{\sigma} \text{ must diverge}\\
\end{align*}

These can be straightforwardly converted into rules for program logics.

\begin{align*}
	\infer[]{
	\forall \sigma\in P\mydot \sem{\WHILEDO{\guard}{C}}{\sigma} \text{ may diverge}		}{
			\tcl{P}{C}{P}\qquad P\subseteq \guard 
		}
	&\quad
	\infer[]{
		\exists \sigma\in P\mydot \sem{\WHILEDO{\guard}{C}}{\sigma} \text{ may diverge}
			}{
		\not\pcl{P}{C}{\neg P} \qquad P\subseteq \guard 
			}\\
	\infer[]{
			\forall \sigma\in P\mydot \sem{\WHILEDO{\guard}{C}}{\sigma} \text{ must diverge}
		}{
			\pcl{P}{C}{P} \qquad P\subseteq \guard
	}
	&\quad 
	\infer[]{
		\exists \sigma\in P\mydot \sem{\WHILEDO{\guard}{C}}{\sigma} \text{ must diverge}
			}{
		\not\tcl{P}{C}{\neg P} \qquad P\subseteq \guard 
			}
\end{align*}
The duality in this context is twofold: moving from left to right, total correctness aligns with the falsification of partial correctness (by~\Cref*{cor:falsify-triples}, essentially capturing the duality between $\forall$ and $\exists$). On the other hand, from top to bottom, the duality is determined by the choices made in our interpretation of nondeterminism and bears resemblance to the one highlighted in~\cite{ZK22}.

As pointed in~\Cref*{tab:other-non-termination}, angelic partial correctness and demonic total correctness have a key role in proving may-nontermination and must-termination. It is thus surprising that~\cite{Raad24} chose to combine (angelic) total correctness and total incorrectness logics for their sound and complete proof system that allows to prove may-nontermination.

In this section, we show how a standard angelic partial correctness proof system relates with the rules in~\cite{Raad24}. We consider guarded imperative languages with nondeterministic choices (i.e., with while constructs instead of Kleene star), and the rules for angelic partial correctness as analogous to those for standard partial correctness, except for the nondeterministic choice~\cite[Definition 4.5]{benni_diss}. In particular, it is well known that by coinduction, the following rule holds:
\begin{align*}
	\infer[]{
		\apcl{P}{\WHILEDO{\guard}{C}}{\neg \guard \land P}}{
			\apcl{P\land \guard}{C}{P}
			}
\end{align*}

We shall observe that angelic partial correctness is a complete proof system (for guarded imperative languages), and this already means that every may-nontermination triple can be proved. However, let us show how we can derive simpler rules (analogous to those in~\cite{Raad24}) without the need to add explicit rules for may-nontermination.

\begin{theorem}
	The following rules are valid in angelic partial correctness logic:
	\begin{align*}
		\infer[]{
			\apcl{P}{\COMPOSE{C_1}{C_2}}{\false}}{
				\apcl{P}{C_1}{\false}
				}
		&\quad 		
		\infer[]{
			\apcl{P}{\COMPOSE{C_1}{C_2}}{\false}}{
				\apcl{P}{C_1}{Q} \quad \apcl{Q}{C_2}{\false}
			}\\
		\infer[]{
				\apcl{P}{\NDCHOICE{C_1}{C_2}}{\false}}{
					\apcl{P}{C_i}{\false} \text{ for some } i\in\{1,2\}
				}
		&\quad \infer[]{
			\apcl{P\land \guard}{\WHILEDO{\guard}{C}}{\false}}{
				\apcl{P\land \guard}{C}{P\land \guard}
				}\\
	\end{align*}
\end{theorem}

The rules above resemble to those in~\cite{Raad24}, but again we stress that here we are not developing a new complex logic. It is also easy to show that the loop rule for while loops in~\cite{Raad24} can be very easily proved:
\begin{align*}
	\infer[]{
			\apcl{P\land \guard}{\WHILEDO{\guard}{C}}{\false}}{
				\infer[]{
					\apcl{P\land \guard}{C}{P\land \guard}}{
						\tcl{P\land \guard}{C}{P\land \guard}
						}
				}
\end{align*}

\subsection{Nontermination and Unreachability}
It's worth noting that in all four rules, we are concerned with correctness triples rather than incorrectness ones. This emphasis is due to our focus on the termination of the forward semantics. Analogous rules for partial incorrectness and total incorrectness triples would facilitate the identification of nonterminating states in the backward semantics. For instance, we can establish:
\begin{align*}
	\infer[]{
		\forall \sigma\in P\mydot \invsem{\STAR{C}}{\sigma} \text{ may diverge}		}{
				\til{P}{C}{P}
			}
	&\quad 
	\infer[]{
		\exists \sigma\in  P\mydot \invsem{\STAR{C}}{\sigma} \text{ may diverge}
			}{
		\not\pil{\neg P}{C}{ P}
			}
\end{align*}
The rules can be used in the context of program inversion to assess whether one could compute the pre-image by simply executing the inverted program.

The correlation between nontermination and unreachability, as highlighted in~\cite{ZK22}, may lead one to question whether proving states as unreachable is related to demonstrating nontermination. However, when considering backward semantics, a single nonterminating trace doesn't provide enough information to establish unreachability. It is essential for all backward traces to be nonterminating, aligning with the concept of must-termination in backward semantics, precisely corresponding to what is conventionally meant by unreachability. This insight strengthens the connection described in~\cite{ZK22}, where their dualities between nontermination and unreachability arise from the resolution of nondeterministic choices. In other words, when~\cite{ZK22} refers to nontermination, they essentially mean must-nontermination.

\paragraph*{Backward Must-Nontermination}
Again, when reasoning about \emph{must-nontermination} on $\STAR{C}$, it is trivially false for the backward semantics as well. To make it worse, we argue that it is trivial for $\WHILEDO{\guard}{C}$ as well: if our final state $\tau\mmodels\guard$, then it is clearly unreachable and otherwise it is reachable (in 0 iterations).

\section{Full calculations and examples omitted from the main text}
\label{app:examples}

\subsection{Full calculations of~\Cref{ex:variance}}
To compute $\whp{\LOOP{\ASSIGN{x}{x+1}}{\frac 1 2}{\frac 1 2}}{\E[x^2]}$, we compute subsequent Kleene's iterates obtaining:

\begin{align*}
	W_{ 0.5}^0 (\E[x^2]\odot 0.5)&\eeq \E[x^2]\odot 0.5\\
	W_{ 0.5}^1 (\E[x^2]\odot 0.5)&\eeq 
	\whp{\ASSIGN{x}{x+1}}{\E[x^2]\odot 0.5} \odot 0.5 \eeq \E[(x+1)^2]\odot 0.5^2\\
	W_{ 0.5}^2 (\E[x^2]\odot 0.5)&\eeq \E[(x+2)^2]\odot 0.5^3\\
	&\vdots\\
	W_{ 0.5}^n (\E[x^2]\odot 0.5)&\eeq \E[(x+n)^2]\odot 0.5^{n+1}
\end{align*}%
This leads to:
\begin{align*}
	&\whp{\LOOP{\ASSIGN{x}{x+1}}{\frac 1 2}{\frac 1 2}}{\E[x^2]} \\
	&\eeq \bigoplus_{n\in\Nats}~W_{0.5}^n (\E[x^2]\odot 0.5)
	\\
	&\eeq \bigoplus_{n\in\Nats}~\E[(x+n)^2]\odot 0.5^{n+1}\\
\end{align*}%

\noindent
To compute $\whp{\LOOP{\ASSIGN{x}{x+1}}{\frac 1 2}{\frac 1 2}}{\E[x]}$, we compute subsequent Kleene's iterates obtaining:

\begin{align*}
	W_{ 0.5}^0 (\E[x]\odot 0.5)&\eeq \E[x]\odot 0.5\\
	W_{ 0.5}^1 (\E[x]\odot 0.5)&\eeq 
	\whp{\ASSIGN{x}{x+1}}{\E[x]\odot 0.5} \odot 0.5 \eeq \E[x+1]\odot 0.5^2\\
	W_{ 0.5}^2 (\E[x]\odot 0.5)&\eeq \E[x+2]\odot 0.5^3\\
	&\vdots\\
	W_{ 0.5}^n (\E[x]\odot 0.5)&\eeq \E[x+n]\odot 0.5^{n+1}
\end{align*}%
This leads to:
\begin{align*}
	&\whp{\LOOP{\ASSIGN{x}{x+1}}{\frac 1 2}{\frac 1 2}}{\E[x]^2} \\
	&\eeq 
	\big(\bigoplus_{n\in\Nats}~W_{0.5}^n (\E[x]\odot 0.5)\big)^2
	\\
	&\eeq 
	\big(\bigoplus_{n\in\Nats}~\E[x+n]\odot 0.5^{n+1}\big)^2
\end{align*}%

\subsection{Conditional expected values}
You decide to play a coin-toss game where winning yields $1$, and losing results in a loss of $5$. You plan ahead by adding specially crafted fake coins to your pocket that guarantee a win when tossed. In addition, you ensure you have some genuine fair coins to display to your opponent. How many coins must be in your pocket (at least) to have a non-negative expected return?%
\begin{align*}
	& \annotate{\iverson{c=0}\cdot\E[1]+\iverson{c\neq 0}\cdot(\frac{1}{2}\E[-5]+\frac{1}{2}\E[1])}\\
	& \IF{\mathit{c}=0}\\
	& \qquad \annotate{\E[1]}\\
	& \qquad \ASSIGN{x}{1}\\
	& \qquad \annotate{\E[x]}\\
	& \ELSE \\
	& \qquad \annotate{\frac{1}{2}\E[-5]+\frac{1}{2}\E[1]}\\
	& \qquad \PCHOICE{\ASSIGN{x}{-5}}{\frac{1}{2}}{\ASSIGN{x}{1}}\\
	& \qquad \annotate{\E[x]}\\
	& \}\\
	& \annotate{\E[x]}
\end{align*}%
With an input boolean variable $c$ we represent whether we have a fair or a fake coin. We represent the game with the simple program $C$ above and compute $\whp{C}{\E[x]}$ which yields the expected return for a given input distribution. We observe that the shape of the input distribution must be $\mu=\frac{n-1}{n}\cdot\iversone{c=0}+\frac{1}{n}\cdot\iversone{c=1}$ and solve:
$\whp{C}{\E[x]}(\mu)\geq 0$, leading to:%
\begin{align*}
	&\whp{C}{\E[x]}(\frac{n-1}{n}\cdot\iversone{c=0}+\frac{1}{n}\cdot\iversone{c=1})\geq 0\\
	&\big( \iverson{c=0}\cdot\E[1]
	+ \iverson{c\neq 0}\cdot(\frac{1}{2}\E[-5]+\frac{1}{2}\E[1]) \big)
	\big(\frac{n-1}{n}\cdot\iversone{c=0}+\frac{1}{n}\cdot\iversone{c=1}\big)\geq 0\\
	&\big( \iverson{c=0}\cdot\E[1] \big)
	\big(\frac{n-1}{n}\cdot\iversone{c=0}+\frac{1}{n}\cdot\iversone{c=1}\big)
	+
	\big(\iverson{c\neq 0}\cdot(\frac{1}{2}\E[-5]+\frac{1}{2}\E[1]) \big)
	\big(\frac{n-1}{n}\cdot\iversone{c=0}+\frac{1}{n}\cdot\iversone{c=1}\big)\geq 0\\
	&
	\frac{n-1}{n}
	-\frac{2}{n}\geq 0\\
	&
	\frac{n-3}{n}\geq 0\\
	&
	n\geq 3
\end{align*}%
The result obtained, implies that you need at least 3 coins in your pocket (at least two fake coins and one fair coin) to guarantee a non-negative expected return in this coin-toss game.

\end{document}